\documentclass[aip,jcp,amsmath,amssymb,preprint,floatfix]{revtex4-1}
\allowdisplaybreaks

\usepackage{graphicx}
\usepackage{bm}
\usepackage[utf8]{inputenc}
\usepackage[T1]{fontenc}
\usepackage{mathptmx}
\usepackage{etoolbox}

\usepackage{tikz}
\usepackage[caption=false]{subfig}
\usepackage{booktabs}
\usepackage{amsthm}
\newtheorem{theorem}{Theorem}
\newtheorem{corollary}[theorem]{Corollary}
\theoremstyle{remark}
\newtheorem{remark}{Remark}

\makeatletter
\def\@email#1#2{%
  \endgroup
  \patchcmd{\titleblock@produce}
    {\frontmatter@RRAPformat}
    {\frontmatter@RRAPformat{\produce@RRAP{*#1\href{mailto:#2}{#2}}}\frontmatter@RRAPformat}
    {}{}
}
\makeatother

\begin{document}

\preprint{AIP/123-QED}

\title[]{A Thermodynamically Consistent Model of a Rechargeable Salt-in-Polyelectrolyte Battery\footnote{\textbf{Government Rights Statement}
This manuscript has been authored by UT-Battelle, LLC, under contract DE-AC05-00OR22725 with the US
Department of Energy (DOE). The US government retains and the publisher, by accepting the article
for publication, acknowledges that the US government retains a nonexclusive, paid-up, irrevocable,
worldwide license to publish or reproduce the published form of this manuscript, or allow others to
do so, for US government purposes. DOE will provide public access to these results of federally
sponsored research in accordance with the DOE Public Access Plan
(https://www.energy.gov/doe-public-access-plan).
}}
\author{Charlotte Beckford}
\affiliation{Mathematics Department, University of Tennessee, Knoxville, TN 37996
}
\author{Rajeev Kumar%
\footnote{Corresponding author: kumarr@ornl.gov}}
\affiliation{Center for Nanophase Materials Sciences, Oak Ridge National Laboratory, Oak Ridge, TN 37831
}%

\author{Steven M. Wise\footnote{Corresponding author: swise1@utk.edu}}
\affiliation{Mathematics Department, University of Tennessee, Knoxville, TN 37996
}%

\date{\today}

\begin{abstract}
  We present an isothermal, thermodynamically consistent model of a salt-in-polyelectrolyte battery
  with intercalation and deintercalation reactions of cations occurring at the electrode-electrolyte
  interfaces under galvanostatic conditions. For the electrolyte, we consider polyelectrolytes,
  which are polyanions, with neutralizing cations (i.e., counterions) such as lithium, along with a
  monovalent salt containing the same cations. Models of batteries containing either a
  dual-ion-conductor (DIC) electrolyte (i.e., uncharged polymers with added salt) or a
  single-ion-conductor (SIC) electrolyte (i.e., charged polymers without added salt) emerge as
  limiting cases of this thermodynamically consistent model. Once the model is presented in its most
  general context, we make specific choices so that it captures well-known mechanisms. In
  particular, the governing diffusion equations recover Fick's law of diffusion, and the
  cation-transfer reaction rate is described by the Butler--Volmer equation. The DIC and SIC
  limiting cases provide a framework for comparing the corresponding batteries. One-dimensional
  DIC and SIC charge--discharge simulations at matched current and duration illustrate
  concentration redistribution, polarization, and electrode utilization. Numerical accuracy
  checks quantify sensitivity to discretization; distinct initial inventories qualify the comparison.
\end{abstract}

\maketitle

\section{Introduction}
\label{sec:intro}

There is an increased demand for reliable, safe, and efficient renewable energy capture and storage
options. Improving battery performance metrics is a key pathway to achieving this
goal~\cite{deguenon2023overcoming}. Batteries using polymers as electrolytes are particularly
promising due to their enhanced performance and safety
features~\cite{chattopadhyay2023applications}. Two types of polymer electrolytes, broadly
categorized as dual-ion conductors (DICs) and single-ion conductors (SICs), have been studied
extensively for their use in solid-state batteries. 
This paper presents a
unified, thermodynamically consistent model that includes electrolyte transport, electrodes, and
interfacial reaction kinetics, with DIC and SIC systems arising as limiting cases. Specifically, we
present a thermodynamically consistent, isothermal, physics-based model of a solid
salt-in-polyelectrolyte rechargeable battery that adequately captures the complex chemical reactions
which occur at electrode-electrolyte interfaces and transport in polymer electrolyte. DIC and SIC
electrolyte batteries are represented by limiting cases of this model. This work contributes to
efforts for improving polymer electrolyte batteries. To contextualize this work, we briefly review
major frameworks for electrochemical modeling of rechargeable batteries based on polymer
electrolytes. We emphasize the simplifying assumptions employed and the frequent absence of a clear
demonstration of thermodynamic consistency.

Batteries based on DICs contain polymer electrolytes, which are neutral polymers like poly(ethylene
oxide) mixed with a salt. Charge transport in such polymer electrolytes is characterized by the
motion of cations (such as lithium) and anions of the salt, both of which contribute significantly.
In addition, cations like lithium undergo ion and electron transfer reactions at each electrode. The
first model for a rechargeable polymer battery based on DICs was developed by Doyle, Fuller, and
Newman~\cite{doyle1993modeling} in 1993 by extending a model for dilute liquid electrolytes by West
and co-workers~\cite{west1982modeling}. Charge and discharge of a battery containing a lithium
anode, polymer with added lithium salt, and a composite cathode (modeled as a porous electrode) was
considered in the presence of a constant electrical current (i.e., under galvanostatic conditions).
The transport in the DIC polymer electrolyte was modeled by using concentrated solution theory,
grounded in linear irreversible
thermodynamics~\cite{groot1962non,Doi_2011,kumar2017rayleighian,kumar2024dynamic}, after invoking
local electroneutrality so that salt concentration instead of the concentration of cations and
anions appears as a spatiotemporal variable~\cite{newman2021electrochemical}. Rates of reaction at
each electrode were modeled using Butler-Volmer kinetics after making certain assumptions about the
dependencies of the overpotentials on the lithium concentration in the electrodes. Later, the model
was extended by treating the anode as another porous electrode~\cite{fuller1994simulation} and
compared with experimental data for lithium-ion batteries based on DICs~\cite{doyle1996comparison}.
Since then, these and similar models have been extended and
reviewed~\cite{bazant2005current,bazant2013theory,chu2005electrochemical,danilov2010modeling,landstorfer2011advanced,pang2019experimental,woeppel2020single,brosa2022continuum,mehta2024mathematical,yildiz2024comparative,alkhedher2024electrochemical,habscheid2025finite}.
Predictions of these models for the limiting current are shown to agree with experiments for
batteries containing DICs~\cite{pesko2018comparing,shah2019comparing}.

In contrast, the modeling of rechargeable polymer batteries containing SICs such as polyanions with
lithium cations is still in an infancy stage, especially under galvanostatic conditions. The
galvanostatic models for the DICs cannot be reduced and applied to the SICs in a straightforward
manner due to technical issues such as the need to go beyond the local electroneutrality
approximation and the invalidity of transference number as an independent transport
coefficient~\cite{kumar2024dynamic}. We are aware of one attempt~\cite{wolff2018model} to derive a
model for SICs under galvanostatic conditions in an \textit{ad hoc} manner, where local
electroneutrality was assumed in the electrolyte.
Thermodynamically consistent models for SICs have been presented in
References~\cite{kumar2017rayleighian,kumar2020capacitance,kumar2024dynamic}. However, these models
did not consider the electrodes and the reaction kinetics at each electrode. The main aim of this
work is to incorporate electrodes and reaction kinetics into a thermodynamically consistent model
for rechargeable polymer batteries, thereby allowing DICs and SICs to be compared.

To achieve our aim, we have developed a new model for a rechargeable battery containing
polyelectrolytes (ionic polymers) with added salt i.e., salt-in-polyelectrolyte. The model can be
reduced to study batteries based on DICs and SICs by considering appropriate limits, which are
described below. {We establish thermodynamic consistency by deriving an isothermal free-energy
balance in which the rate of change of the system free energy equals the negative internal
dissipation minus the electrical power delivered through the terminals. The internal dissipation is
nonnegative, while the system free energy may increase when electrical work is supplied during
charging.} The model shows that the Butler-Volmer equations for the rates of cation transfer at each
electrode appear as a specialized case of a generalized reaction
rate~\cite{dickinson2019butler,zhang2025lithium}.

The paper is organized as follows. In Section~\ref{sec:model}, we present the model and prove it is
thermodynamically consistent and can recapture well-known mechanisms. In
Section~\ref{sec:nonDimSimplify}, we recapitulate and nondimensionalize the model. In
Section~\ref{sec:limiting-cases}, we relate the DIC and SIC limiting cases to prior modeling
attempts. In Section~\ref{sec:numerical-simulations}, we present one-dimensional DIC and SIC
charge--discharge cycles at dimensionless current magnitude $1.6$ with matched durations,
examine concentration profiles, polarization, and electrode utilization, and assess
conservation and numerical accuracy while accounting for their different initial inventories.
Finally, we summarize our findings in Section~\ref{sec:conclusion}.

\section{Model}
\label{sec:model}


\begin{figure*}
  \centering
  \subfloat[]{%
  \label{fig:salt_in_polymer}%
  \begin{minipage}[b]{0.32\textwidth}
    \centering
    \begin{tikzpicture}[
    >=stealth,
    font=\normalsize,
    every node/.style={font=\normalsize}
]

\def\height{7.0}
\def\intL{0}
\def\intR{4}
\def\amp{0.25}
\def\freq{2}

\tikzset{
    boundary/.style={line width=1.2pt},
    interface/.style={line width=1.2pt, draw=gray, dashed},
    ionplus/.style={circle, draw=black, line width=1pt, minimum size=1cm},
    ionminusB/.style={circle, draw=gray, line width=1pt, minimum size=14pt},
    ionminusF/.style={circle, draw=black, line width=1pt, minimum size=14pt},
    vacancy/.style={circle, draw=black,line width=1pt, minimum size=1cm}
}

\draw[boundary] (\intL,0) -- (\intR,0);
\draw[boundary] (\intL,\height) -- (\intR,\height);

\draw[boundary] plot[domain=0:\height, samples=150]
    ({\intL + \amp*sin(360*\x/\height*\freq)}, \x);

\draw[boundary] plot[domain=0:\height, samples=150]
    ({\intR + \amp*sin(360*\x/\height*\freq + 180)}, \x);

\draw[interface, domain=0.5:3.5, samples=200]
  plot ({\x}, {0.4*\x + 0.15*sin(360*\x) + 1.8});

\foreach \x in {1.0,3.0}
{
    \node[ionminusB]
    at ({\x}, {0.4*\x+ 0.15*sin(360*\x)  + 2.4})
    {$\mathbf{-}$};
}

\node[ionplus]  at (1.2, 6.4) {Li$^+$};
\node[ionplus] at (3, 4.9) {Li$^{+}$};
\node[ionplus]  at (2, 4.0) {Li$^{+}$};

\node[ionminusF] at (2.5, 2) {$\mathbf{-}$};

\node[vacancy] at (3.0,6.2) {$\mathrm{V}_\mathrm{M}$};
\node[vacancy] at (1.7,1.5) {$\mathrm{V}_\mathrm{M}$};
\node[vacancy] at (0.9,5.0) {$\mathrm{V}_\mathrm{M}$};

\node at (2.0,0.8) {Electrolyte};
\node at (2.0,0.3) {$\Omega_\mathrm{M}$};

\end{tikzpicture}
  \end{minipage}}
  \hfill
  \subfloat[]{%
  \label{fig:DIC_Avenue}%
  \begin{minipage}[b]{0.32\textwidth}
    \centering
    \begin{tikzpicture}[
    >=stealth,
    font=\normalsize,
    every node/.style={font=\normalsize}
]

\def\height{7.0}
\def\intL{0}
\def\intR{4}
\def\amp{0.25}
\def\freq{2}

\tikzset{
    boundary/.style={line width=1.2pt},
    interface/.style={line width=1.2pt, draw=gray, dashed},
    ionplus/.style={circle, draw=black, line width=1pt, minimum size=1cm},
    ionminusB/.style={circle, draw=gray, line width=1pt, minimum size=14pt},
    ionminusF/.style={circle, draw=black, line width=1pt, minimum size=14pt},
    vacancy/.style={circle, draw=black,line width=1pt, minimum size=1cm}
}

\draw[boundary] (\intL,0) -- (\intR,0);
\draw[boundary] (\intL,\height) -- (\intR,\height);

\draw[boundary] plot[domain=0:\height, samples=150]
    ({\intL + \amp*sin(360*\x/\height*\freq)}, \x);

\draw[boundary] plot[domain=0:\height, samples=150]
    ({\intR + \amp*sin(360*\x/\height*\freq + 180)}, \x);

\draw[interface, domain=0.5:3.5, samples=200]
  plot ({\x}, {0.4*\x + 0.15*sin(360*\x) + 1.8});


\node[ionplus]  at (1.2, 6.4) {Li$^+$};

\node[ionminusF] at (2.5, 2) {$\mathbf{-}$};

\node[vacancy] at (3.0,6.2) {$\mathrm{V}_\mathrm{M}$};
\node[vacancy] at (1.7,1.5) {$\mathrm{V}_\mathrm{M}$};
\node[vacancy] at (0.9,5.0) {$\mathrm{V}_\mathrm{M}$};

\node at (2.0,0.8) {Electrolyte};
\node at (2.0,0.3) {$\Omega_\mathrm{M}$};

\end{tikzpicture}
  \end{minipage}}
  \hfill
  \subfloat[]{%
  \label{fig:SIC_Avenue}%
  \begin{minipage}[b]{0.32\textwidth}
    \centering
    \begin{tikzpicture}[
    >=stealth,
    font=\normalsize,
    every node/.style={font=\normalsize}
]

\def\height{7.0}
\def\intL{0}
\def\intR{4}
\def\amp{0.25}
\def\freq{2}

\tikzset{
    boundary/.style={line width=1.2pt},
    interface/.style={line width=1.2pt, draw=gray, dashed},
    ionplus/.style={circle, draw=black, line width=1pt, minimum size=1cm},
    ionminusB/.style={circle, draw=gray, line width=1pt, minimum size=14pt},
    ionminusF/.style={circle, draw=black, line width=1pt, minimum size=14pt},
    vacancy/.style={circle, draw=black,line width=1pt, minimum size=1cm}
}

\draw[boundary] (\intL,0) -- (\intR,0);
\draw[boundary] (\intL,\height) -- (\intR,\height);

\draw[boundary] plot[domain=0:\height, samples=150]
    ({\intL + \amp*sin(360*\x/\height*\freq)}, \x);

\draw[boundary] plot[domain=0:\height, samples=150]
    ({\intR + \amp*sin(360*\x/\height*\freq + 180)}, \x);

\draw[interface, domain=0.5:3.5, samples=200]
  plot ({\x}, {0.4*\x + 0.15*sin(360*\x) + 1.8});

 \foreach \x in {1.0,3.0}
 {
     \node[ionminusB]
     at ({\x}, {0.4*\x+ 0.15*sin(360*\x)  + 2.4})
     {$\mathbf{-}$};
 }

\node[ionplus] at (3, 4.9) {Li$^{+}$};
\node[ionplus]  at (2, 4.0) {Li$^{+}$};

\node[vacancy] at (3.0,6.2) {$\mathrm{V}_\mathrm{M}$};
\node[vacancy] at (1.7,1.5) {$\mathrm{V}_\mathrm{M}$};
\node[vacancy] at (0.9,5.0) {$\mathrm{V}_\mathrm{M}$};

\node at (2.0,0.8) {Electrolyte};
\node at (2.0,0.3) {$\Omega_\mathrm{M}$};

\end{tikzpicture}
  \end{minipage}}
  \caption{Various idealized electrolyte configurations. (a) The electrolyte is a mixture of polymer
  chains (polyanions), their counterions (cations), added salt, and vacancies. Electrolyte vacancies
  are represented as well; they are labeled with $\mathrm{V}_\mathrm{M}$. (b) In the DIC limiting
  case, the polymer is uncharged, so there are no counterions of the polymer. (c) In the SIC
  limiting case, the polymer chain is a polymerized anion and there is no added salt. Cations and anions are not drawn to scale.}
  \label{fig:1}
\end{figure*}
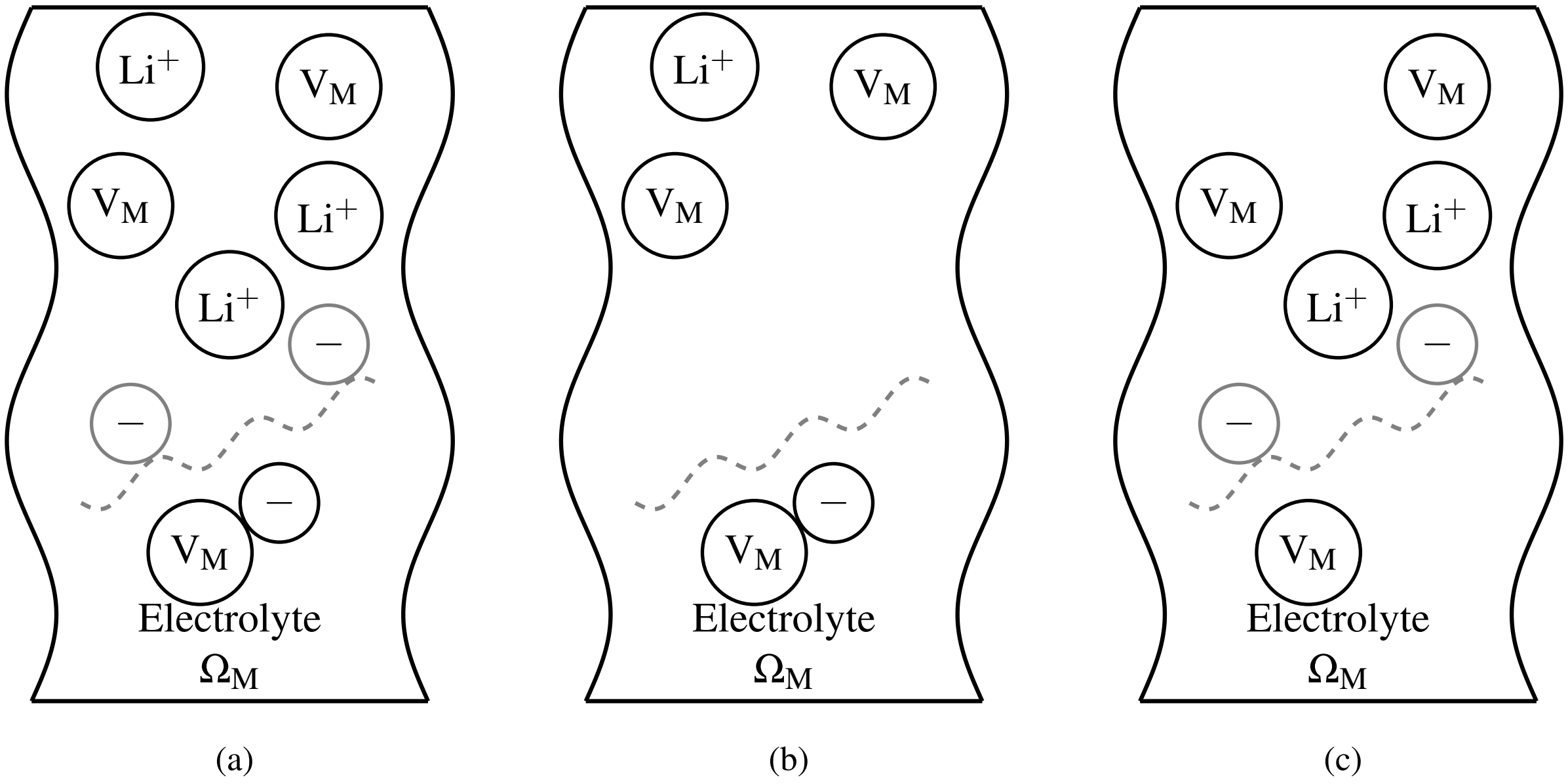

Our idealized salt-in-polyelectrolyte contains negatively charged polymers (polyanions), their
counterions (cations), and added salt (see Figure~\ref{fig:salt_in_polymer}). We restrict the class
of polymers included in our model such that when they are charged, their counterions are identical
to the cation that composes the salt. For example, consider the case of lithium
bis(tri\-fluoro\-methane\-sulfonyl)imide (LiTFSI) salt being added to a lithium conducting polymer
like lithium poly(methacrylate
bis(tri\-fluoro\-methane\-sulfonyl)imide)~\cite{zhu2024using,singh2025revisiting,li2025emergent}. Each polymer chain consists of $N$ statistical segments of signed valency $z_p$, which will be
chosen to be $0$ for DICs and -1 for SICs. {The electrolyte contains four physical
constituents---cations, anions, polymer segments, and vacancies---but only three independent
composition fields because local incompressibility determines the vacancy fraction algebraically.
Similarly, each electrode contains intercalated species and vacancies.} The inclusion of electrolyte vacancies is explained in
Section~\ref{subsec:modelFormulation}. We consider only one oxidation state of the electrolyte ionic
species. Since we also consider the intercalated metal in the electrodes formed by the cation being
reduced, we consider the monovalent case, so that valencies of cations and anions are chosen to be
$z_+ = 1$ and $z_- = -1$, respectively. Figure~\ref{fig:salt_in_polymer} depicts this system where
the polymer is a polymerized anion, that is, anions are covalently bonded to the polymer backbone.
Figure~\ref{fig:DIC_Avenue} depicts the DIC limiting case. To model a SIC battery, we consider the
limiting case where there is no added salt. This leaves only the negatively charged polymers (i.e.,
$z_p = -1$), its counterions (the cations), and the vacancies in the electrolyte.
Figure~\ref{fig:SIC_Avenue} depicts this limiting case where we have specified that the charged
polymer is a polymerized anion. Generalizations of the model to consider multivalent cations and
polycations for DICs and SICs, respectively, are straightforward but not attempted here.

\subsection{Model Formulation}
\label{subsec:modelFormulation}
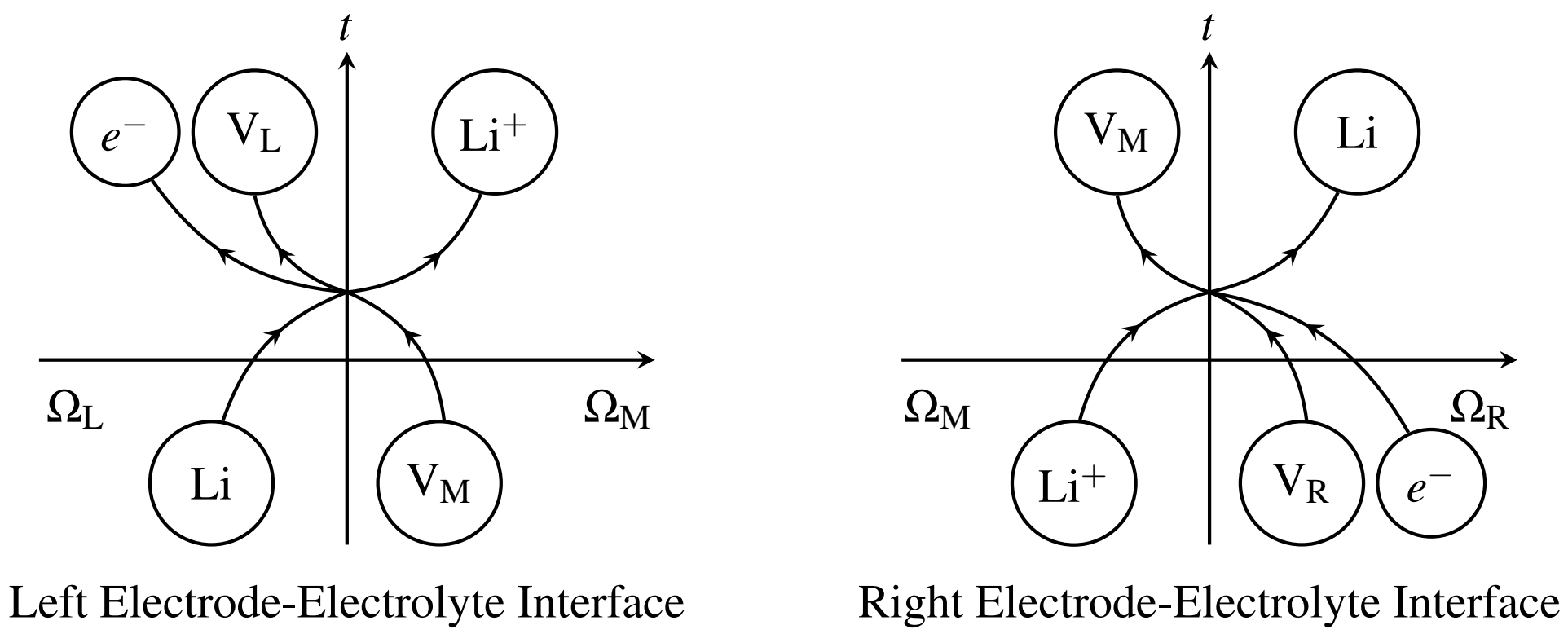
\begin{figure}[t!]
  \centering
  \usetikzlibrary{decorations.markings, arrows.meta}

\begin{tikzpicture}[>=stealth, thick]

\begin{scope}
\draw[->] (-2.5,0) -- (2.5,0);
\draw[->] (0,-1.5) -- (0,2.5);
\node at (0,2.7) {$t$};
\node at (-2.2,-0.4) {$\Omega_\mathrm{L}$};
\node at (2.2,-0.4) {$\Omega_\mathrm{M}$};

\node[draw,circle] (eL) at (-1.8,1.85) {$e^-$};
\node[draw,circle, minimum size=1cm] (vL) at (-0.75,1.85) {$\mathrm{V}_\mathrm{L}$};
\node[draw,circle, minimum size=1cm] (LiPlus) at (1.2,1.85) {Li$^+$};
\node[draw,circle, minimum size=1cm] (Li) at (-1.1,-1) {Li};
\node[draw,circle, minimum size=1cm] (vM) at (0.75,-1) {$\mathrm{V}_\mathrm{M}$};

\coordinate (J) at (0,0.55);

\tikzset{
  midarrow/.style={
    postaction={
      decorate,
      decoration={
        markings,
        mark=at position 0.6 with {\arrow{>}}
      }
    }
  }
}

\draw[thick, midarrow] (J) to[bend left=25] (eL);
\draw[thick, midarrow] (J) to[bend left=30] (vL);
\draw[thick, midarrow] (Li) to[bend left=25] (J);

\draw[thick, midarrow] (J) to[bend left=-30] (LiPlus);
\draw[thick, midarrow] (vM) to[bend left=-30] (J);

\node[black] at (0,-2) {Left Electrode-Electrolyte Interface};
\end{scope}

\begin{scope}[xshift=7cm]
\draw[->] (-2.5,0) -- (2.5,0);
\draw[->] (0,-1.5) -- (0,2.5);
\node at (0,2.7) {$t$};
\node at (-2.2,-0.4) {$\Omega_\mathrm{M}$};
\node at (2.2,-0.4) {$\Omega_\mathrm{R}$};

\node[draw,circle, minimum size=1cm] (vM) at (-0.75,1.85) {$\mathrm{V}_\mathrm{M}$};
\node[draw,circle, minimum size=1cm] (Li) at (1.2,1.85) {Li};
\node[draw,circle, minimum size=1cm] (LiPlus) at (-1.1,-1) {Li$^+$};
\node[draw,circle, minimum size=1cm] (vR) at (0.75,-1) {$\mathrm{V}_\mathrm{R}$};
\node[draw,circle] (eL) at (1.8,-1) {$e^-$};

\coordinate (J) at (0,0.55);

\tikzset{
  midarrow/.style={
    postaction={
      decorate,
      decoration={
        markings,
        mark=at position 0.6 with {\arrow{>}}
      }
    }
  }
}

\draw[thick, midarrow] (eL) to[bend left=-25] (J);
\draw[thick, midarrow] (J) to[bend left=30] (vM);
\draw[thick, midarrow] (J) to[bend left=-25] (Li);

\draw[thick, midarrow] (LiPlus) to[bend left=30] (J);
\draw[thick, midarrow] (vR) to[bend left=-30] (J);

\node[black] at (0,-2) {Right Electrode-Electrolyte Interface};
\end{scope}

\end{tikzpicture}
  \caption{Reaction scheme for the example of the forward reactions occurring at the left and right
  electrode-electrolyte interfaces during a discharge cycle. During a charge cycle, the reactions
  depicted by the schema would be the backward reactions.}
  \label{fig:reactionScheme}
\end{figure}

{We use negative and positive as fixed electrode labels: the left electrode is negative and the
right electrode is positive. The kinetic labels anode and cathode depend on operating mode, as
summarized in Table~\ref{tab:electrodeRoles}.}
\begin{table}

  \caption{\label{tab:electrodeRoles}Fixed electrode labels and operating-mode-dependent reaction
  roles.}
  \begin{ruledtabular}
    \begin{tabular}{ccc}
      Operating mode & Left/negative electrode & Right/positive electrode\\
      \midrule
      Discharge & Anode; oxidation & Cathode; reduction\\
      Charge & Cathode; reduction & Anode; oxidation
    \end{tabular}
  \end{ruledtabular}
\end{table}
We restrict the class of electrochemical reactions occurring at the electrode-electrolyte interfaces
to the intercalation and deintercalation of the cations (like Li$^+$) into and out of the
electrodes. These reactions necessitate the use of an additional species in the electrolyte and
electrodes which occupy the same amount of volume as the intercalated species and the cations,
respectively. We elect to use the simplest option in our model, electrode and electrolyte vacancies.
We denote the neutral vacancies in the left and right electrodes by $\mathrm{V}_\mathrm{L}$ and
$\mathrm{V}_\mathrm{R}$, respectively, and the vacancies in the electrolyte by
$\mathrm{V}_\mathrm{M}$; as shown in Figure~\ref{fig:reactionScheme}. For
example, let's consider the case of the LiTFSI salt added to a lithium conducting polymer during a
discharge cycle. The reactions at the left and right electrode-electrolyte interfaces, respectively,
would be
\begin{eqnarray}
  \mathrm{Li} + \mathrm{V}_\mathrm{M} \rightleftharpoons \mathrm{Li}^+ + e^- + \mathrm{V}_\mathrm{L}\phantom{,}\label{leftLiReaction}\\
  \mathrm{Li}^+ + e^- + \mathrm{V}_\mathrm{R}\rightleftharpoons
  \mathrm{Li} + \mathrm{V}_\mathrm{M}
  \label{rightLiReaction}.
\end{eqnarray} That is, during a discharge cycle the forward reaction at the left electrode-electrolyte interface is the oxidation of lithium to form lithium cation and the forward reaction at the right electrode-electrolyte interface is the reduction of lithium cation to form lithium. During charging, the expressions for the net rates $\nu_\mathrm d$ are the negative of those for the discharge-oriented reactions. The forward reaction scheme of this example reaction during a discharge cycle is in Figure~\ref{fig:reactionScheme}.

\begin{figure}[t!]
  \centering
  \usetikzlibrary{decorations.markings}

\begin{tikzpicture}[
    >=stealth,
    font=\normalsize,
    every node/.style={font=\normalsize}
]

\def\width{9}
\def\height{7}

\def\intL{2.5}
\def\intR{6.5}
\def\amp{0.35}
\def\freq{2}  

\draw[thick] (0,0) rectangle (\width,\height);

\draw[thick] plot[domain=0:\height, samples=100] 
    ({\intL + \amp*sin(360*\x/\height*\freq)}, \x);

\draw[thick] plot[domain=0:\height, samples=100] 
    ({\intR + \amp*sin(360*\x/\height*\freq + 180)}, \x);

\node[above] at (\width/2, \height) {\textit{periodic b.c.}};
\node[below] at (\width/2, 0) {\textit{periodic b.c.}};

\node at (-0.7, \height+0.3) {$\Gamma_\mathrm{L}$};
\draw[->, bend right=25] (-0.5, \height+0.1) to (0.05, \height-0.5);

\draw[thick, ->] (0, \height/2) -- (-0.8, \height/2);
\node[left] at (-0.8, \height/2) {$\boldsymbol{n}_\mathrm{L}$};

\node at (\width+0.7, \height+0.3) {$\Gamma_\mathrm{R}$};
\draw[->, bend left=25] (\width+0.5, \height+0.1) to (\width-0.05, \height-0.5);

\draw[thick, ->] (\width, \height/2) -- (\width+0.8, \height/2);
\node[right] at (\width+0.8, \height/2) {$\boldsymbol{n}_\mathrm{R}$};

\node[above] at ({\intL - 1.5}, \height-1) {$\Sigma_\mathrm{L}$};
\draw[->, bend left=20] ({\intL - 1.2}, \height-0.5) to ({\intL + \amp*sin(360*6.5/\height*\freq)}, 6.5);

\node[above] at ({\intR + 1.5}, \height-1.5) {$\Sigma_\mathrm{R}$};
\draw[->, bend right=20] ({\intR + 1.2}, \height-1) to ({\intR + \amp*sin(360*6.5/\height*\freq + 180)}, 6.5);

\node[align=center] at (1.25, \height/2-0.8) {Left};
\node[align=center] at (1.25, \height/2-1.3) {Electrode};
\node[align=center] at (1.25, \height/2-1.8) {(Anode)};
\node at (1.5, 1.0) {$\Omega_\mathrm{L}$};

\node at (\width/2, \height/2 - 1.3) {Electrolyte};
\node at (\width/2, 1.0) {$\Omega_\mathrm{M}$};

\node[align=center] at (7.75, \height/2-0.8) {Right};
\node[align=center] at (7.75, \height/2-1.3) { Electrode};
\node[align=center] at (7.75, \height/2-1.8) {(Cathode)};
\node at (7.75, 1.0) {$\Omega_\mathrm{R}$};

\node[draw, circle, inner sep=2pt] at (5, 5.2) {Li$^+$};
\draw[->] (5.4, 5.2) -- (5.9, 5.2);

\pgfmathsetmacro{\yNL}{5.2}
\pgfmathsetmacro{\omegaL}{360*\freq/\height}
\pgfmathsetmacro{\dxdyNL}{\amp*\omegaL*cos(\omegaL*\yNL)/57.2958}  
\pgfmathsetmacro{\xNL}{\intL + \amp*sin(\omegaL*\yNL)}
\pgfmathsetmacro{\magNL}{sqrt(1 + \dxdyNL*\dxdyNL)}
\pgfmathsetmacro{\nxNL}{1/\magNL}
\pgfmathsetmacro{\nyNL}{-\dxdyNL/\magNL}
\draw[thick, ->] (\xNL, \yNL) -- ({\xNL + 0.8*\nxNL}, {\yNL + 0.8*\nyNL});
\node[right] at ({\xNL + 0.85*\nxNL}, {\yNL + 0.85*\nyNL}) {$\boldsymbol{n}_\mathrm{L}$};

\pgfmathsetmacro{\yNMl}{3.0}
\pgfmathsetmacro{\dxdyNMl}{\amp*\omegaL*cos(\omegaL*\yNMl)/57.2958}
\pgfmathsetmacro{\xNMl}{\intL + \amp*sin(\omegaL*\yNMl)}
\pgfmathsetmacro{\magNMl}{sqrt(1 + \dxdyNMl*\dxdyNMl)}
\pgfmathsetmacro{\nxNMl}{-1/\magNMl}
\pgfmathsetmacro{\nyNMl}{\dxdyNMl/\magNMl}
\draw[thick, ->] (\xNMl, \yNMl) -- ({\xNMl + 0.8*\nxNMl}, {\yNMl + 0.8*\nyNMl});
\node[left] at ({\xNMl + 0.85*\nxNMl}, {\yNMl + 0.85*\nyNMl}) {$\boldsymbol{n}_\mathrm{M}$};

\pgfmathsetmacro{\yNMr}{5.2}
\pgfmathsetmacro{\dxdyNMr}{\amp*\omegaL*cos(\omegaL*\yNMr + 180)/57.2958}
\pgfmathsetmacro{\xNMr}{\intR + \amp*sin(\omegaL*\yNMr + 180)}
\pgfmathsetmacro{\magNMr}{sqrt(1 + \dxdyNMr*\dxdyNMr)}
\pgfmathsetmacro{\nxNMr}{1/\magNMr}
\pgfmathsetmacro{\nyNMr}{-\dxdyNMr/\magNMr}
\draw[thick, ->] (\xNMr, \yNMr) -- ({\xNMr + 0.8*\nxNMr}, {\yNMr + 0.8*\nyNMr});
\node[right] at ({\xNMr + 0.85*\nxNMr}, {\yNMr + 0.85*\nyNMr}) {$\boldsymbol{n}_\mathrm{M}$};
\pgfmathsetmacro{\yNR}{3.0}
\pgfmathsetmacro{\dxdyNR}{\amp*\omegaL*cos(\omegaL*\yNR + 180)/57.2958}
\pgfmathsetmacro{\xNR}{\intR + \amp*sin(\omegaL*\yNR + 180)}
\pgfmathsetmacro{\magNR}{sqrt(1 + \dxdyNR*\dxdyNR)}
\pgfmathsetmacro{\nxNR}{-1/\magNR}
\pgfmathsetmacro{\nyNR}{\dxdyNR/\magNR}
\draw[thick, ->] (\xNR, \yNR) -- ({\xNR + 0.8*\nxNR}, {\yNR + 0.8*\nyNR});
\node[left] at ({\xNR + 0.85*\nxNR}, {\yNR + 0.85*\nyNR}) {$\boldsymbol{n}_\mathrm{R}$};

\draw[thick, rounded corners=10pt]
(0, \height/2-1)
-- (-1.7, \height/2-1)
-- (-1.7, \height+1.3)
-- (\width+1.7, \height+1.3)
-- (\width+1.7, \height/2-1)
-- (\width, \height/2-1);

 \draw[->, thick] (0, \height/2-1) -- (-1.2, \height/2-1);
 \draw[->, thick] (-1.7, \height/2) -- (-1.7, \height-1.5);
 \draw[->, thick] (-1.2, \height+1.3) -- (0.5, \height+1.3);
\draw[->, thick] (\width/2 - 1, \height+1.3) -- (\width/2 + 1, \height+1.3);
 \draw[->, thick] (\width-0.5, \height+1.3) -- (\width+1.2, \height+1.3);
 \draw[->, thick] (\width+1.7, \height) -- (\width+1.7, \height-1.5);
  \draw[->, thick]  (\width+1.2, \height/2-1)
-- (\width, \height/2-1);

\node[above] at (\width/2, \height+1.3) {External Circuit};
\node[below] at (\width/2, \height+1.3) {$e^-$};
\node[right] at (-1.7, \height-1.5) {$e^-$};
\node[left] at (\width+1.75, \height-1.5) {$e^-$};

\draw[thick,fill=white] (\width/2+1.5,\height+0.8) rectangle (\width/2+4.5, \height+1.8);
\node[above] at (\width/2+3,\height+1.0) {Load/Charger};

\end{tikzpicture}
  \caption{Diagram of the modeled system during a discharge cycle. Outward-facing normals and the
  direction of cation and electron flow are indicated. During a charge cycle, these flows would
  reverse direction. The configuration is assumed to be very tall. For simplicity, this is modeled
  by imposing periodic boundary conditions on the top and bottom of
  $\Omega_\mathrm{L}\cup\Omega_\mathrm{M}\cup\Omega_\mathrm{R}$. By convention, the direction of
  electrical current is opposite to the flow of electrons. In our preliminary numerical
  implementation, we reduce the domain to one dimensional, letting $L$ and $L_{e_\mathrm{d}}$ be the
  widths of $\Omega_\mathrm{M}$ and $\Omega_\mathrm{d}$, respectively.}
  \label{fig:modeledDomain}
\end{figure}
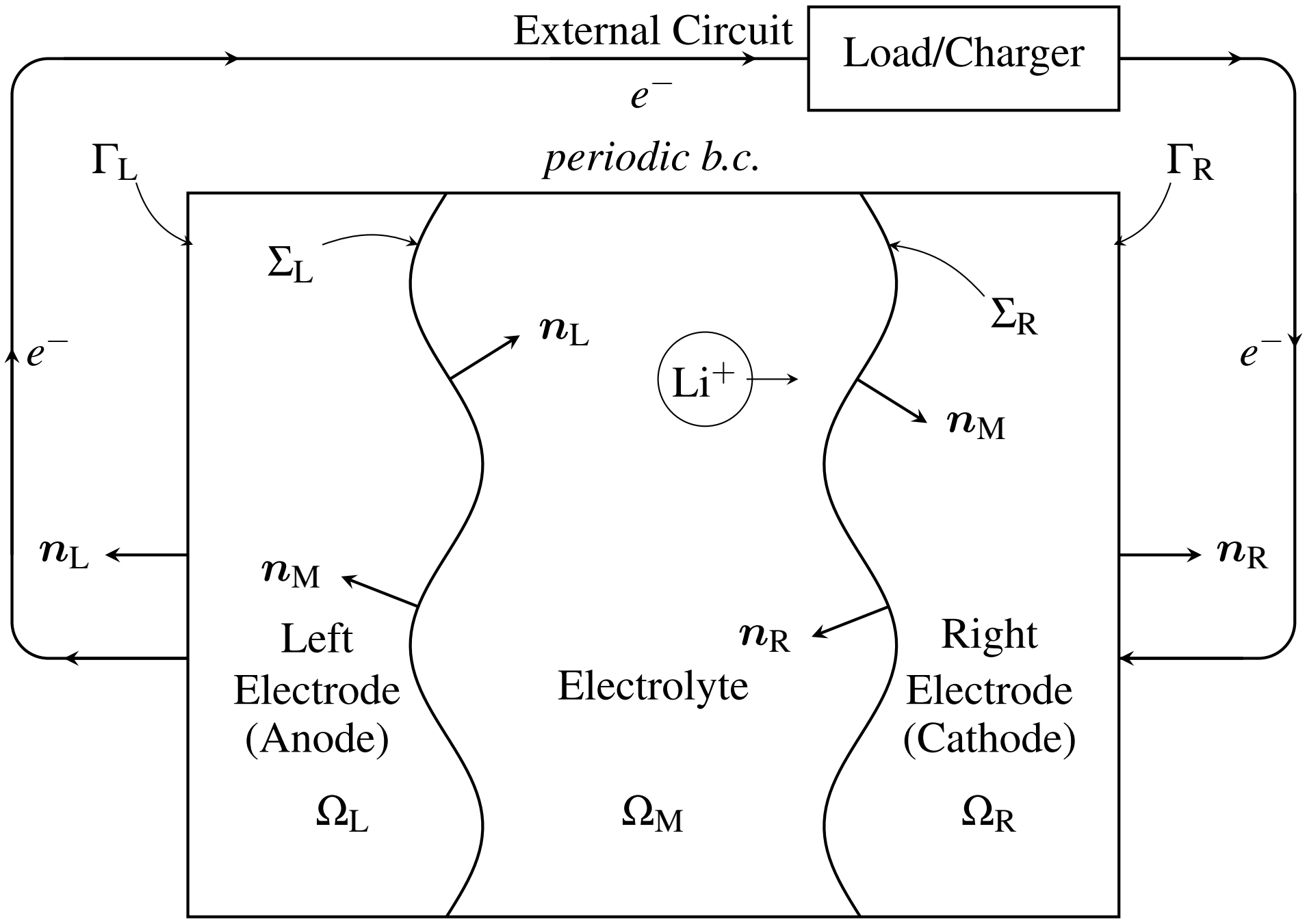

The modeled domain is depicted in Figure~\ref{fig:modeledDomain}. This domain comprises an
electrolyte sandwiched between two electrodes connected by an external circuit. We let
$\Omega_{\mathrm{d}}$, $\mathrm{d}\in\left\{\mathrm{L},\mathrm{R}\right\}$, be the left and right
electrodes, respectively, and $\Omega_\mathrm{M}$ be the electrolyte. The external circuit is
represented through the terminal current conditions; periodic conditions are imposed only on paired
lateral boundaries. We let $\Sigma_\mathrm{L}$ be the boundary between the electrolyte and the left
electrode and $\Sigma_\mathrm{R}$ be the boundary between the electrolyte and the right electrode.
We let $\Gamma_\mathrm{L}$ be the far boundary of the left electrode and $\Gamma_\mathrm{R}$ be the
far boundary of the right electrode.

Preceding one of the domain symbols with $\partial$ indicates the boundary of that domain; for
example, $\partial \Omega_\mathrm{M}$ represents the boundary of the electrolyte. We observe that,
for example, $\Sigma_\mathrm{L}\cup \Sigma_\mathrm{R}\subseteq \partial \Omega_\mathrm{M}$. We
assume that our system is much taller than it is wide and impose periodic boundary conditions on the
top and bottom boundaries for all species. Under this constraint, it is reasonable to restrict the
model to a one-dimensional version, which we later do. We use $\bm n_{{\mathrm{d}}}$ to represent
the outward facing normal with respect to ${\Omega}_{{\mathrm{d}}}$, for ${\mathrm{d}}\in
\left\{\mathrm{L},\mathrm{M},\mathrm{R}\right\}$.

In the model, generically, number densities are represented by $\rho_m$ and electrostatic potentials
are denoted by $\psi_d$, where $m$ and $d$ represent carefully chosen subscripts. The primary
variables in the model are listed in Table~\ref{tab:table1}.

\begin{table}
  \caption{\label{tab:table1}Variables present in the model. All variables are functions of position
  $\bm r= (x, y, z)\in \mathbb{R}^3$ and time $t$. Number densities have units of
  $\frac{1}{\left(\mathrm{Length}\right)^3}$ and electrostatic potentials have units of Volts.}
  \begin{ruledtabular}
    \begin{tabular}{lc}
      Symbol & Definition
      \\
      \midrule
      $\rho_+\left(\bm r,t\right)$ & Local number density of cations in $\Omega_\mathrm{M}$
      \\
      $\rho_-\left(\bm r,t\right)$& Local number density of anions in $\Omega_\mathrm{M}$\\
      $\rho_p\left(\bm r,t\right)$ & Local number density of statistical segments
      \\
      & of polymer chains in $\Omega_\mathrm{M}$
      \\
      $\rho_{i_\mathrm{L}}\left(\bm r,t\right)$ & Local number density of intercalated metal in $\Omega_\mathrm{L}$
      \\
      $\rho_{i_\mathrm{R}}\left(\bm r,t\right)$ & Local number density of intercalated metal in $\Omega_\mathrm{R}$
      \\
      $\rho_{e_\mathrm{L}}\left(\bm r,t\right)$ & Local number density of electrons in $\Omega_\mathrm{L}$
      \\
      $\rho_{e_\mathrm{R}}\left(\bm r,t\right)$ & Local number density of electrons in $\Omega_\mathrm{R}$
      \\
      $\rho_{\mathrm{V}_\mathrm{L}}\left(\bm r,t\right)$ & Local number density of vacancies in $\Omega_\mathrm{L}$
      \\
      $\rho_{\mathrm{V}_\mathrm{M}}\left(\bm r,t\right)$ & Local number density of vacancies in $\Omega_\mathrm{M}$
      \\
      $\rho_{\mathrm{V}_\mathrm{R}}\left(\bm r,t\right)$ & Local number density of vacancies in $\Omega_\mathrm{R}$
      \\
      $\psi_\mathrm{L}\left(\bm r,t\right)$ & Electrostatic potential in $\Omega_\mathrm{L}$
      \\
      $\psi_\mathrm{M}\left(\bm r,t\right)$ & Electrostatic potential in $\Omega_\mathrm{M}$
      \\
      $\psi_\mathrm{R}\left(\bm r,t\right)$ & Electrostatic potential in $\Omega_\mathrm{R}$
      \\
    \end{tabular}
  \end{ruledtabular}
\end{table}

For ${m}\in\mathcal{V} :=
\left\{i_\mathrm{L},i_\mathrm{R},\mathrm{V}_\mathrm{L},\mathrm{V}_\mathrm{R}, +, -, p,
\mathrm{V}_\mathrm{M}\right\}$, we let $1/\rho_{{m}}^o$ denote the volume of one particle of type
${m}$. For the polymer, $1/\rho_p^o$ is the reference volume of one polymer statistical segment.
Thus $\rho_p$ and $\rho_p^o$ are segment number densities. The polymer-chain density is
$\rho_{\mathrm{chain}}=\rho_p/N$, and $\rho_p^o=N\rho_{\mathrm{chain}}^o$. For simplicity, we assume
metal atoms and their cations which occupy vacancy sites and the vacancies themselves have precisely
the same volume, i.e.,
\[
1/\rho_{i_\mathrm{d}}^o = 1/\rho_{\mathrm{V}_\mathrm{d}}^o, \quad  \mbox{for} \ \mathrm{d} \in \{\mathrm{L}, \mathrm{R}\}, \quad \mbox{and} \quad 1/\rho_{+}^o = 1/\rho_{\mathrm{V}_\mathrm{M}}^o.
\]
The unitless volume fractions of particles of type ${m}$ is then defined as
\begin{equation}
  \phi_{m}\left(\bm r,t\right) := \frac{\rho_{m}\left(\bm r,t\right)}{\rho_{m}^o}\label{phiM}.
\end{equation}

In each electrode and in the electrolyte, we assume that the sum of volume fractions of all
components, except for the electrons, is always unity at all locations and times. That is, for
$\mathrm{d}\in \left\{\mathrm{L},\mathrm{R}\right\},$
\begin{eqnarray}
  \sum_{{m}\in\left\{i_\mathrm{d},\mathrm{V}_\mathrm{d}\right\}}\phi_{{m}}\left(\bm r,t\right) = 1, \quad  \bm r \in \Omega_\mathrm{d},
  \label{localincompresselectrodes}
  \\
  \sum_{m \in \left\{+, -, p, \mathrm{V}_\mathrm{M}\right\}} \phi_{m}\left(\bm r,t\right) = 1, \quad  \bm r \in \Omega_\mathrm{M}.
  \label{localincompresselectrolyte}
\end{eqnarray}
In this way there are no voids in the mixture. We observe that an analogous no-voids
assumption was employed in~\cite{kumar2017rayleighian} and in many other works.

Since the volume of an electron is negligible, we use $\rho_+^o$ as the
reference density for electrons and define
\begin{equation}
  \phi_{e_\mathrm d}:=\frac{\rho_{e_\mathrm d}}{\rho_+^o},
  \qquad \mathrm d\in\{\mathrm L,\mathrm R\}.
  \label{phiElectron}
\end{equation}
The symbol $\phi_{e_\mathrm d}$ is a dimensionless electron number density,
not a volume or site fraction. It is not a member of $\mathcal V$ and does not
enter either incompressibility constraint. Let $\mu_n$ denote the chemical potential excluding the
electrostatic contribution, and $\mu_n^\circ$ its constant standard-state contribution. We write
$\tilde\mu_n=\mu_n+z_n e\psi$ for the electrochemical potential, with electron valence $z_{e_\mathrm
d}=-1$ and the potential evaluated in the species' domain. Provided that the electrode vacancies and
intercalated species are uncharged, the electrochemical potential of species ${n}$ is
\begin{equation}
  \frac{\tilde{\mu}_{{n}}}{k_{\mathrm{B}}T} =
  \begin{cases}
    \frac{\mu_{e_\mathrm{d}}^\circ}{k_{\mathrm{B}}T}+\ln \left(\frac{\rho_{e_\mathrm d}}{\rho_+^o}\right) + 1 - \frac{e\psi_\mathrm{d}}{k_{\mathrm{B}}T}, &{n}= e_\mathrm{d},
    \\
    \frac{\mu_n^\circ}{k_{\mathrm{B}}T}+\ln \left( \frac{\rho_{{n}}}{\rho_{{n}}^o}\right) + 1,&{n} \in \left\{ i_\mathrm{d},\mathrm{V}_\mathrm{d},  \mathrm{V}_\mathrm{M}\right\},
    \\
    \frac{\mu_n^\circ}{k_{\mathrm{B}}T}+\ln \left( \frac{\rho_{{n}} }{\rho_{{n}}^o}\right) + 1 + z_{{n}} \frac{e\psi_\mathrm{M}}{k_{\mathrm{B}}T}, &{n} \in \left\{+, -\right\},
    \\
    \frac{\mu_p^\circ}{k_{\mathrm{B}}T}+\frac{1}{N} \left[ \ln \left( \frac{\rho_p }{
    \rho_{p}^o
    }\right) + 1+Nz_p \frac{e\psi_\mathrm{M}}{k_{\mathrm{B}}T}\right], & {n} = p,
  \end{cases}
  \label{chem_potentials}
\end{equation}
for $\mathrm{d} \in \left\{\mathrm{L}, \mathrm{R}\right\}$. Here and throughout, $k_{\mathrm{B}}$ is
the Boltzmann constant and $T$ is the constant, spatially uniform absolute temperature. A list of
model parameters and outputs are defined in Table~\ref{tab:table2}. Additional symbols are defined
in Table~\ref{tab:addnlSymbols} in Section~\ref{sec:listofSymbols}.

For clarity, $\tilde{\mu}_p$ is the electrochemical potential of the statistical segments. Only
stoichiometric combinations of the constants $\mu_n^\circ$ (i.e., the linear combination of these constants where the coefficients are the corresponding stoichiometric coefficients) affect reaction equilibrium.  Because of
the local no voids conditions, the exchange electrochemical potential of species ${n}$ is needed:
\begin{equation}
  \frac{\tilde{\mu}_{{n}}^*}{k_{\mathrm{B}}T} =
  \begin{cases}
    \frac{\tilde{\mu}_{e_\mathrm{d}}}{k_{\mathrm{B}}T}, &{n} = e_\mathrm{d},
    \\
    \frac{\tilde{\mu}_{i_\mathrm{d}}}{k_{\mathrm{B}}T} - \frac{\rho_{\mathrm{V}_\mathrm{d}}^o }{\rho_{i_\mathrm{d}}^o }\frac{\tilde{\mu}_{\mathrm{V}_\mathrm{d}}}{k_{\mathrm{B}}T}, & {n} = i_\mathrm{d},
    \\
    \frac{\tilde{\mu}_{{n}}}{k_{\mathrm{B}}T} - \frac{\rho_{\mathrm{V}_\mathrm{M}}^o}{\rho_{{n}}^o}\frac{\tilde{\mu}_{\mathrm{V}_\mathrm{M}}}{k_{\mathrm{B}}T}, & n \in \left\{+, -, p\right\},
    \\
    0, & {n} \in \left\{\mathrm{V}_\mathrm{d}, \mathrm{V}_\mathrm{M}\right\},
  \end{cases}
  \label{exchange_chem_potentials}
\end{equation}
for $\mathrm{d} \in \left\{\mathrm{L}, \mathrm{R}\right\}$.

\begin{table*}
  \caption{\label{tab:table2}Definitions of model parameters and outputs, as well as their units.
  Where indicated, ${m}\in\left\{ i_\mathrm{L}, i_\mathrm{R}, \mathrm{V}_\mathrm{L},
  \mathrm{V}_\mathrm{R}, +, -, p, \mathrm{V}_\mathrm{M}\right\}$, $\mathrm{d} \in \{\mathrm{L},
  \mathrm{R}\}$, ${n}\in\left\{e_\mathrm{L}, e_\mathrm{R},i_\mathrm{L}, i_\mathrm{R},
  \mathrm{V}_\mathrm{L}, \mathrm{V}_\mathrm{R}, +, -, p, \mathrm{V}_\mathrm{M}\right\}$, and $\ell
  \in \left\{e_\mathrm{L}, e_\mathrm{R},i_\mathrm{L}, i_\mathrm{R},+, -, p\right\}$.}
  \small
  \begin{ruledtabular}
    \begin{tabular}{ccc}
      Symbol & Definition & Units\\ \midrule
      \multicolumn{1}{c}{}&
      \multicolumn{1}{c}{{Parameters}}&
      \multicolumn{1}{c}{}
      \\
      $N$ & Number of statistical segments per polymer chain & Unitless\\
      $z_p$ & Valency of a statistical segment, assumed to be  & Unitless\\
      & 0 for DIC case and -1 for SIC case & \\
      $z_+$ & Cation valency, assumed to be +1& Unitless\\
      $z_-$ & Anion valency, assumed to be -1 & Unitless \\
      $1/\rho_{m}^o$ & The volume of a particle of type ${m}$ & ${\left( \mathrm{Length}\right)^3}$\\
      $\mathrm{q}_\mathrm{d}$ & Fixed number density of charges, uniformly distributed in $\Omega_\mathrm{d}$ &$\frac{1}{\left(\mathrm{Length}\right)^3}$\\
      $\epsilon_\mathrm{M}$ & {Effective background relative permittivity of $\Omega_\mathrm{M}$} & Unitless \\
      $\epsilon_\mathrm{d}$ & {Effective background relative permittivity of $\Omega_\mathrm{d}$} & Unitless \\
      {$\epsilon_0$} & {Vacuum permittivity} & {$\frac{\mathrm{Farads}}{\mathrm{Length}}$}\\
      $e$ & An elementary unit of charge & Coulomb\\
      $l_\mathrm{B}$ & Bjerrum length in vacuum & ${\mathrm{Length}}$\\
      $k_{\mathrm{B}}$ & Boltzmann constant & $\frac{\mathrm{Joules}}{\mathrm{Kelvin}}$\\
      $T$ & Constant and uniform system temperature & $\mathrm{Kelvin}$\\
      $N_A$ & {Avogadro constant} & {$\frac{1}{\mathrm{Mole}}$} \\
      $s_{{n},\mathrm{d}}$ & Stoichiometric coefficient of species ${n}$ on $\Sigma_\mathrm{d}$& Unitless \\
      $k_\mathrm d$ & Interface-dependent mass transfer coefficient & {$\frac{\mathrm{Moles}}{(\mathrm{Length})^2 \mathrm{Time}}$}\\
      ${I_{\mathrm{app}}}$ & {Applied terminal current density} & $\frac{\mathrm{Amperes}}{\left(\mathrm{Length}\right)^2}$\\
      $D_{\ell}$ &{Mobility parameter of species ${\ell}$, with diffusivity units} & $\frac{\left(\mathrm{Length}\right)^2}{\mathrm{Time}}$\\
      $\alpha_\mathrm{d}$ & Charge transfer coefficient for the reaction on $\Sigma_\mathrm{d}$ & Unitless \\
      \multicolumn{1}{c}{}&
      \multicolumn{1}{c}{{Outputs}}&
      \multicolumn{1}{c}{}
      \\
      $V(t)$ & System voltage & Volts\\
    \end{tabular}
  \end{ruledtabular}
\end{table*}

Having defined the system geometry and identified constituent species, we now present the governing
equations and boundary conditions. For every transported species $\ell$, let $M_\ell\geq0$ denote
its nonnegative mobility function, which may depend on the local state; it has units of
$1/(\mathrm{Length}\cdot\mathrm{Time})$. In $\Omega_\mathrm{d}$, $\mathrm{d}\in
\left\{\mathrm{L},\mathrm{R}\right\}$, for ${\ell}\in \left\{e_\mathrm{d}, i_\mathrm{d}\right\}$,
\begin{eqnarray}
  \frac{\partial \rho_{{\ell}}}{\partial t} &=& \nabla \!\cdot\! \left[ M_{{\ell}} \nabla \left( \frac{\tilde{\mu}_{{\ell}}^*}{k_{\mathrm{B}}T}\right) \right]  ,
  \label{ellHatGovEqn}
  \\
  - \nabla \!\cdot\! \left(\frac{\epsilon_\mathrm{d}}{4 \pi l_\mathrm{B}} \nabla \frac{e\psi_\mathrm{d}}{k_{\mathrm{B}}T} \right) &=& \mathrm{q}_\mathrm{d}-\rho_{e_\mathrm{d}},
  \label{poissonElectrodes}
\end{eqnarray}
where
\begin{equation}
  \mathrm{q}_\mathrm{d} := \frac{1}{\left|\Omega_\mathrm{d}\right|}\int_{\Omega_\mathrm{d}} \rho_{e_\mathrm{d}}\left(\bm r,0\right) \, \mathrm{d}\bm r \label{varrho},
\end{equation}
is a constant background charge density and $l_B = e^2/4\pi \epsilon_0 k_B T$ so that $\epsilon_0$ is the permittivity of vacuum. The constant $\mathrm q_\mathrm d$ represents the immobile
positive background charge density in $\Omega_\mathrm d$ and is chosen so that each electrode is
globally neutral initially. Individual-electrode electro-neutrality is not imposed for $t>0$:
terminal electron transport and interfacial reaction may change an electrode's net charge, while
balanced terminal currents and charge-conserving reactions preserve the global charge of the
complete cell. In a highly conducting metal, physical excess charge is confined to a short screening
length, so computed electrode space-charge profiles should not be over-interpreted unless that
screening length scale is resolved.

In $\Omega_\mathrm{M}$, for $\ell \in \left\{+, -, p\right\}$,
\begin{eqnarray}
  \frac{\partial \rho_\ell}{\partial t} &=& \nabla \!\cdot\! \left[ M_{{\ell}} \nabla \left( \frac{\tilde{\mu}_{{\ell}}^*}{k_{\mathrm{B}}T}\right) \right],
  \label{ellGovEqn}
  \\
  - \nabla \!\cdot\! \left(\frac{\epsilon_\mathrm{M}}{4 \pi l_\mathrm{B}} \nabla \frac{e\psi_\mathrm{M}}{k_{\mathrm{B}}T} \right) &=&
  \sum_{\ell \in \{+, -,p\}} z_\ell \rho_\ell.
  \label{poissonElectrolyte}
\end{eqnarray}

{In this model, we assume a linear dielectric response in the weak-field regime and represent each
domain by a field-independent effective background permittivity. The effective permittivity includes
bound-polarization mechanisms that relax rapidly relative to the modeled cell dynamics, but it
excludes the redistribution of mobile charged species, which is represented explicitly by the
transport equations. A high-frequency permittivity would instead correspond to polarization modes
that are frozen on the modeled time scale. Spatial, compositional, and temporal variations of the
background permittivity are neglected~\cite{marcus1956electrostatic,kumar2025ferro}.}

We now state the flux--reaction boundary conditions at the electrode--electrolyte interfaces. They
are chosen to enforce mass and charge conservation and to be compatible with the free-energy balance
proved in Section~\ref{sec:modelProperties}. For $\ell \in
\left\{e_\mathrm{d},i_\mathrm{d}\right\}$,
\begin{equation}
  \left(-M_\ell \nabla \left( \frac{\tilde{\mu}_\ell^*}{k_{\mathrm{B}}T}\right) \!\cdot\!\bm n_\mathrm{d} + N_As_{\ell,\mathrm{d}}\nu_\mathrm{d}\right)\Big|_{\Sigma_\mathrm{d}} =0,
  \label{ellHat_BC}
\end{equation}
and, for $\ell \in \left\{+, -, p\right\}$,
\begin{equation}
  \left(-M_\ell \nabla \left( \frac{\tilde{\mu}_\ell^*}{k_{\mathrm{B}}T}\right) \!\cdot\!\bm n_\mathrm{M} + N_As_{\ell,\mathrm{d}}\nu_\mathrm{d}\right)\Big|_{\Sigma_\mathrm{d}} =0,
  \label{ell_BC}
\end{equation}
where $\mathrm{d}\in\left\{\mathrm{L},\mathrm{R}\right\}$, $\nu_\mathrm{d}$ is the net reaction rate
on $\Sigma_\mathrm{d}$, and $s_{\ell,\mathrm{d}}$ is the stoichiometric coefficient of species
$\ell$ on $\Sigma_\mathrm{d}$. These conditions make each normal species flux proportional to the
interfacial reaction rate, with the sign and magnitude set by its stoichiometric coefficient. The
factor $N_A$ converts the molar reaction rate per unit area into the number flux used in the species
balances.

We make a constitutive choice for the form of $\nu_\mathrm{d}$ so that, as we later prove, these
interfacial reactions are free energy non-increasing. First, define the reacting-species index set
\begin{equation}
  \mathcal{I} := \left\{e_\mathrm{L},e_\mathrm{R},i_\mathrm{L},i_\mathrm{R},\mathrm{V}_\mathrm{L},\mathrm{V}_\mathrm{R},+,-,p,\mathrm{V}_\mathrm{M}\right\}.\label{idxSet}
\end{equation}
Suppose that $f_\mathrm{inc}(x)$ is any differentiable function with the property that
$f_\mathrm{inc}'(x) \ge 0$, and $\tilde{\mu}_{\ddagger, \mathrm{d}}$ is any parameter with units of
Joules. Set
\begin{eqnarray}
  \nu_\mathrm{d} = k_\mathrm d\left[  f_\mathrm{inc}\left(\frac{A^{\mathrm{r}}_{\mathrm{d}}-\tilde{\mu}_{\ddagger,\mathrm{d}}}{k_{\mathrm{B}}T}\right)-f_\mathrm{inc}\left(\frac{A^{\mathrm{p}}_{\mathrm{d}}-\tilde{\mu}_{\ddagger,\mathrm{d}}}{k_{\mathrm{B}}T}\right)\right],\label{r_transitionstatenotsubbed}
  \\
  A^{\mathrm{r}}_{\mathrm{d}} :={-}\sum_{\substack{n\in\mathcal{I},s_{{{n}},\mathrm{d}}< 0}}\tilde{\mu}_{{{n}}}^*s_{{n},\mathrm{d}},
  \label{Ar}
  \\
  A^{\mathrm{p}}_{\mathrm{d}} :=\phantom{-}\sum_{\substack{n\in\mathcal{I},s_{{{n}},\mathrm{d}}\geq 0}}\tilde{\mu}_{{{n}}}^*s_{{n},\mathrm{d}},
  \label{Ap}
\end{eqnarray}
where $k_\mathrm d\geq 0$ is the mass transfer coefficient for interface $\Sigma_\mathrm d$. This
choice of net reaction rate leads to a free energy dissipative process, as we see in
Section~\ref{sec:modelProperties}. Note that the general model permits $k_\mathrm L\ne k_\mathrm R$;
equality is later imposed only for the symmetric illustrative cell.

This chosen form of $\nu_\mathrm{d}$ is motivated by the transition state theory. The quantities
$A^{\mathrm{r}}_{\mathrm{d}}$ and $A^{\mathrm{p}}_{\mathrm{d}}$ are decompositions of the chemical
affinity of the reaction on $\Sigma_\mathrm{d}$, $A_\mathrm{d}$, into the contributions from the
reactants and products~\cite{prigogine1948affinity}:
\[
A_\mathrm{d} = A^{\mathrm{r}}_{\mathrm{d}}-A^{\mathrm{p}}_{\mathrm{d}}.
\]
If $\tilde{\mu}_{\ddagger, \mathrm{d}}$ represents the excess electrochemical potential of the
transition state of the chemical reaction occurring on $\Sigma_\mathrm{d}$, then the arguments of
$f_\mathrm{inc}(x)$ in~\eqref{r_transitionstatenotsubbed} correspond to the dimensionless
free-energy differences between the transition state and the reactant and product states.

This reaction rate expression can capture well-known reaction expressions. To reproduce the
Butler-Volmer equation, we set $f_\mathrm{inc}(x) = \exp(x)$ and pick a form of
$\tilde{\mu}_{\ddagger, \mathrm{d}}$ motivated by~\cite{bazant2013theory}. This is detailed in
Section~\ref{sec:recapturingKnownDynamics}.

The electrostatic potential satisfies the following transmission boundary conditions on the
electrode-electrolyte interfaces:
\begin{eqnarray}
  \left(\psi_\mathrm{d}-\psi_\mathrm{M}\right)\Big|_{\Sigma_\mathrm{d}} = 0 \label{transmission_cont}, \\
  \left(\epsilon_\mathrm{d}\nabla \psi_\mathrm{d}\!\cdot\! \bm n_\mathrm{d}+       \epsilon_\mathrm{M}\nabla \psi_\mathrm{M}\!\cdot\! \bm n_\mathrm{M}\right)\Big|_{\Sigma_\mathrm{d}}=0\label{transmission_jump}.
\end{eqnarray}
{Boundary condition~\eqref{transmission_cont} excludes an unresolved compact-layer potential jump or
interfacial dipole sheet, but it does not exclude diffuse space-charge layers resolved within the
electrode and electrolyte domains. Boundary condition~\eqref{transmission_jump} is the standard
transmission condition between two dielectric media in the absence of a singular free surface-charge
density.}

\begin{remark}
  {A model permitting $\psi_\mathrm{d}\ne\psi_\mathrm{M}$ on $\Sigma_\mathrm{d}$ would require
  additional interfacial physics. For example, one could introduce an interfacial surface-charge or
  compact-layer state, a constitutive relation between that state and the potential jump, and a
  corresponding interfacial electrostatic free energy. The transmission conditions and the
  free-energy dissipation proof would then need to be modified consistently.}
\end{remark}

Because the Poisson and transport equations are invariant under a spatially
uniform shift of electrostatic potential, a reference value must be selected. We set that reference at the left terminal and
retain a homogeneous normal-displacement condition at the right terminal:
\begin{eqnarray}
  \psi_\mathrm L\big|_{\Gamma_\mathrm L}&=&0,
  \label{psiDirichletBCLeft}\\
  \nabla\psi_\mathrm R\cdot\bm n_\mathrm R\big|_{\Gamma_\mathrm R}&=&0.
  \label{psiNeumannBCRight}
\end{eqnarray}
The value in Equation~\eqref{psiDirichletBCLeft} is a reference potential,
not an independently imposed cell voltage.  For globally neutral data, the
integrated Poisson equation recovers zero normal displacement at the left
terminal.

Charge conservation (of electrons) in the external circuit equates the integrated
terminal currents rather than pointwise flux densities on two different surfaces. With $\boldsymbol
j_{e,\mathrm d}:=-M_{e_\mathrm{d}}\nabla(\tilde\mu_{e_\mathrm{d}}^*/k_{\mathrm{B}}T)$, we impose
\begin{equation}
  \int_{\Gamma_\mathrm L}\boldsymbol j_{e,\mathrm L}\!\cdot\!\boldsymbol n_\mathrm L\,\mathrm dS
  +\int_{\Gamma_\mathrm R}\boldsymbol j_{e,\mathrm R}\!\cdot\!\boldsymbol n_\mathrm R\,\mathrm dS=0.\label{fluxECBC}
\end{equation}
{For the one-dimensional, constant-cross-section geometry used in the simulations, the terminal
current densities are uniform and equal in magnitude. The integral condition therefore reduces to
the following local galvanostatic conditions:}
\begin{eqnarray}
  -M_{e_\mathrm{L}}\nabla \left( \frac{\tilde{\mu}^*_{e_\mathrm{L}}}{k_{\mathrm{B}}T}\right)\!\cdot\!\bm n_\mathrm{L}\Big|_{\Gamma_\mathrm{L}} & = \frac{I_\mathrm{app}}{e},
  \label{currentBCL}
  \\
  M_{e_\mathrm{R}}\nabla \left( \frac{\tilde{\mu}^*_{e_\mathrm{R}}}{k_{\mathrm{B}}T}\right)\!\cdot\!\bm n_\mathrm{R}\Big|_{\Gamma_\mathrm{R}} & = \frac{I_\mathrm{app}}{e},
  \label{currentBCR}
\end{eqnarray}
where $I_\mathrm{app}$ is the applied current density; it is negative during a charge cycle and
positive during a discharge cycle. This convention is determined so that during a charge cycle
electrons move out of the right electrode at $\Gamma_\mathrm{R}$ and into the left electrode at
$\Gamma_\mathrm{L}$. This implies that during the charge cycle there are electrons being produced at
the right electrode, so that oxidation is occurring, and being consumed at the left electrode, so
that reduction is occurring. {Thus, during charge the right/positive electrode is the anode and the
left/negative electrode is the cathode, consistent with Table~\ref{tab:electrodeRoles}.}

The total electric current is the sum of conduction and dielectric displacement
currents~\cite{kumar2017rayleighian}.  
The mixed Dirichlet--Neumann Poisson problem is well posed without a neutrality
compatibility condition; here neutrality additionally implies, through the
integrated Poisson equation, that the net left-boundary dielectric displacement is zero.
Displacement current may nevertheless remain nonzero inside the cell during
a transient.  All reported electrostatic
voltages are differences relative to the left-terminal reference in
Equation~\eqref{psiDirichletBCLeft}; the reference value does not change
electric fields, reaction affinities, or transport fluxes.

\begin{remark}
  Other electrostatic boundary conditions can be thermodynamically consistent when their boundary
  energy and power are included. Prescribed surface charge gives nonhomogeneous Neumann data; an
  evolving surface charge requires a corresponding surface free energy. A time-dependent prescribed
  electrode potential gives Dirichlet data and may contribute electrical boundary power. In the
  present formulation the left Dirichlet value is the constant zero reference, so its electrostatic
  boundary contribution vanishes because $\dot\psi_\mathrm L=0$; galvanostatic work instead enters
  through the terminal electron electrochemical-potential flux. Let $C_\mathrm{dl}$ denote the
  interfacial double-layer capacitance per unit area, with units of $\mathrm{F\,m^{-2}}$. A
  compact-layer or capacitive model may use a potential-jump or Robin condition together with an
  interfacial energy such as
  \[
  \frac{1}{2} \int_{\Sigma_\mathrm{d}} C_\mathrm{dl}(\psi_\mathrm{d}-\psi_\mathrm{M})^2\,\mathrm{d}S.
  \]
  Mixed conditions are likewise admissible provided that the electrostatic boundary term generated
  by integration by parts either vanishes or is identified with stored interfacial energy or
  external electrical work.
\end{remark}

Due to the closed nature of the system, the intercalated metal in the electrodes satisfies no-flux
boundary conditions on the electrode-external circuit interfaces:
\begin{equation}
  -M_{i_\mathrm{d}}\nabla \left( \frac{\tilde{\mu}^*_{i_\mathrm{d}}}{k_{\mathrm{B}}T}\right)\!\cdot\!\bm n_\mathrm{d}\Big|_{\Gamma_\mathrm{d}
  } = 0. \label{noFluxI}
\end{equation}

On the remaining electrode boundaries, $\partial \Omega_\mathrm{d}\setminus
\left\{\Sigma_\mathrm{d}\cup \Gamma_\mathrm{d}\right\}$, $\rho_{{\ell}}$ and $\psi_\mathrm{d}$
satisfy periodic boundary conditions, for $\ell \in \left\{e_\mathrm{d},i_\mathrm{d}\right\}$.
Similarly, on the remaining electrolyte boundaries, $\partial \Omega_\mathrm{M}\setminus
\left\{\Sigma_\mathrm{L}\cup \Sigma_\mathrm{R}\right\}$, $\rho_\ell$ and $\psi_\mathrm{M}$ satisfy
periodic boundary conditions, for $\ell \in \left\{+, -, p\right\}$.

We idealize each external current collector as equipotential with respect to electron
electrochemical potential, $\tilde\mu_{e_\mathrm{d}}^*=\tilde\mu_{e_\mathrm{d},\Gamma}^*(t)$ on
$\Gamma_\mathrm d$. This condition is automatic in the one-dimensional model, where a terminal is a
point. In reality, a standard voltmeter measures the difference in electrochemical potential between
two points of a system~\cite{boettcher2020potentially}. To mimic this in the model, in a
galvanostatic setting, we output the quantity
\begin{equation}
  V(t) = \frac{1}{e}\left(\tilde{\mu}_{e_\mathrm{L}}^*\vline_{\Gamma_\mathrm{L}} - \tilde{\mu}_{e_\mathrm{R}}^*\vline_{\Gamma_\mathrm{R}}\right).\label{voltBC}
\end{equation}

Because $I_\mathrm{app}$ in boundary conditions~\eqref{currentBCL}--\eqref{currentBCR} is a current
density, $I_\mathrm{app}V$ is the terminal power per unit area in the constant-cross-section model.
In general geometry, the signed total applied current is
$I_\mathrm{app}^{\mathrm{tot}}:=\int_{\Gamma_\mathrm R} I_\mathrm{app}\,\mathrm dS$, and the total
terminal power is $I_\mathrm{app}^{\mathrm{tot}}V$ under the ideal-current-collector and
balanced-current assumptions. For a uniform current density on a collector of area $A_\Gamma$,
$I_\mathrm{app}^{\mathrm{tot}}=A_\Gamma I_\mathrm{app}$. The ratio $V/I_\mathrm{app}$, when defined,
is an apparent whole-cell area-specific resistance; it is not a constitutive contact resistance
because $V$ includes equilibrium chemical electromotive force and polarization. We therefore retain
the terminal power explicitly rather than imposing an Ohmic relation $V=I_\mathrm{app}R$.

\subsection{Model Properties}
\label{sec:modelProperties}

\subsubsection{Thermodynamic Consistency}
\label{sec:thermodynamicConsistency}

Due to the class of reactions under consideration (i.e., Equations~\eqref{leftLiReaction}
and~\eqref{rightLiReaction}), the stoichiometric coefficients ($s_{n,\mathrm{L}}$ on
$\Sigma_\mathrm{L}$ and $s_{n,\mathrm{R}}$ on $\Sigma_\mathrm{R}$) satisfy
\begin{equation}
  s_{{n},\mathrm{L}} =
  \begin{cases}
    -s_{+,\mathrm{L}}, & {n} \in \left\{i_\mathrm{L}, \mathrm{V}_\mathrm{M}\right\},
    \\
    \phantom{-}s_{+,\mathrm{L}}, & {n} \in \left\{e_\mathrm{L}, \mathrm{V}_\mathrm{L},+\right\},
    \\
    \phantom{-}0, & {n} \in \left\{e_{\mathrm{R}}, i_{\mathrm{R}}, \mathrm{V}_{\mathrm{R}},-, p \right\},
  \end{cases}
  \quad s_{{n},\mathrm{R}} =
  \begin{cases}
    -s_{+,\mathrm{R}}, & {n} \in \left\{i_\mathrm{R}, \mathrm{V}_\mathrm{M}\right\},
    \\
    \phantom{-}s_{+,\mathrm{R}}, & {n} \in \left\{e_\mathrm{R}, \mathrm{V}_\mathrm{R},+\right\},
    \\
    \phantom{-}0, & {n} \in \left\{e_{\mathrm{L}}, i_{\mathrm{L}}, \mathrm{V}_{\mathrm{L}},-, p \right\},
  \end{cases}
  \label{sk}
\end{equation}
where $|s_{+,\mathrm{d}}| = 1$. See Section~\ref{sec:exProp} for an illustrative discussion.

\begin{theorem}[Thermodynamic consistency]\label{thmThermoConsist}
  Suppose the system stays near enough to equilibrium that the equilibrium free energy expressions
  below are valid at all times, as in~\cite{kumar2024dynamic}. Consider a sufficiently smooth
  solution of the governing equations and boundary and transmission conditions in
  Section~\ref{subsec:modelFormulation}, with strictly positive densities wherever a logarithm is
  evaluated. Assume $M_\ell\geq 0$, $k_\mathrm d\geq 0$, that the function $f_\mathrm{inc}$
  in~\eqref{r_transitionstatenotsubbed} is nondecreasing, that the interfacial stoichiometry
  conserves charge, and that the integrated terminal electron currents satisfy~\eqref{fluxECBC}.
  Suppose the total free energy is
  \begin{eqnarray}
    F = F_{\mathrm{L}} + F_{\mathrm{M}}+ F_{\mathrm{R}} ,
    \label{H}
  \end{eqnarray}
  where the electrode energies are
  \begin{eqnarray}
    \frac{F_\mathrm{d}}{k_{\mathrm{B}}T} = \int_{\Omega_\mathrm{d}} \left\{\rho_{e_\mathrm{d}}\left[\ln \phi_{e_\mathrm d}+\frac{\mu_{e_\mathrm{d}}^\circ}{k_{\mathrm{B}}T}\right] + \sum_{{\ell}\in\left\{i_\mathrm{d}, \mathrm{V}_\mathrm{d}\right\}}\rho_{{\ell}}\left[\ln\left(\frac{\rho_{{\ell}}}{\rho_{{\ell}}^o}\right)+\frac{\mu_\ell^\circ}{k_{\mathrm{B}}T}\right] \right\} \, \mathrm{d}\bm r
    \nonumber
    \\
    + \int_{\Omega_\mathrm{d}}\left\{\left( \mathrm{q}_\mathrm{d} - \rho_{e_\mathrm{d}}\right)\frac{e\psi_\mathrm{d}}{k_{\mathrm{B}}T}  - \frac{\epsilon_\mathrm{d}}{8 \pi l_\mathrm{B}} \left|\nabla \frac{e\psi_\mathrm{d}}{k_{\mathrm{B}}T} \right|^2 \right\}\, \mathrm{d}\bm r
    \label{eq:H_d},
  \end{eqnarray}
  for $\mathrm{d} \in \{\mathrm{L}, \mathrm{R}\}$, and the electrolyte energy is
  \begin{eqnarray}
    \frac{F_\mathrm{M}}{k_{\mathrm{B}}T} = \int_{\Omega_\mathrm{M}} \left\{\sum_{\ell \in \left\{+, -, \mathrm{V}_\mathrm{M}\right\}} \rho_\ell \left[\ln \left(\frac{\rho_\ell}{\rho_\ell^o} \right)+\frac{\mu_\ell^\circ}{k_{\mathrm{B}}T}\right] + \frac{\rho_p}{N}\ln\left(\frac{\rho_p}{\rho_p^o}\right)+\rho_p\frac{\mu_p^\circ}{k_{\mathrm{B}}T}\right\} \, \mathrm{d}\bm r
    \nonumber
    \\
    + \int_{\Omega_\mathrm{M}}\left\{\sum_{\ell \in \left\{+, -, p\right\}}z_\ell \rho_\ell \frac{e\psi_\mathrm{M}}{k_{\mathrm{B}}T}  - \frac{\epsilon_\mathrm{M}}{8 \pi l_\mathrm{B}} \left|\nabla \frac{e\psi_\mathrm{M}}{k_{\mathrm{B}}T} \right|^2 \right\}\, \mathrm{d}\bm r
    \label{eq:H_M}.
  \end{eqnarray}
  The galvanostatic model presented in Section~\ref{subsec:modelFormulation} is thermodynamically
  consistent under isothermal conditions in the sense that the following properties hold.
  \begin{enumerate}
    \item
    The free energy satisfies the open-system balance
    \begin{equation}
      \dot F+P_\mathrm{term}=-\mathcal D_\mathrm{int}\leq 0,
      \label{freeEnergyBalance}
    \end{equation}
    where $\mathcal D_\mathrm{int}\geq0$ is the internal bulk-plus-interfacial dissipation rate and
    $P_\mathrm{term}$ is the total terminal power. In the constant-cross-section 
    model,
    \begin{equation}
      P_\mathrm{term}=\int_{\Gamma_\mathrm R}I_\mathrm{app}V\,\mathrm dS.
    \end{equation}
    Consequently, the free energy need not decrease during charging.
    \item
    The model conserves the total number of lithium nuclei (counting both mobile cations and
    intercalated lithium), the numbers of free anions and polymer segments, and the total number of
    vacancies summed over the electrolyte and both electrodes.
    \item
    The total charge of the complete cell is conserved for all time as a consequence of both
    charge-conserving interfacial reactions and balanced integrated terminal currents.
    \item
    If the electrolyte is initially globally electro-neutral, then the complete cell is globally
    electro-neutral for all time.
  \end{enumerate}
\end{theorem}

The proof is given in Section~\ref{sec:proofThmThermoConsist}.

\begin{remark}
  The logarithmic electron energy term is retained as an idealized convex electronic free energy, or
  regularized electronic compressibility, rather than as a claim that metallic electrons form a
  classical dilute ideal gas. More generally it may be replaced by an electronic free-energy density
  $f_{e,\mathrm d}(\rho_{e_\mathrm{d}})$ with $f_{e,\mathrm d}''>0$, using
  $\tilde\mu_{e_\mathrm{d}}=f_{e,\mathrm d}'-e\psi_\mathrm{d}$ throughout; the primes denote derivatives
  with respect to electron number density. In the high-conductivity regime the electron density is
  nearly uniform, which reduces sensitivity to this closure; quantitative screening and absolute
  electrode voltages require a density-of-states-based electronic free energy that we do not pursue
  here.
\end{remark}

\begin{remark}
  A potentiostatic extension would instead prescribe the difference in terminal electron
  electrochemical potentials and solve for the terminal current. It obeys the same open-system
  free-energy balance~\eqref{freeEnergyBalance}, with $P_\mathrm{term}$ equal to terminal current
  times voltage; galvanostatic and potentiostatic control differ in which of these two quantities is
  prescribed. Because the potentiostatic boundary-value problem is not formulated here, the theorem
  is stated only for the galvanostatic problem analyzed in this work.
\end{remark}

\subsubsection{Divergence-Free Total Electric Current Density}
\label{sec:uniformI}
The charge density $\rho_c$ at location $\bm r$ and time
$t$ is
\begin{equation}
  \rho_c\left(\bm r,t\right) = \begin{cases}
    e\left(\mathrm{q}_\mathrm{d}-\rho_{e_\mathrm{d}}\right), & \bm r \in \Omega_\mathrm{d},\\
    \sum_{\ell \in \left\{+,-, p\right\}} ez_\ell \rho_\ell, & \bm r \in \Omega_\mathrm{M}.
  \end{cases}
\end{equation}
Differentiating the charge density with respect to time, we see that it satisfies the conservation
equation $\dot{\rho}_c = -\nabla \cdot \boldsymbol{j}_{\rho_c}$ where
\begin{equation}
  \boldsymbol{j}_{\rho_c} =
  \begin{cases}
    eM_{e_\mathrm{d}}\nabla \left( \frac{\tilde{\mu}^*_{e_\mathrm{d}}}{k_{\mathrm{B}}T}\right), & \bm r \in \Omega_\mathrm{d},
    \\
    -e\sum_{\ell \in \left\{+, -, p\right\}}z_\ell M_\ell \nabla \left( \frac{\tilde{\mu}_\ell^*}{k_{\mathrm{B}}T}\right), & \bm r \in \Omega_\mathrm{M}.
  \end{cases}
  \label{jrhoC}
\end{equation}
We call $\boldsymbol j_{\rho_c}$ the conduction-current density: it is electronic in the electrodes
and ionic in the electrolyte. We observe that, by Equations~\eqref{poissonElectrodes}
and~\eqref{poissonElectrolyte},
\begin{equation}
  \dot{\rho}_c =
  \begin{cases}
    -\nabla \cdot \left( \frac{\epsilon_\mathrm{d}}{4 \pi l_\mathrm{B}}\frac{e^2}{k_{\mathrm{B}}T} \frac{\partial \nabla \psi_\mathrm{d}}{\partial t}\right), &\bm r \in \Omega_\mathrm{d},
    \\
    -\nabla \cdot \left( \frac{\epsilon_\mathrm{M}}{4 \pi l_\mathrm{B}}\frac{e^2}{k_{\mathrm{B}}T} \frac{\partial \nabla \psi_\mathrm{M}}{\partial t}\right), &\bm r \in \Omega_\mathrm{M}.
  \end{cases}
  \label{dotRhoCAlt}
\end{equation}

With $\bm E_\mathrm{d}:=-\nabla\psi_\mathrm{d}$ and $\bm E_\mathrm{M}:=-\nabla\psi_\mathrm{M}$
denoting the electrode and electrolyte electric fields, respectively, we define the total electric
current density, as in~\cite{kumar2017rayleighian}, by
\begin{equation}
  \bm I_\mathrm{tot} = \begin{cases}
    \frac{\epsilon_\mathrm{d}}{4 \pi l_\mathrm{B}}\frac{e^2}{k_{\mathrm{B}}T} \frac{\partial \bm E_\mathrm{d}}{\partial t}+ eM_{e_\mathrm{d}}\nabla \left( \frac{\tilde{\mu}_{e_\mathrm{d}}^*}{k_{\mathrm{B}}T}\right) , & \bm r \in \Omega_\mathrm{d},\\
    \frac{\epsilon_\mathrm{M}}{4 \pi l_\mathrm{B}}\frac{e^2}{k_{\mathrm{B}}T} \frac{\partial \bm E_\mathrm{M}}{\partial t}- e\sum_{\ell \in \left\{+, -, p\right\}}z_\ell M_{\ell}\nabla \left( \frac{\tilde{\mu}_{\ell}^*}{k_{\mathrm{B}}T}\right), & \bm r \in \Omega_\mathrm{M}.
  \end{cases}\label{totalElectricCurrentDensity}
\end{equation}
At the right terminal, Equation~\eqref{psiNeumannBCRight} makes the normal displacement current
vanish, and Equation~\eqref{currentBCR} gives $\boldsymbol{I}_\mathrm{tot}\cdot\boldsymbol n_\mathrm
R=I_\mathrm{app}$. At the left terminal the fixed value $\psi_\mathrm L=0$ does not by itself
prescribe the normal displacement. Instead, global neutrality and the integrated Poisson equation,
together with the right Neumann condition, give zero net left-terminal displacement flux. Hence the
integrated total currents satisfy
\begin{equation}
  \int_{\Gamma_\mathrm L}\boldsymbol I_\mathrm{tot}\cdot\boldsymbol n_\mathrm L\,\mathrm dS=-A_\Gamma I_\mathrm{app},\qquad
  \int_{\Gamma_\mathrm R}\boldsymbol I_\mathrm{tot}\cdot\boldsymbol n_\mathrm R\,\mathrm dS=A_\Gamma I_\mathrm{app}.
\end{equation}
In the one-dimensional constant-cross-section model these integrated relations reduce to the
corresponding pointwise terminal current densities.

\begin{theorem}[Divergence-free total electric current density]\label{thmdivFreeI}
  {The total electric current density is divergence-free in each bulk domain. Its normal component
  is continuous across each electrode--electrolyte interface, so it is divergence-free across the
  complete cell in the distributional sense.}
\end{theorem}
The proof is given in Section~\ref{sec:proofThmDivFreeI}.

\begin{corollary}
  { In the one-dimensional model, $\boldsymbol{I}_\mathrm{tot} = I_{\mathrm{tot},x}\hat{i}$ and
  $\partial_x I_{\mathrm{tot},x}=0$. Interfacial current continuity implies that the axial component
  $I_{\mathrm{tot},x}$ is spatially constant throughout the connected cell and is fixed by the
  applied terminal current. In more than one dimension, divergence-free current need not be
  spatially uniform.}
\end{corollary}

\subsection{Mobilities and Reaction Kinetics}
\label{sec:recapturingKnownDynamics}

For recovering Fickian-type diffusion, we let $M_\ell = D_\ell \rho_\ell $ for ${\ell} \in
\left\{e_\mathrm{L}, e_\mathrm{R}, i_\mathrm{L}, i_\mathrm{R}, +, -\right\}$. Because $\rho_p$ is a
segment density so that each of the $N$ segments can be charged, we instead set
\begin{equation}
  M_p=N D_p^{\mathrm{chain}}\rho_p=:D_p\rho_p,
\end{equation}
where $D_p^{\mathrm{chain}}$ is the polymer-chain center-of-mass diffusivity and the coefficient
used in the transport equations is $D_p:=N D_p^{\mathrm{chain}}$. This convention prevents a
spurious second factor of $1/N$ in the dilute polymer diffusion rate.
Substituting this choice of $M_{\ell}$ and Equations~\eqref{chem_potentials}
and~\eqref{exchange_chem_potentials} into~\eqref{ellHatGovEqn} and~\eqref{ellGovEqn}, the diffusing
species' governing equations are as follows.

For $\mathrm{d} \in \left\{\mathrm{L},\mathrm{R}\right\}$, in $\Omega_\mathrm{d}$
\begin{eqnarray}
  \frac{ \partial \rho_{e_\mathrm{d}} }{\partial t}
  &=& \nabla \!\cdot\!\left[D_{e_\mathrm{d}}\left( \nabla \rho_{e_\mathrm{d}} - \rho_{e_\mathrm{d}} \nabla \left( \frac{e\psi_\mathrm{d}}{k_{\mathrm{B}}T}\right)\right) \right]
  \label{electronGovEqn}\\
  \frac{\partial \rho_{i_\mathrm{d}}}{\partial t}
  &=& \nabla \!\cdot\! \left[ \frac{D_{i_\mathrm{d}}}{1-\frac{\rho_{i_\mathrm{d}}}{\rho_{i_\mathrm{d}}^o}} \nabla \rho_{i_\mathrm{d}}\right] \label{iGovEqn}.
\end{eqnarray}

In $\Omega_\mathrm{M}$, for $\ell \in \{+, -\}$,
\begin{equation}
  \frac{ \partial \rho_\ell}{\partial t}
  = \nabla \!\cdot\! \left[ D_\ell \left( \nabla \rho_\ell + z_\ell \rho_\ell\nabla \left( \frac{e\psi_\mathrm{M}}{k_{\mathrm{B}}T}\right)-\frac{\rho_+^o}{\rho_{\ell }^o}\frac{\rho_{\ell}}{\rho_{\mathrm{V}_\mathrm{M}}}\nabla \rho_{\mathrm{V}_\mathrm{M}}\right)\right],\label{ell_gov_eqn_non_dilute}
\end{equation}
and
\begin{equation}
  \frac{ \partial \rho_p}{\partial t}
  = \nabla \!\cdot\! \left[ D_p \left( \frac{1}{N}\nabla \rho_p + z_p \rho_p\nabla \left( \frac{e\psi_\mathrm{M}}{k_{\mathrm{B}}T}\right)-\frac{\rho_+^o}{\rho_{p }^o}\frac{\rho_{p}}{\rho_{\mathrm{V}_\mathrm{M}}}\nabla \rho_{\mathrm{V}_\mathrm{M}}\right)\right].\label{p_gov_eqn_non_dilute}
\end{equation}

Mathematical rearrangements are given in Section~\ref{FickianDiffSI} of the Supplementary
Information. {In the absence of an electric field, Equations~\eqref{ell_gov_eqn_non_dilute}
and~\eqref{p_gov_eqn_non_dilute} reduce to Fick's second law only when the electrolyte is dilute or
the vacancy fraction is effectively spatially uniform. Otherwise, the flux of each species depends
on both its own density gradient and the vacancy-density gradient, so the species diffuse in a
coupled manner. Consequently, $D_\ell$ sets the scale of the species mobility but does not by itself
give the measurable rate at which a concentration profile spreads. 

\begin{remark}
  {The mobility choices above are intended for states in the interior of the admissible composition
  set. For a classical solution on the time interval considered, we assume uniform positive lower
  bounds for $\rho_{e_\mathrm{d}}$, $\phi_{i_\mathrm{d}}$,
  $1-\phi_{i_\mathrm{d}}=\phi_{\mathrm{V}_\mathrm{d}}$, and every electrolyte fraction retained in
  the applicable limiting model, including $\phi_{\mathrm{V}_\mathrm{M}}=1-\phi_+-\phi_--\phi_p$. A
  species absent in a limiting model is removed before any logarithmic chemical potential is
  evaluated. The dissipation theorem is conditional on remaining in this admissible region; it does
  not itself prove positivity or exclude saturation. A vacancy-mediated alternative could use
  nonnegative mobilities such as
  $M_{i_\mathrm{d}}=D_{i_\mathrm{d}}\rho_{i_\mathrm{d}}\phi_{\mathrm{V}_\mathrm{d}}$ and
  $M_\ell=D_\ell\rho_\ell\phi_{\mathrm{V}_\mathrm{M}}$. Such choices preserve the dissipation
  structure but lead to different transport equations, conductivity relations, parameter values, and
  numerical results, and are therefore left for future work.}
\end{remark}

For recovering a reaction rate expression similar to the Butler-Volmer kinetics, we let
$f_\mathrm{inc}(x) = \exp\left(x\right)$ and consider the reactions at each interface described in
Equations~\eqref{leftLiReaction} and~\eqref{rightLiReaction}. Let $\gamma_n$ denote the activity
coefficient of species $n\in\mathcal I$, and let $\gamma_{\ddagger,\mathrm d}$ denote the activity
coefficient of the transition state on $\Sigma_\mathrm d$. Define the dimensionless activities
\begin{equation}
  a_{e_\mathrm{d}} := \gamma_{e_\mathrm{d}}\phi_{e_\mathrm d},\qquad
  a_m := \gamma_m\frac{\rho_m}{\rho_m^o}=\gamma_m\phi_m,\quad m\in\mathcal V.
\end{equation}
When an activity carries an additional interface subscript $\mathrm d$, it denotes the value of that
activity on $\Sigma_\mathrm d$. We further define the composite oxidized- and reduced-state
electrochemical potentials and activities by
\begin{eqnarray}
  \tilde{\mu}_{\mathit{O},\mathrm{d}}  = {\tilde{\mu}}_{\mathrm{V}_\mathrm{d}}^* +{\tilde{\mu}}_+^*\label{mu_O}\\
  \tilde{\mu}_{\mathit{R},\mathrm{d}} = {\tilde{\mu}}_{i_\mathrm{d}}^*+ {\tilde{\mu}}_{\mathrm{V}_\mathrm{M}}^*\label{mu_R}\\
  a_{\mathit{O},\mathrm{d}}= a_{\mathrm{V}_\mathrm{d},\mathrm{d}}a_{+,\mathrm{d}}
  \\
  a_{\mathit{R},\mathrm{d}} = a_{i_\mathrm{d},\mathrm{d}}a_{\mathrm{V}_\mathrm{M},\mathrm{d}}
  .  \label{a_R}
\end{eqnarray}
In this scenario, we postulate the exchange chemical potential of the transition state at
$\Sigma_\mathrm{d}$, $\mathrm{d}\in \left\{\mathrm{L},\mathrm{R}\right\}$, to be
\begin{equation}
  \tilde{\mu}_{\ddagger, \mathrm{d}}
  = k_{\mathrm{B}}T \ln\left(\frac{\gamma_{\ddagger, \mathrm{d}}}{\left(a_{e_\mathrm{d}}a_{\mathit{O},\mathrm{d}}\right)^{1-\alpha_\mathrm{d}}a_{\mathit{R},\mathrm{d}}^{\alpha_\mathrm{d}}}\right) + \left(1-\alpha_\mathrm{d}\right) \left( {\tilde{\mu}}_{e_\mathrm{d}}^*+\tilde{\mu}_{\mathit{O},\mathrm{d}}\right) +\alpha_\mathrm{d} \tilde{\mu}_{\mathit{R},\mathrm{d}},\label{transition_state}
\end{equation}
where $\alpha_\mathrm{d}$ is as defined in Table~\ref{tab:table2}.
From Equations~\eqref{sk},~\eqref{mu_O}, and~\eqref{mu_R}, for this choice of $f_\mathrm{inc}(x)$,
Equation~\eqref{r_transitionstatenotsubbed} becomes
\begin{equation}
  \nu_\mathrm{d}
  = -\mathrm{sgn}\left(s_{+,\mathrm{d}}\right)k_\mathrm d\exp \left( -\frac{\tilde{\mu}_{\ddagger, \mathrm{d}}}{k_{\mathrm{B}}T}\right)\left[\exp \left( \frac{{\tilde{\mu}}_{e_\mathrm{d}}^*+{\tilde{\mu}}_{\mathit{O},\mathrm{d}}}{k_{\mathrm{B}}T}\right)- \exp \left( \frac{\tilde{\mu}_{\mathit{R},\mathrm{d}}}{k_{\mathrm{B}}T}\right) \right].\label{firstEqnRmrk2}
\end{equation}

At equilibrium,
\begin{align}
  \nu_\mathrm{d} &= 0
\end{align}
which implies
\begin{align}
  \tilde{\mu}_{e_\mathrm{d}}^* + \tilde{\mu}_{\mathit{O},\mathrm{d}} &= \tilde{\mu}_{\mathit{R},\mathrm{d}}.
\end{align}
Out of equilibrium, the reaction rate is controlled by the difference in electrochemical potentials
of reactants and products across the interface (i.e., $\frac{\tilde{\mu}_{\mathit{R},\mathrm{d}} -
{\tilde{\mu}}_{e_\mathrm{d}}^*- \tilde{\mu}_{\mathit{O},\mathrm{d} }}{e}$ ), which we denote as the
overpotential, $\eta_\mathrm{d}$. Define the local, generally state-dependent exchange-current
density by
\begin{equation}
  I_{\mathrm{ex}, \mathrm{d}} := e N_A k_\mathrm d \frac{\left(a_{e_\mathrm{d}}a_{\mathit{O},\mathrm{d}}\right)^{1-\alpha_\mathrm{d}}a_{\mathit{R},\mathrm{d}}^{\alpha_\mathrm{d}}}{\gamma_{\ddagger, \mathrm{d}}}.\label{def:I0}
\end{equation}
Substituting Equation~\eqref{transition_state} into~\eqref{firstEqnRmrk2} and using the definition
of $\eta_\mathrm{d}$,
\begin{equation}
  \nu_\mathrm{d} =-\mathrm{sgn}\left(s_{+,\mathrm{d}}\right)\frac{ I_{\mathrm{ex}, \mathrm{d}}}{eN_A}
  \left[\exp\left(\frac{-\alpha_\mathrm{d} e \eta_\mathrm{d} }{k_{\mathrm{B}}T}\right)-\exp\left(\frac{\left(1-\alpha_\mathrm{d}\right) e\eta_\mathrm{d}}{k_{\mathrm{B}}T}\right)\right] \label{reactionRateBV}
\end{equation}
{To fix the sign independently of which direction is operating, define $\Delta G_\mathrm d^\circ$
for the reduction direction $e_\mathrm d+\mathit O_\mathrm d\rightarrow\mathit R_\mathrm d$, using
the composite states defined above:
\[
\Delta G_\mathrm d^\circ
:=\mu_{i_\mathrm d}^\circ+\mu_{\mathrm{V}_\mathrm M}^\circ
-\mu_{e_\mathrm d}^\circ-\mu_+^\circ-\mu_{\mathrm{V}_\mathrm d}^\circ.
\]
The standard reduction potential is then $E_\mathrm d^\circ:=-\Delta G_\mathrm d^\circ/e$. This
definition remains tied to the reduction direction even when the net reaction reverses during
operation. Because the ideal free-energy densities in~\eqref{eq:H_d}--\eqref{eq:H_M} are written as
$\rho\ln(\rho/\rho^o)$ rather than $\rho[\ln(\rho/\rho^o)-1]$, their derivatives produce a
convention-dependent constant $k_{\mathrm{B}}T/e$. We absorb it into the effective standard
potential $\widehat E_\mathrm d^\circ:=E_\mathrm d^\circ+k_{\mathrm{B}}T/e$. The illustrative
symmetric-cell calculations use the same effective standard-state reference at both electrodes, so
this constant cancels from the full-cell voltage; the plotted voltage is therefore relative to that
reference.}
{The simulations use the ideal-mixture closure $\gamma_n=1$ and treat $\gamma_{\ddagger,\mathrm d}$
as constant. We therefore define the single effective rate constant
\[
k_\mathrm d^{\mathrm{eff}}:=\frac{k_\mathrm d}{\gamma_{\ddagger,\mathrm d}},
\]
so Equation~\eqref{def:I0} may be written as $I_{\mathrm{ex},\mathrm d}=eN_Ak_\mathrm
d^{\mathrm{eff}}(a_{e_\mathrm d}a_{\mathit O,\mathrm d})^{1-\alpha_\mathrm d}a_{\mathit R,\mathrm
d}^{\alpha_\mathrm d}$. Composition-dependent activity coefficients may be introduced only together
with a compatible nonideal free-energy density; inserting them solely into the kinetic law would not
preserve the stated thermodynamic structure.}
{The reaction is driven by the electrochemical affinity of the reactants and products. In the
present transmission model, this affinity does not contain an independent compact-layer potential
jump.}

Letting
\begin{equation}
  \Delta \psi_\mathrm{d} = \psi_\mathrm{d} - \psi_\mathrm{M}\label{interfacial_voltage}
\end{equation}
and
\begin{equation}
  \Delta \psi_\mathrm{d}^\mathrm{eq} = \widehat E_\mathrm d^\circ+\frac{k_{\mathrm{B}} T}{e}\ln \left[\left(
  \frac{\rho_+}{\rho_+^o-\rho_+-\frac{\rho_+^o}{\rho_-^o}\rho_--\frac{\rho_+^o}{\rho_p^o}\rho_p}\right)\phi_{e_\mathrm d}\frac{\rho_{i_\mathrm{d}}^o-\rho_{i_\mathrm{d}}}{\rho_{i_\mathrm{d}}}
  \right], \label{eq:eleceq}
\end{equation}
for $\mathrm{d}\in\left\{\mathrm{L},\mathrm{R}\right\}$, then
\begin{equation}
  \eta_\mathrm{d} =
  \frac{\tilde{\mu}_{\mathit{R},\mathrm{d}} - {\tilde{\mu}}_{e_\mathrm{d}}^*- \tilde{\mu}_{\mathit{O},\mathrm{d} }}{e}= \Delta \psi_\mathrm{d} - \Delta \psi_\mathrm{d}^\mathrm{eq}.
  \label{eta}
\end{equation}
{Boundary
condition~\eqref{transmission_cont} implies $\Delta\psi_\mathrm{d}=0$, and hence
$\eta_\mathrm{d}=-\Delta\psi_\mathrm{d}^\mathrm{eq}$. The nonequilibrium driving force is supplied
through the composition and electron-density dependence of $\Delta\psi_\mathrm{d}^\mathrm{eq}$
rather than through an independent electrostatic-potential jump. The resulting
difference-of-exponentials retains the Butler--Volmer functional form.}
In particular, defining
\begin{equation}
  g_{\mathrm{BV}, \mathrm{d}}:=\left[\exp \left(\frac{-\alpha_\mathrm{d} e \eta_\mathrm{d} }{k_{\mathrm{B}}T}\right) -  \exp \left(\frac{\left(1-\alpha_\mathrm{d}\right) e \eta_\mathrm{d}}{k_{\mathrm{B}}T}\right) \right],
\end{equation}
$\nu_\mathrm d=-\mathrm{sgn}\left(s_{+,\mathrm d}\right)I_{\mathrm{ex},\mathrm
d}g_{\mathrm{BV},\mathrm d}/(eN_A)$, so the interfacial free-energy contribution $-N_A s_{+,\mathrm
d}e\eta_\mathrm d\nu_\mathrm d=I_{\mathrm{ex},\mathrm d}\eta_\mathrm d g_{\mathrm{BV},\mathrm d}$, which is
nonpositive.

Hence, for $\mathrm{d} \in \left\{\mathrm{L, R}\right\}$, using boundary
conditions~\eqref{ellHat_BC} and~\eqref{ell_BC} and Equation~\eqref{sk} (specifically that
$|s_{+,\mathrm{d}}| = 1$), we have, on $\Sigma_\mathrm{d}$,
\begin{eqnarray}
  M_{e_\mathrm{d}} \nabla \left( \frac{\tilde{\mu}_{e_\mathrm{d}}^*}{k_{\mathrm{B}}T}\right)\!\cdot\! \bm n_\mathrm{d}
  = -\frac{I_{\mathrm{ex}, \mathrm{d}}}{e}
  g_{\mathrm{BV}, \mathrm{d}}\label{electron_flux_reaction}\\
  M_+ \nabla \left( \frac{\tilde{\mu}_+^*}{k_{\mathrm{B}}T}\right)\!\cdot\!\bm n_\mathrm{M}
  = -\frac{I_{\mathrm{ex}, \mathrm{d}}}{e}
  g_{\mathrm{BV}, \mathrm{d}}.\label{cation_flux_reaction}
\end{eqnarray}
This is the Butler-Volmer equation with the state-dependent exchange current $I_{\mathrm{ex}, \mathrm{d}}$ from
Equation~\eqref{def:I0}.
{A tabulated value $I_{\mathrm{ex},\mathrm d}^{\mathrm{ref}}$ must be
understood at a specified reference state. Solving the preceding exchange-current relation at that
state gives the calibration formula for the already-defined effective rate constant,
\begin{equation}
  k_\mathrm d^{\mathrm{eff}}=\frac{I_{\mathrm{ex},\mathrm d}^{\mathrm{ref}}}{eN_A\left(a_{e_\mathrm{d}}^{\mathrm{ref}}a_{O,\mathrm d}^{\mathrm{ref}}\right)^{1-\alpha_\mathrm d}\left(a_{R,\mathrm d}^{\mathrm{ref}}\right)^{\alpha_\mathrm d}}.
\end{equation}
The local exchange current is then evaluated from the evolving activities.}

\begin{remark}
  Equation~\eqref{transition_state} is the model of the transition state
  from~\cite{bazant2013theory}, with the exchange electrochemical potentials used in place of the
  electrochemical potentials.
\end{remark}

\begin{remark}
  The notation of the stoichiometric coefficients
  is different from that in~\cite{bazant2013theory}, where the stoichiometric coefficients are
  positive. This is inferred from that paper's statement that $\sum_i
  s_{i,\mathit{O}}z_{i,\mathit{O}}-n = z_\mathit{O} -n = z_\mathit{R} = \sum_j
  s_{j,\mathit{R}}z_{j,\mathit{R}}$ for the general Faradaic reaction, where $n$ is the number of
  electrons reacting in the reduction process. In our context, $s_{i,\mathit{O}} \leq 0$ and
  $s_{j,\mathit{R}} \geq 0$ when the forward reaction is reduction.
  Considering the reaction $\mathrm{Cu}^{2+} + e^- \rightarrow \mathrm{Cu}^{+}$, for the statement
  to be true, $s_{\mathrm{Cu}^{2+},\mathit{O}} = 1$ and $s_{\mathrm{Cu}^{+}, \mathit{R}} = 1$, but
  in our context, $s_{\mathrm{Cu}^{2+},\mathit{O}} = -1$ and $s_{\mathrm{Cu}^{+}, \mathit{R}} = 1$.
\end{remark}

\begin{remark}
  By the extension of the no voids assumption in the electrodes to the electrode-electrolyte
  interface, define the effective oxidized- and reduced-state kinetic prefactors
  \begin{eqnarray}
    k_\mathit{O}^\mathrm{d}:= \frac{k_\mathrm d \gamma_{e_\mathrm{d}}\gamma_+\gamma_{\mathrm{V}_\mathrm{d}}}{\gamma_{\ddagger, \mathrm{d}}},\\
    k_\mathit{R}^\mathrm{d}:=\frac{k_\mathrm d\gamma_{i_\mathrm{d}}\gamma_{\mathrm{V}_\mathrm{M}}}{\gamma_{\ddagger, \mathrm{d}}}.
  \end{eqnarray}
  Then we re-write $I_{\mathrm{ex}, \mathrm{d}}$ as follows.
  \begin{equation}
    \frac{I_{\mathrm{ex}, \mathrm{d}}}{eN_A}
    =  \left(k_\mathit{O}^\mathrm{d}\phi_{e_\mathrm d}\phi_+ \left(1-\phi_{i_\mathrm{d}}\right)\right)^{1-\alpha_\mathrm{d}}\left(k_\mathit{R}^\mathrm{d}\phi_{\mathrm{V}_\mathrm{M}} \phi_{i_\mathrm{d}}\right)^{\alpha_\mathrm{d}}, \label{rewrittenI0}
  \end{equation}
  In Equation~\eqref{rewrittenI0}, $\phi_{i_\mathrm{d}}$ is the volume fraction of intercalated
  cation, or of filled electrode vacancy sites, and $1-\phi_{i_\mathrm{d}}$ is the volume fraction
  of electrode vacancies, or of unfilled electrode vacancy sites. Then the term
  \begin{eqnarray}
    \left(1-\phi_{i_\mathrm{d}}\right)^{1-\alpha_\mathrm{d}} \phi_{i_\mathrm{d}}^{\alpha_\mathrm{d}}
  \end{eqnarray}
  can be thought of as serving a similar function to the term
  \begin{eqnarray}\left(c_{\max} -c \right)^{\alpha_{c1}}c^{\alpha_{a1}}.
  \end{eqnarray} in Equation~(8) of Reference~\cite{doyle1993modeling}. Here $c$ is the electrolyte salt molar concentration, $c_{\max}$ is its polymer-lattice site capacity (salt solubility limit), and $\alpha_{c1}$ and $\alpha_{a1}$ are the cathodic and anodic charge-transfer coefficients at the lithium interface. These borrowed symbols are distinct from our number densities and volume fractions.
\end{remark}

\begin{remark}
  Recent work relates intercalation current to vacancy availability~\cite{zhang2025lithium}. Holding
  the other activities fixed, the exchange current in this model contains the electrode factor
  \[
  (1-\phi_{i_\mathrm d})^{1-\alpha_\mathrm d}\phi_{i_\mathrm d}^{\alpha_\mathrm d}.
  \]
  This factor depends on both vacant and occupied sites and is generally nonmonotonic: its maximum occurs at
  vacancy fraction $1-\alpha_\mathrm d$, and it vanishes at either occupancy endpoint. An operating
  current also depends on overpotential through the full Butler--Volmer relation; comparison with an
  experimental vacancy dependence must account for both factors.
\end{remark}

\section{Model Recapitulation and Non-Dimensionalization}
\label{sec:nonDimSimplify}
For clarity, we first re-capitulate our complete working model with all constitutive choices
established, including the specifics of the domains, before non-dimensionalizing it for simulation
purposes.

\subsection{Model Recapitulation}
\label{sec:recap}
Suppose that the three domains have a common finite rectangular cross-section $A$, with paired
periodic lateral boundaries. The electrodes have lengths $L_e$, and the electrolyte occupies
$0<x<L$. In the one-dimensional specialization all fields are independent of the transverse
coordinates.\\
In $\Omega_\mathrm{d}$, $\mathrm{d} \in \left\{\mathrm{L}, \mathrm{R}\right\}$,
\begin{align}
  \frac{\partial \rho_{e_\mathrm{d}}}{\partial t} &= \nabla \cdot \left[D_{e_\mathrm{d}}\left( \nabla \rho_{e_\mathrm{d}} - \rho_{e_\mathrm{d}} \nabla  \frac{e\psi_\mathrm{d}}{k_{\mathrm{B}}T}\right) \right] \\
  \frac{\partial \rho_{i_\mathrm{d}}}{\partial t} &= \nabla \cdot \left[ \frac{D_{i_\mathrm{d}} \rho_{i_\mathrm{d}}^o}{\rho_{i_\mathrm{d}}^o-\rho_{i_\mathrm{d}}} {\nabla \rho_{i_\mathrm{d}}}\right] \label{dimrhoI}\\
  -\nabla \cdot \left(\frac{\epsilon_\mathrm{d}}{4\pi l_\mathrm{B}}\nabla \frac{e\psi_\mathrm{d}}{k_{\mathrm{B}}T} \right) &= \mathrm{q}_\mathrm{d}-\rho_{e_\mathrm{d}},
\end{align}
where $\mathrm{q}_\mathrm{d}$ is given by Equation~\eqref{varrho}.\\
In $\Omega_\mathrm{M}$, for $\ell \in \left\{+, -\right\}$,
\begin{align}
  \frac{\partial \rho_\ell}{\partial t} &= \nabla\cdot\left[D_\ell \left( \nabla \rho_\ell + z_\ell \rho_\ell \nabla\frac{e\psi_\mathrm{M}}{k_{\mathrm{B}}T} - \frac{\rho_+^o}{\rho_\ell^o} \rho_\ell \frac{\nabla \phi_{\mathrm{V}_\mathrm{M}}}{\phi_{\mathrm{V}_\mathrm{M}}}\right)\right]\\
  \frac{\partial \rho_p}{\partial t} &= \nabla\cdot\left[D_p \left( \frac{1}{N}\nabla \rho_p + z_p \rho_p \nabla\frac{e\psi_\mathrm{M}}{k_{\mathrm{B}}T} - \frac{\rho_+^o}{\rho_p^o}\rho_p \frac{\nabla \phi_{\mathrm{V}_\mathrm{M}}}{\phi_{\mathrm{V}_\mathrm{M}}}\right)\right]\\
  -\nabla \cdot \left(\frac{\epsilon_\mathrm{M}}{4\pi l_\mathrm{B}}\nabla \frac{e\psi_\mathrm{M}}{k_{\mathrm{B}}T} \right) &= \sum_{\ell \in \left\{+, -, p\right\}}z_\ell \rho_\ell,
\end{align}
where
\begin{align}
  \phi_{\mathrm{V}_\mathrm{M}} &= 1-\frac{\rho_+}{\rho_+^o}-\frac{\rho_-}{\rho_-^o}-\frac{\rho_p}{\rho_p^o}.
\end{align}
Letting $\ell \in \left\{+, -\right\}$ and $\mathrm{d} \in \left\{\mathrm{L},\mathrm{R}\right\}$,
the boundary conditions are as follows.
On ${\Gamma}_\mathrm{d}$,
\begin{equation}
  - \frac{{D}_{i_\mathrm{d}}\rho_{i_\mathrm{d}}^o}{\rho_{i_\mathrm{d}}^o-\rho_{i_\mathrm{d}}} \left( {\nabla}\rho_{i_\mathrm{d}}\right)\!\cdot\!\bm n_\mathrm{d} = 0.
\end{equation}
On the electrode-external circuit interfaces:
\begin{eqnarray}
  -
  {D}_{e_\mathrm{L}}\left({\nabla} {\rho_{e_\mathrm{L}}} -{\rho_{e_\mathrm{L}}} {\nabla}\frac{e{\psi}_\mathrm{L}}{k_{\mathrm{B}}T} \right) \!\cdot\!\bm n_\mathrm{L}\Big|_{{\Gamma}_\mathrm{L}
  }&=&\frac{{I}_\mathrm{app}}{e}\\
  {D}_{e_\mathrm{R}}\left({\nabla} {\rho_{e_\mathrm{R}}} -{\rho_{e_\mathrm{R}}} {\nabla}\frac{e{\psi}_\mathrm{R}}{k_{\mathrm{B}}T} \right) \!\cdot\!\bm n_\mathrm{R}\Big|_{{\Gamma}_\mathrm{R}
  }&=&\frac{{I}_\mathrm{app}}{e}\\
  \psi_\mathrm{L}\big|_{\Gamma_\mathrm{L}} &=&0\\
  {\nabla \psi_\mathrm{R} \cdot \bm n_\mathrm{R}\Big|_{\Gamma_\mathrm{R}}} &=& {0}.
\end{eqnarray}
{The left terminal fixes the electrostatic reference potential, while the
right terminal retains homogeneous normal-displacement data.}
On ${\Sigma}_\mathrm{d}$ the electrostatic potential satisfies
\begin{eqnarray}
  \psi_\mathrm{d}-\psi_\mathrm{M} = 0, \\
  \epsilon_\mathrm{d}\nabla \psi_\mathrm{d}\!\cdot\! \bm n_\mathrm{d}+       \epsilon_\mathrm{M}\nabla \psi_\mathrm{M}\!\cdot\! \bm n_\mathrm{M}=0,
\end{eqnarray}
and the electrode species satisfy
\begin{eqnarray}
  -
  {D}_{e_\mathrm{d}}\left({\nabla} {\rho_{e_\mathrm{d}}} -{\rho_{e_\mathrm{d}}} {\nabla}\frac{e{\psi}_\mathrm{d}}{k_{\mathrm{B}}T} \right) \!\cdot\!\bm n_\mathrm{d}+N_A s_{e_\mathrm{d},\mathrm{d}}\nu_\mathrm{d}
  = 0 \\
  - \frac{{D}_{i_\mathrm{d}}\rho_{i_\mathrm{d}}^o}{\rho_{i_\mathrm{d}}^o-\rho_{i_\mathrm{d}}} \left( {\nabla}\rho_{i_\mathrm{d}}\right)\!\cdot\!\bm n_\mathrm{d}+N_A s_{i_\mathrm{d},\mathrm{d}}\nu_\mathrm{d}
  = 0 \\
\end{eqnarray}
and the electrolyte species satisfy
\begin{eqnarray}
  -D_\ell \left( \nabla \rho_\ell + z_\ell \rho_\ell \nabla\frac{e\psi_\mathrm{M}}{k_{\mathrm{B}}T} - \frac{\rho_+^o}{\rho_\ell^o} \rho_\ell \frac{\nabla \phi_{\mathrm{V}_\mathrm{M}}}{\phi_{\mathrm{V}_\mathrm{M}}}\right)\!\cdot\!\bm n_\mathrm{M}+N_A s_{\ell,\mathrm{d}}\nu_\mathrm{d} = 0\\
  -D_p \left( \frac{1}{N}\nabla \rho_p + z_p \rho_p \nabla\frac{e\psi_\mathrm{M}}{k_{\mathrm{B}}T} - \frac{\rho_+^o}{\rho_p^o}\rho_p \frac{\nabla \phi_{\mathrm{V}_\mathrm{M}}}{\phi_{\mathrm{V}_\mathrm{M}}}\right)\!\cdot\! \bm n_\mathrm{M}+  N_A s_{p,\mathrm{d}}{\nu}_\mathrm{d} =0,
\end{eqnarray}
where $\ell \in \left\{+, -\right\}$,
\begin{align}
  {\nu}_\mathrm{d} &= {k_\mathrm d}\left[ f_\mathrm{inc}\left(\frac{A^{\mathrm{r}}_{\mathrm{d}}-\tilde{\mu}_{\ddagger, \mathrm{d}}}{k_{\mathrm{B}}T}\right)-f_\mathrm{inc}\left(\frac{A^{\mathrm{p}}_{\mathrm{d}}-\tilde{\mu}_{\ddagger, \mathrm{d}}}{k_{\mathrm{B}}T}\right)\right],
\end{align} and $A^{\mathrm{r}}_{\mathrm{d}}, A^{\mathrm{p}}_{\mathrm{d}}$ are as in Equations~\eqref{Ar} and~\eqref{Ap}.

We also have the outputs
\begin{equation}
  V\left(t\right) = \frac{\mu_{e_\mathrm L}^\circ-\mu_{e_\mathrm R}^\circ}{e}+\frac{k_{\mathrm{B}} T}{e}\left[\left( \ln \phi_{e_\mathrm L} - \frac{e\psi_\mathrm{L}}{k_{\mathrm{B}}T}\right)\Big|_{\Gamma_\mathrm{L}}-\left( \ln \phi_{e_\mathrm R} -\frac{e\psi_\mathrm{R}}{k_{\mathrm{B}}T} \right)\Big|_{\Gamma_\mathrm{R}} \right].
\end{equation}
{For the symmetric illustrative cell, $\mu_{e_\mathrm L}^\circ=\mu_{e_\mathrm R}^\circ$, so this
constant offset vanishes.}
{When $I_\mathrm{app}\ne0$, the diagnostic ratio $V/I_\mathrm{app}$ may be reported as an apparent
whole-cell area-specific resistance, but it is not used as a constitutive law and is not interpreted
as a contact resistance.}

\subsection{Non-Dimensionalization}
\label{sec:nondimensionalization}
We use the lithium-ion diffusion time scale $\tau = L^2/D_+$ and let $\delta = L_e/L$. Thus $\hat{D}_+=1$. We introduce
the scalings
\begin{equation}
  \hat{\bm r}:=\frac{\bm r}{L},\qquad
  \hat t:=\frac{t}{\tau},\qquad
  \hat\nabla_{\hat{\bm r}}:=L\nabla_{\bm r},\qquad
  \hat\psi_\mathrm d:= \frac{e\psi_\mathrm d}{k_{\mathrm{B}}T},\qquad
  \hat\nu_\mathrm d:=\frac{N_A L}{D_+\rho_+^o}\nu_\mathrm d,
\end{equation}
for $\mathrm d\in\{\mathrm L,\mathrm M,\mathrm R\}$ where applicable. The remaining nondimensional
parameter definitions and values used below are collected in Table~\ref{tab:table3}. Let
$\hat{\Omega}_{\mathrm{L}} = \{\hat{\bm r}: -\delta <\hat x<0,\ (\hat y,\hat z)\in\hat A\}$,
$\hat{\Omega}_{\mathrm{M}} = \{\hat{\bm r}: 0<\hat x<1,\ (\hat y,\hat z)\in\hat A\}$, and
$\hat{\Omega}_{\mathrm{R}} = \{\hat{\bm r}: 1<\hat x<1+\delta,\ (\hat y,\hat z)\in\hat A\}$. Here
$\hat A=A/L^2$ is a finite rectangular cross-section with paired periodic lateral boundaries; the
one-dimensional specialization assumes all fields depend only on $\hat x$. Let
$\hat{\Gamma}_{{\mathrm{d}}}$ be the far boundary of $\hat{\Omega}_{\mathrm{d}}$, and let
$\hat{\Sigma}_{{\mathrm{d}}}$ be the interface between $\hat{\Omega}_{\mathrm{d}}$ and
$\hat{\Omega}_{\mathrm{M}}$. Then, the non-dimensionalized model is as follows.
In $\hat{\Omega}_{\mathrm{d}}$, $\mathrm{d}\in \left\{\mathrm{L},\mathrm{R}\right\}$,
\begin{eqnarray}
  \frac{\partial \phi_{e_\mathrm d}}{\partial \hat{t}} = \hat{\nabla}_{\hat{\bm r}} \!\cdot\! \left[\hat{D}_{e_\mathrm{d}}\left(\hat{\nabla}_{\hat{\bm r}} \phi_{e_\mathrm d} -\phi_{e_\mathrm d} \hat{\nabla}_{\hat{\bm r}}\hat{\psi}_\mathrm{d} \right) \right]\label{electronGovEqnNonDim}\\
  \frac{\partial \phi_{i_\mathrm{d}}}{\partial \hat{t}} = \hat{\nabla}_{\hat{\bm r}} \!\cdot\! \left[ \frac{\hat{D}_{i_\mathrm{d}}}{1-\phi_{i_\mathrm{d}}} \left( \hat{\nabla}_{\hat{\bm r}}\phi_{i_\mathrm{d}}\right)\right] \label{iGovEqnNonDim}\\
  - \hat{\nabla}_{\hat{\bm r}} \!\cdot\! \left(\hat{\epsilon}_\mathrm{d}^2\hat{\nabla}_{\hat{\bm r}} \hat{\psi}_\mathrm{d} \right) =
  \frac{1}{2}\left( \frac{\mathrm{q}_\mathrm{d}}{\rho_+^o}-\phi_{e_\mathrm d}\right),\label{poissonElectrodesNonDim}
\end{eqnarray} where $\mathrm{q}_\mathrm{d}$ is given by Equation~\eqref{varrho}.

In $\hat{\Omega}_{\mathrm{M}}$, for $\ell \in \left\{+, -\right\}$,
\begin{eqnarray}
  \frac{\partial \phi_\ell}{\partial \hat{t}} = \hat{\nabla}_{\hat{\bm r}} \!\cdot\! \left[ \hat{D}_\ell \left(\hat{\nabla}_{\hat{\bm r}}\phi_\ell + z_\ell\phi_\ell \hat{\nabla}_{\hat{\bm r}} \hat{\psi}_\mathrm{M} - \hat{\rho}_\ell^o \frac{\phi_\ell}{ \phi_{\mathrm{V}_\mathrm{M}}}\hat{\nabla}_{\hat{\bm r}}\phi_{\mathrm{V}_\mathrm{M}}\right) \right]\label{ellGovEqnNonDim}\\
  \frac{\partial \phi_p}{\partial \hat{t}} = \hat{\nabla}_{\hat{\bm r}} \!\cdot\! \left[ \hat{D}_p \left( \frac{1}{N}\hat{\nabla}_{\hat{\bm r}}\phi_p + z_p\phi_p\hat{\nabla}_{\hat{\bm r}} \hat{\psi}_\mathrm{M}-\hat{\rho}_p^o\frac{\phi_p}{\phi_{\mathrm{V}_\mathrm{M}}} \hat{\nabla}_{\hat{\bm r}}\phi_{\mathrm{V}_\mathrm{M}} \right) \right]\phantom{++}\label{pGovEqnNonDim}\\
  - \hat{\nabla}_{\hat{\bm r}} \!\cdot\! \left(\hat{\epsilon}_\mathrm{M}^2 \hat{\nabla}_{\hat{\bm r}} \hat{\psi}_\mathrm{M}\right) =
  \frac{1}{2} \sum_{\ell \in \{+, -, p\}} \frac{z_\ell}{\hat{\rho}_\ell^o} \phi_\ell,\label{poissonElectrolyteNonDim}
\end{eqnarray}
where
\begin{equation}
  \phi_{\mathrm{V}_\mathrm{M}} = 1-\phi_+-\phi_--\phi_p.
\end{equation}

Letting $\ell \in \left\{+, -\right\}$ and $\mathrm{d} \in \left\{\mathrm{L},\mathrm{R}\right\}$,
the boundary conditions are as follows.
On $\hat{\Gamma}_\mathrm{d}$,
\begin{equation}
  - \frac{\hat{D}_{i_\mathrm{d}}}{1-\phi_{i_\mathrm{d}}} \left( \hat{\nabla}_{\hat{\bm r}}\phi_{i_\mathrm{d}}\right)\!\cdot\!\bm n_\mathrm{d} = 0. \label{noFluxINonDim}
\end{equation}
On the electrode-external circuit interfaces:
\begin{eqnarray}
  -
  \hat{D}_{e_\mathrm{L}}\left(\hat{\nabla}_{\hat{\bm r}} \phi_{e_\mathrm L} -\phi_{e_\mathrm L} \hat{\nabla}_{\hat{\bm r}}\hat{\psi}_\mathrm{L} \right) \!\cdot\!\bm n_\mathrm{L}\Big|_{\hat{\Gamma}_\mathrm{L}
  }&=& \hat{I}_\mathrm{app}\label{currentElectrodeNonDim}\\
  \hat{D}_{e_\mathrm{R}}\left(\hat{\nabla}_{\hat{\bm r}} \phi_{e_\mathrm R} -\phi_{e_\mathrm R} \hat{\nabla}_{\hat{\bm r}}\hat{\psi}_\mathrm{R} \right) \!\cdot\!\bm n_\mathrm{R}\Big|_{\hat{\Gamma}_\mathrm{R}
  }&=&\hat{I}_\mathrm{app}\label{fluxECBCNonDim}\\
  \hat{\psi}_\mathrm{L}\big|_{\hat{\Gamma}_\mathrm{L}} &=&0 \label{psiDirichletBCLeftNonDim}\\
  {\hat{\nabla}_{\hat{\bm r}}\hat{\psi}_\mathrm{R}\cdot\bm n_{\mathrm{R}} \Big|_{\hat{\Gamma}_\mathrm{R}}} &= &{0}\label{psiNeumannBCRightNonDim}
\end{eqnarray}
{The left Dirichlet value fixes the nondimensional potential reference; no
additional zero-mean constraint is imposed.}
On $\hat{\Sigma}_\mathrm{d}$ the electrostatic potential satisfies
\begin{eqnarray}
  \hat{\psi}_\mathrm{d}-\hat{\psi}_\mathrm{M}= 0 \label{transmission_contNonDim} \\
  \hat{\epsilon}_\mathrm{d}^2\hat{\nabla}_{\hat{\bm r}}  \hat{\psi}_\mathrm{d}\!\cdot\! \bm n_\mathrm{d}+      \hat{\epsilon}_\mathrm{M}^2\hat{\nabla}_{\hat{\bm r}}  \hat{\psi}_\mathrm{M}\!\cdot\! \bm n_\mathrm{M}=0,\label{transmission_jumpNonDim}
\end{eqnarray}
the electrode species satisfy
\begin{eqnarray}
  -\hat{D}_{e_\mathrm{d}}\left( \hat{\nabla}_{\hat{\bm r}}  \phi_{e_\mathrm d}-\phi_{e_\mathrm d}\hat{\nabla}_{\hat{\bm r}} \hat{\psi}_\mathrm{d}\right)\!\cdot\! \bm n_\mathrm{d}+s_{e_\mathrm{d},\mathrm{d}} \hat{\nu}_\mathrm{d}
  = 0
  \label{eBCNonDim} \\
  -\frac{\hat{D}_{i_\mathrm{d}}}{1-\phi_{i_\mathrm{d}}} \hat{\nabla}_{\hat{\bm r}}  \phi_{i_\mathrm{d}}\!\cdot\! \bm n_\mathrm{d}+\hat{\rho}_{i_\mathrm{d}}^o s_{i_{\mathrm{d}},\mathrm{d}} \hat{\nu}_\mathrm{d}
  = 0,
  \label{iBCNonDim}
\end{eqnarray}
and the electrolyte species satisfy
\begin{eqnarray}
  -\hat{D}_\ell \left( \hat{\nabla}_{\hat{\bm r}} \phi_\ell +z_\ell\phi_\ell\hat{\nabla}_{\hat{\bm r}} \hat{\psi}_\mathrm{M}-\hat{\rho}_\ell^o\frac{\phi_\ell}{\phi_{\mathrm{V}_\mathrm{M}}} \hat{\nabla}_{\hat{\bm r}} \phi_{\mathrm{V}_\mathrm{M}}\right) \!\cdot\! \bm n_\mathrm{M}+ \hat{\rho}_\ell^os_{\ell,\mathrm{d}}\hat{\nu}_\mathrm{d} =0\label{ellBCNonDim}\\
  -\hat{D}_p \left( \frac{1}{N}\hat{\nabla}_{\hat{\bm r}} \phi_p +z_p\phi_p\hat{\nabla}_{\hat{\bm r}} \hat{\psi}_\mathrm{M}-\hat{\rho}_p^o\frac{\phi_p}{\phi_{\mathrm{V}_\mathrm{M}}}\hat{\nabla}_{\hat{\bm r}} \phi_{\mathrm{V}_\mathrm{M}} \right) \!\cdot\! \bm n_\mathrm{M}+  \hat{\rho}_p^o s_{p,\mathrm{d}}\hat{\nu}_\mathrm{d} =0\label{pBCNonDim},
\end{eqnarray}
where $\ell \in \left\{+, -\right\}$,
\begin{align}
  \hat{\nu}_\mathrm{d} &= \hat{k}_\mathrm d\left[ f_\mathrm{inc}\left(\frac{A^{\mathrm{r}}_{\mathrm{d}}-\tilde{\mu}_{\ddagger, \mathrm{d}}}{k_{\mathrm{B}}T}\right)-f_\mathrm{inc}\left(\frac{A^{\mathrm{p}}_{\mathrm{d}}-\tilde{\mu}_{\ddagger, \mathrm{d}}}{k_{\mathrm{B}}T}\right)\right],
\end{align} and $A^{\mathrm{r}}_{\mathrm{d}}, A^{\mathrm{p}}_{\mathrm{d}}$ are as in Equations~\eqref{Ar} and~\eqref{Ap}.
{Thus the nondimensional working model retains the general increasing function $f_\mathrm{inc}$. The
specialization $f_\mathrm{inc}(x)=\exp(x)$ yields the Butler--Volmer law and its calibration.
Simulations must report the function actually implemented.}

We also have the outputs
\begin{equation}
  \hat V_e^\circ+\left(\ln \phi_{e_\mathrm L}-\hat{\psi}_\mathrm{L} \right)\Big|_{\hat{\Gamma}_\mathrm{L}} -  \left(\ln \phi_{e_\mathrm R}-\hat{\psi}_\mathrm{R}\right)\Big|_{\hat{\Gamma}_\mathrm{R}}=\hat{V}\left(\hat{t}\right),\qquad \hat V_e^\circ:=\frac{\mu_{e_\mathrm L}^\circ-\mu_{e_\mathrm R}^\circ}{k_{\mathrm{B}}T}.\label{valueECBCNonDim}
\end{equation}

\begin{table}
  \caption{\label{tab:table3}Definitions of non-dimensionalized parameters, where ${\ell} \in
  \left\{e_\mathrm{L}, e_\mathrm{R}, i_\mathrm{L}, i_\mathrm{R}, +, -, p\right\}$, and $\mathrm{d}
  \in \left\{ \mathrm{L, R}\right\}$.}
  \begin{ruledtabular}
    \begin{tabular}{cc}
      Symbol & Definition\\
      $\delta$ & $\frac{L_e}{L}$\\
      $\hat{D}_{{\ell}}$ &$\frac{D_\ell}{D_+}$\\
      $\hat{\epsilon}_\mathrm{d}$ & $\frac{1}{L}\sqrt{\frac{\epsilon_\mathrm{d}}{8 \pi l_\mathrm{B}}\frac{1}{\rho_{+}^o}}$\\
      $\hat{\rho}_\ell^o$ & $\frac{\rho_+^o}{\rho_\ell^o}$\\
      $\hat{\epsilon}_\mathrm{M}$ & $\frac{1}{L}\sqrt{\frac{\epsilon_\mathrm{M}}{8 \pi l_\mathrm{B}}\frac{1}{\rho_{+}^o}}$\\
      $\hat{I}_\mathrm{app}$ & $\frac{L}{D_+\rho_+^o}\frac{I_\mathrm{app}}{e}$\\
      $\hat{k}_\mathrm d $&$\frac{L}{D_+\rho_+^o}N_Ak_\mathrm d $\\
      $\hat{V}(\hat{t})$ & $\frac{eV\left( \hat{t}\right)}{k_{\mathrm{B}}T}$\\
    \end{tabular}
  \end{ruledtabular}
\end{table}

\section{Limiting Cases}
\label{sec:limiting-cases}
In this section, we analyze two limiting cases of the model, which represent batteries based on
either a dual ion conductor (DIC) or a single ion conductor (SIC). In order to make a distinction
between the DIC and the SIC, we first present expressions used for physical interpretations of the
transport coefficients of the model in terms of experimentally measurable quantities.

\subsection{Physical Interpretation of the Transport Coefficients}
\label{sec:transportCoeffs}
Here $D_\ell$ sets the mobility scale through $M_\ell=D_\ell\rho_\ell$. For an ideal dilute solution
in the absence of electric-field driving, $\nabla\mu_\ell/(k_\mathrm{B}T)=\nabla\ln\rho_\ell$, and
the resulting flux is $-D_\ell\nabla\rho_\ell$. Thus, only in this limit can $D_\ell$ be interpreted
directly as the usual dilute tracer diffusivity. At finite concentration, the chemical-potential
gradient also depends on composition, so the coefficient governing the spreading of a concentration
perturbation (the collective diffusivity) generally differs from $D_\ell$. A connection between the
experimentally measurable quantities (ionic conductivity, diffusion constant of the salt,
transference number of cations, and a thermodynamic factor) of a DIC and transport coefficients of
the model can be established by following an approach outlined in Reference~\cite{kumar2024dynamic},
where the electrolyte vacancies and galvanostatic conditions were not considered. For constructing a
model of the DIC, we consider the case of $z_p = 0$ without any cations of the polyelectrolytes, so
that neutral polymers with added salt are being modeled. In Reference~\cite{kumar2024dynamic}, the
nonionic fraction is a single constituent, denoted here by $\phi_\mathrm{M}:=1-\phi_+-\phi_-$, to
which a reference density $\rho_\mathrm M^o$ may be assigned. {In the present four-component
electrolyte, however, $1-\phi_+-\phi_-=\phi_p+\phi_{\mathrm{V}_\mathrm{M}}$ combines polymer
segments and vacancies with generally different reference volumes. We therefore use $\rho_\mathrm
M^o$ only when quoting the earlier single-constituent theory or in the special polymer-free limit
$\phi_p=0$, for which $\phi_\mathrm M=\phi_{\mathrm{V}_\mathrm{M}}$ and $\rho_\mathrm
M^o=\rho_{\mathrm{V}_\mathrm{M}}^o=\rho_+^o$. We do not assign an aggregate number density to
$\phi_p+\phi_{\mathrm{V}_\mathrm{M}}$ in the finite-polymer model.}
In this polymer-free limit, define $\rho_\mathrm M:=\rho_\mathrm M^o\phi_\mathrm M=\rho_{\mathrm
V_M}$. From Equations~\eqref{chem_potentials} and~\eqref{H}, we see that $\tilde{\mu}_\ell=
\frac{\delta F}{\delta \rho_\ell}$ for $\ell \in \left\{+, -\right\}$. This in conjunction with
Equation~\eqref{ellGovEqn}, and our choice of $M_\ell = D_\ell\rho_\ell $, for $\ell \in \left\{+,
-\right\}$, implies that, in the limiting case $\phi_p \equiv 0$, in our model,
\begin{eqnarray}
  \frac{\partial \rho_\ell}{\partial t} = \nabla \!\cdot\! \left[ D_\ell \phi_\ell \nabla \left[ \frac{\rho_{\ell }^o}{k_{\mathrm{B}}T} \frac{\delta F}{\delta \rho_\ell}-\frac{\rho_{\mathrm{M}}^o}{k_{\mathrm{B}}T} \frac{\delta F}{\delta \rho_\mathrm{M}}\right]\right] \nonumber \\
  = \nabla \!\cdot\! \left[ D_\ell \phi_\ell\nabla \left[ \frac{1}{k_{\mathrm{B}}T} \frac{\delta F}{\delta \phi_\ell}-\frac{1}{k_{\mathrm{B}}T} \frac{\delta F}{\delta \phi_\mathrm{M}}\right]\right],
\end{eqnarray}
so
\begin{equation}
  \frac{ \partial\phi_\ell}{\partial t} = \nabla \!\cdot\! \left[ D_\ell \frac{\phi_\ell }{\rho_{\ell }^o} \nabla \left[ \frac{1}{k_{\mathrm{B}}T} \frac{\delta F}{\delta \phi_\ell}-\frac{1}{k_{\mathrm{B}}T} \frac{\delta F}{\delta \phi_\mathrm{M}}\right]\right].\label{eqn105}
\end{equation}

For comparison with Reference~\cite{kumar2024dynamic}, let $\boldsymbol j_\ell:=\boldsymbol
J_\ell/\rho_\ell^o$ be the volume-fraction flux, where $\boldsymbol
J_\ell=-M_\ell\nabla(\tilde\mu_\ell^*/k_{\mathrm B}T)$ is the species number flux, so
$\partial_t\phi_\ell=-\nabla\cdot\boldsymbol j_\ell$. The symmetric transport matrix
$\widetilde{\boldsymbol L}^{-1}$ maps exchange-potential gradients to these fluxes; we write its
entries as $\tilde L_{ij}^{-1}:=(\widetilde{\boldsymbol L}^{-1})_{ij}$, not as reciprocals of
individual matrix entries. The ionic fluxes satisfy
\begin{eqnarray}
  \begin{bmatrix}
    -\bm j_+ \\
    -\bm j_-
  \end{bmatrix} &=&
  \begin{bmatrix}
    \tilde{L}_{++}^{-1} &  \tilde{L}_{+-}^{-1} \\
    \tilde{L}_{+-}^{-1} &  \tilde{L}_{--}^{-1}
  \end{bmatrix}
  \begin{bmatrix}
    \nabla\left[ \frac{\delta F}{\delta \phi_+}-\frac{\delta F}{\delta \phi_\mathrm{M}}\right]\\
    \nabla\left[ \frac{\delta F}{\delta \phi_-}-\frac{\delta F}{\delta \phi_\mathrm{M}}\right]
  \end{bmatrix}
\end{eqnarray}

In this paper,
\begin{eqnarray}
  \tilde{L}_{+-}^{-1} &=& 0 \label{LpmSubbed}\\
  \tilde{L}_{\ell \ell}^{-1} &=& D_\ell \frac{\phi_\ell}{\rho_\ell^o k_{\mathrm{B}}T},\label{Lellell}
\end{eqnarray}
for $\ell \in \left\{+, -\right\}$. Substituting Equations~\eqref{LpmSubbed} and~\eqref{Lellell}
into Equations 37, 40, 43, and 44 in~\cite{kumar2024dynamic}, we obtain the transport quantities
below. The subscript $\mathrm V$ denotes the vacancy reference frame used by our exchange-potential
transport law; $v_p$ denotes the polymer velocity, so $t_{+,v_p}$ is evaluated in the frame moving
with the polymer. Frame conversion is needed before comparing these two transference numbers. We
find
\begin{eqnarray}
  \sigma = \frac{e^2}{k_{\mathrm{B}}T}\left(D_+\rho_++D_-\rho_-\right)\label{conductivity}\\
  t_{+,\mathrm{V}}
  = \frac{D_+\rho_+}{D_+\rho_++D_-\rho_-}\label{transference_number}\\
  t_{+,v_p} =      t_{+,\mathrm{V}}+\frac{\phi_+}{\phi_\mathrm{M}}\left( \frac{D_+\phi_+-D_-\phi_-}{D_+\phi_++\frac{\rho_-^o}{\rho_+^o}D_-\phi_-}\right)\\
  {D_{\mathrm{salt}}}
  = \frac{D_+D_-\rho_+\rho_-}{D_+\rho_++D_-\rho_-}\left( \frac{1}{\rho_+}+\frac{1}{\rho_-} + \frac{\rho_\mathrm{M}^o}{\phi_\mathrm{M}}\left(\frac{1}{\rho_+^o}+\frac{1}{\rho_-^o} \right)^2\right)\label{ratio_salt_diff_thermodynamic_factor},
\end{eqnarray}
where $t_{+,v_p}$ denotes
the cation transference number in the polymer-related reference frame used in
Reference~\cite{kumar2024dynamic} (see section S6 in the Supporting Information for the derivation).

\begin{remark}
  Since $D_+,D_-, \rho_+,\rho_-\geq 0$, provided we are in the non-trivial case, so $D_+\rho_+> 0$
  or $D_- \rho_->0$, $t_{+,\mathrm{V}}\in [0, 1]$.
\end{remark}


\subsection{Model for DICs in Galvanostatic Conditions and connection to the Doyle-Fuller-Newman (DFN) model}
\label{sec:compDFN}
Here, we construct a model for DICs (i.e., uncharged polymers like poly(ethylene oxide) with added
salt) under galvanostatic conditions by considering uncharged polymers (i.e., $z_p = 0$). To
highlight the implications of the local electroneutrality assumption and the effects of vacancies,
we rewrite the underlying equations in terms of the local density of salt, the charge density, and
the overpotential. This not only facilitates a comparison with the DFN
model~\cite{doyle1993modeling} (see Table~\ref{tab:Neumanntab1} for a recapitulation of this model), but also reveals the effects of the vacancies, dielectric mismatch,
and the local electroneutrality assumption.

{For the DICs, define the total mobile-ion number density $\rho_\mathrm{salt}:=\rho_++\rho_-$. For a locally electroneutral 1:1 electrolyte,
$\rho_\mathrm{salt}^{\mathrm{fu}}=\rho_\mathrm{ion}/2$ and
$c_\mathrm{salt}=\rho_\mathrm{ion}/(2N_A)$. Comparisons with DFN concentrations and coefficients
must include this normalization.}
Let $t_{-,\mathrm{V}}:=1-t_{+,\mathrm{V}}$, where $t_{+,\mathrm{V}}$ is given by
Equation~\eqref{transference_number}, and define the state-dependent coefficients
\begin{eqnarray}
  D_\mathrm{salt}^{\mathrm{eff}} := \frac{D_+D_-\rho_+\rho_-}{D_+\rho_++D_-\rho_-}
  \left[\frac{1}{\rho_+}+\frac{1}{\rho_-}
  +\frac{\rho_+^o}{\phi_{\mathrm{V}_\mathrm{M}}}
  \left(\frac{1}{\rho_+^o}+\frac{1}{\rho_-^o}\right)^2\right]\\
  \xi_\mathrm{DIC} := \frac{1}{\rho_+}-\frac{1}{\rho_-}  + \frac{1}{\rho_{\mathrm{V}_\mathrm{M}}}\left[1-\left(\frac{\rho_+^o}{\rho_-^o}\right)^2\right]\\
\chi^*_{\mathrm{DIC}} := \frac{2D_+D_-\rho_+\rho_-}{D_+\rho_++D_-\rho_-}\frac{\rho_+^o}{\rho_p^o}\frac{1}{\phi_{\mathrm{V}_\mathrm{M}}}\left(\frac{1}{\rho_+^o}+\frac{1}{\rho_-^o}\right)
\end{eqnarray}
An equation for this density can be derived (see Section~\ref{DFNComp} in the SI for the derivation),
\begin{eqnarray}
  \frac{\partial \rho_\mathrm{salt}}{\partial t} = \nabla \cdot \left[ {D_\mathrm{salt}^{\mathrm{eff}}} \nabla \rho_\mathrm{salt} + \xi_\mathrm{DIC} t_{+,\mathrm{V}} t_{-,\mathrm{V}} \frac{k_{\mathrm{B}}T}{e^2}\sigma \nabla \left(\frac{\rho_c}{e}\right) + \chi_\mathrm{DIC}^{*} t_{+,\mathrm{V}} t_{-,\mathrm{V}} \frac{k_{\mathrm{B}}T}{e^2}\sigma \nabla \rho_p \right.\nonumber\\- \left.\left( t_{+,\mathrm{V}}-t_{-,\mathrm{V}}\right)\left( \frac{\boldsymbol{I}_\mathrm{tot}}{e}-\frac{\epsilon_\mathrm{M}}{4\pi l_\mathrm{B}}\frac{e}{k_{\mathrm{B}}T}\frac{\partial \boldsymbol{E}_\mathrm{M}}{\partial t}\right)\right],\label{saltDIC}
\end{eqnarray}
where $\rho_c = e(\rho_+ - \rho_-)$ is the local charge density.
{The coefficient $D_\mathrm{salt}^{\mathrm{eff}}$ is obtained directly from the two ionic fluxes and
remains well defined when polymer and vacancy reference volumes differ. It is a state-dependent
collective coefficient of the present diagonal four-component model, not automatically the
experimentally measured salt diffusivity of Reference~\cite{kumar2024dynamic}. In the polymer-free specialization ($\rho_p=0$) it reduces to the corresponding coefficient of the earlier
single-neutral-constituent theory, Equation~\eqref{ratio_salt_diff_thermodynamic_factor}.}
{Throughout the DIC and SIC reductions, every quantity $\rho_\ell^{\mathrm{eq}}>0$ denotes a
spatially uniform bulk reference density, so $\nabla\rho_\ell^{\mathrm{eq}}=0$ and
$\nabla\ln(\rho_\ell/\rho_\ell^{\mathrm{eq}})=\nabla\ln\rho_\ell$. These constants provide
dimensionless normalizations and are not the generally nonuniform equilibrium profiles of the full
Poisson problem near an interface. The same convention applies to derived reference densities such
as $\rho_\mathrm{salt}^{\mathrm{eq}}$.}

Furthermore, the governing equations for the charge density, polymer density, and electric field in
the electrolyte are as follows
\begin{align}
  \frac{\partial \rho_c}{\partial t} &= \nabla \cdot \left[
  \boldsymbol{I}_\mathrm{tot} -\sigma \boldsymbol{E}_\mathrm{M} + \sigma \frac{k_{\mathrm{B}}T}{e} \frac{\nabla \rho_p}{\rho_p^o\phi_{\mathrm{V}_\mathrm{M}}} \left( t_{+,\mathrm{V}}-\frac{\rho_+^o}{\rho_-^o}t_{-,\mathrm{V}}\right)\right.\nonumber\\
  &\qquad+ \sigma \frac{k_{\mathrm{B}}T}{e} \frac{\nabla \rho_\mathrm{salt}}{2}\left(
  \frac{t_{+,\mathrm{V}}}{\rho_+}-\frac{t_{-,\mathrm{V}}}{\rho_-}+\frac{1}{\rho_{\mathrm{V}_\mathrm{M}}}\left( 1+\frac{\rho_+^o}{\rho_-^o}\right) \left(t_{+,\mathrm{V}}-\frac{\rho_+^o}{\rho_-^o}t_{-,\mathrm{V}} \right)\right) \nonumber\\
  &\qquad\left.
  + \sigma \frac{k_{\mathrm{B}}T}{e} \frac{\nabla (\rho_c/e)}{2}\left(
  \frac{t_{+,\mathrm{V}}}{\rho_+}+\frac{t_{-,\mathrm{V}}}{\rho_-}+\frac{1}{\rho_{\mathrm{V}_\mathrm{M}}}\left( 1-\frac{\rho_+^o}{\rho_-^o}\right) \left(t_{+,\mathrm{V}}-\frac{\rho_+^o}{\rho_-^o}t_{-,\mathrm{V}} \right)
  \right)
  \right]
\end{align}
\begin{eqnarray}
  \frac{\partial \rho_p}{\partial t} = \nabla \cdot \left[ \frac{D_p\phi_p}{2\phi_{\mathrm{V}_\mathrm{M}}}\left(\left(1+\frac{\rho_+^o}{\rho_-^o}\right)
  \nabla \rho_\mathrm{salt}+ 
  \left(1-\frac{\rho_+^o}{\rho_-^o}\right)\nabla \left( \frac{\rho_c}{e}\right) \right)+D_p \left( \frac{1}{N}+\frac{\rho_+^o}{\rho_p^o}\frac{\phi_p}{\phi_{\mathrm{V}_\mathrm{M}}}\right) \nabla \rho_p\right]
\end{eqnarray}
\begin{equation}
  \nabla \cdot \left[ \frac{\epsilon_\mathrm{M}}{4\pi l_\mathrm{B}}\frac{e^2}{k_{\mathrm{B}}T}\boldsymbol{E}_\mathrm{M}\right] = \rho_c.
\end{equation}
Because electrons are the only charged species in either electrode,
$t_{e_\mathrm{d},\mathrm{V}}:=1$. Define the electrode conductivity by
\begin{equation}
  \sigma_\mathrm{d} := \frac{e^2}{k_{\mathrm{B}}T}D_{e_\mathrm{d}}\rho_{e_\mathrm{d}}.\label{sigmaD}
\end{equation}
The bulk electric fields satisfy separate relations on their respective domains. In the electrode,
the conduction-current law gives
\begin{equation}
  \boldsymbol E_\mathrm d=\frac{\boldsymbol I_{\mathrm{tot},\mathrm d}}{\sigma_\mathrm d}
  -\frac{k_{\mathrm B}T}{e}\nabla\ln\frac{\phi_{e_\mathrm d}}{\phi_{e_\mathrm d}^{\mathrm{eq}}}
  -\frac{\epsilon_0\epsilon_\mathrm d}{\sigma_\mathrm d}\partial_t\boldsymbol E_\mathrm d.
  \label{bulkElectrodeField}
\end{equation}
In the DIC electrolyte, it gives
\begin{align}
  \boldsymbol E_\mathrm M&=\frac{\boldsymbol I_{\mathrm{tot},\mathrm M}}{\sigma}
  +\frac{k_{\mathrm B}T}{e}\left[t_{+,\mathrm V}\nabla\ln\frac{\rho_+}{\rho_+^{\mathrm{eq}}}-t_{-,\mathrm V}\nabla\ln\frac{\rho_-}{\rho_-^{\mathrm{eq}}}
  -\left(t_{+,\mathrm V}-\frac{\rho_+^o}{\rho_-^o}t_{-,\mathrm V}\right)\nabla\ln\frac{\rho_{\mathrm V_M}}{\rho_{\mathrm V_M}^{\mathrm{eq}}}\right]\nonumber\\
  &\quad-\frac{\epsilon_0\epsilon_\mathrm M}{\sigma}\partial_t\boldsymbol E_\mathrm M.\label{etaDIC}
\end{align}
Here $\epsilon_0$ is the permittivity of vacuum. Under local bulk electroneutrality,
$\rho_+=\rho_-=\rho_\mathrm{salt}/2$, the electrolyte relation becomes
\begin{align}
  \boldsymbol E_\mathrm M&=\frac{\boldsymbol I_{\mathrm{tot},\mathrm M}}{\sigma}
  +\frac{k_{\mathrm B}T}{e}\left[(t_{+,\mathrm V}-t_{-,\mathrm V})\nabla\ln\frac{\rho_\mathrm{salt}}{\rho_\mathrm{salt}^{\mathrm{eq}}}
  -\left(t_{+,\mathrm V}-\frac{\rho_+^o}{\rho_-^o}t_{-,\mathrm V}\right)\nabla\ln\frac{\rho_{\mathrm V_M}}{\rho_{\mathrm V_M}^{\mathrm{eq}}}\right]\nonumber\\
  &\quad-\frac{\epsilon_0\epsilon_\mathrm M}{\sigma}\partial_t\boldsymbol E_\mathrm M.\label{etaElectroNeut}
\end{align}
These bulk equations are integrated within each domain to determine potential differences. The
interfacial overpotential $\eta_\mathrm d$ is evaluated only on $\Sigma_\mathrm d$ from
Equation~\eqref{eta} and supplies the reaction law. No subtraction of gradients defined on disjoint
domains is required. Note that time scales, $\epsilon_0 \epsilon_d/\sigma_d$ and $\epsilon_0 \epsilon_M/\sigma$ appear, which characterize the time scales of the dielectric relaxations. 
The remaining governing equations and boundary conditions to complete the system are enumerated in Table~\ref{tab:DICElectrolytetab}. 

\subsection{Model for SICs in Galvanostatic Conditions}
\label{sec:SICModelGalvano}
{In this SIC reduction we specialize to monovalent charge carriers, $z_+=1$ and $z_p=-1$.}
Define $r:=\rho_+^o/\rho_p^o$ and the positive composition factors
\begin{align}
  \mathcal A&:=\frac1{\rho_+}+\frac{1+r}{\rho_+^o\phi_{\mathrm V_M}},&
  \mathcal B&:=\frac1{N\rho_p}+\frac{r(1+r)}{\rho_+^o\phi_{\mathrm V_M}},\\
  \mathcal K&:=\frac{D_+D_p\rho_+\rho_p}{D_+\rho_++D_p\rho_p}.
\end{align}
The SIC transport coefficients and transference numbers are
\begin{align}
  D_\mathrm{salt,mac}&:=\mathcal K(\mathcal A+\mathcal B),&
  \xi_\mathrm{SIC}&:=\frac{\mathcal A-\mathcal B}{\mathcal A+\mathcal B},\\
  t_{+,\mathrm V}^{\mathrm{SIC}}&:=\frac{D_+\rho_+}{D_+\rho_++D_p\rho_p},&
  t_{p,\mathrm V}^{\mathrm{SIC}}&:=1-t_{+,\mathrm V}^{\mathrm{SIC}}.\label{SICtransference_number}
\end{align}
Define the corresponding SIC conductivity by
\begin{equation}
  \sigma_\mathrm{SIC}:= \frac{e^2}{k_{\mathrm{B}}T} \left[{ D_+\rho_+ +D_{p} \rho_{p}} \right]. \label{SICconductivity}
\end{equation}
An equation for the total mobile-cation-plus-charged-segment density in the SIC model, for which we
retain the legacy symbol $\rho_\mathrm{salt,mac} = \rho_+ + \rho_p$, can be derived (see Section~\ref{SISIC} in the SI for the derivation),
\begin{eqnarray}
  \frac{\partial \rho_\mathrm{salt,mac}}{\partial t} = \nabla \cdot \left[ {D}_\mathrm{salt,mac}\nabla \rho_{\mathrm{salt,mac}}+\xi_\mathrm{SIC}D_\mathrm{salt,mac}\nabla \left(\frac{\rho_c}{e}\right) \nonumber\right.\\
  \left.- \left(t_{+,\mathrm{V}}^\mathrm{SIC}-t_{p,\mathrm{V}}^\mathrm{SIC}\right)
  \left( \frac{\boldsymbol{I}_\mathrm{tot}}{e}-\frac{\epsilon_\mathrm{M}}{4\pi l_\mathrm{B}}\frac{e}{k_{\mathrm{B}}T} \frac{\partial \boldsymbol{E}_\mathrm{M}}{\partial t}\right)
  \right],
\end{eqnarray}
where $\rho_c = e\left(\rho_+-\rho_p\right)$ is the local charge density.
{Under monovalent local electroneutrality, $\rho_+=\rho_p$ and hence
$\rho_\mathrm{salt,mac}=2\rho_+=2\rho_p$. Thus $\rho_\mathrm{salt,mac}$ is a total
charged-constituent density, while the analogous electroneutral formula-unit density is
$\rho_\mathrm{salt,mac}^{\mathrm{fu}}=\rho_\mathrm{salt,mac}/2$. This is only charge-stoichiometric
bookkeeping: the polymer segments remain chemically connected and are not freely diffusing ion-pair
constituents.}
Furthermore, the governing equations for the charge density and polymer density in the electrolyte
are as follows
\begingroup
\small
\begin{align}
  \frac{\partial \rho_c}{\partial t} &= \nabla \cdot \left[
  \boldsymbol{I}_\mathrm{tot}-\sigma_\mathrm{SIC}\boldsymbol{E}_\mathrm{M}\right. \nonumber\\
  &- \frac{k_{\mathrm{B}}T}{e}\sigma_\mathrm{SIC}\left(
  \frac{t_{p,\mathrm{V}}^\mathrm{SIC}}{2}\mathcal{B}-\frac{t_{+,\mathrm{V}}^\mathrm{SIC}}{2}\mathcal{A}
  \right)\nabla \rho_\mathrm{salt,mac} \nonumber\\
  &- \left.\frac{k_{\mathrm{B}}T}{e}\sigma_\mathrm{SIC}\left(
  \frac{t_{p,\mathrm{V}}^\mathrm{SIC}}{2}\left( -\frac{1}{\rho_p}\frac{1}{N}+\frac{1}{\rho_p^o}\frac{1}{\phi_{\mathrm{V}_\mathrm{M}}}\left[ 1-r\right]\right) - \frac{t_{+,\mathrm{V}}^\mathrm{SIC}}{2}
  \left( \frac{1}{\rho_+}+\frac{1}{\rho_+^o}\frac{1}{\phi_{\mathrm{V}_\mathrm{M}}}\left[ 1-r\right]\right) \right)\nabla \left(\frac{\rho_c}{e}\right)
  \right]
\end{align}
\begin{align}
  \frac{\partial \rho_p}{\partial t}=\nabla\cdot\left[\frac{\mathcal K}{2}(\mathcal A+\mathcal B)\nabla\rho_\mathrm{salt,mac}
  +\frac{\mathcal K}{2}(\mathcal A-\mathcal B)\nabla(\rho_c/e)
  +t_{p,\mathrm V}^{\mathrm{SIC}}\frac{\boldsymbol j_{\rho_c}}e\right].
\end{align}
\endgroup
The SIC electrolyte electric field satisfies
\begin{align}
  \boldsymbol E_\mathrm M&=\frac{\boldsymbol I_{\mathrm{tot},\mathrm M}}{\sigma_\mathrm{SIC}}
  +\frac{k_{\mathrm B}T}{e}\left[t_{+,\mathrm V}^{\mathrm{SIC}}\nabla\ln\frac{\rho_+}{\rho_+^{\mathrm{eq}}}
  -\frac{t_{p,\mathrm V}^{\mathrm{SIC}}}{N}\nabla\ln\frac{\rho_p}{\rho_p^{\mathrm{eq}}}
  -(t_{+,\mathrm V}^{\mathrm{SIC}}-r t_{p,\mathrm V}^{\mathrm{SIC}})\nabla\ln\frac{\rho_{\mathrm V_M}}{\rho_{\mathrm V_M}^{\mathrm{eq}}}\right]\nonumber\\
  &\quad-\frac{\epsilon_0\epsilon_\mathrm M}{\sigma_\mathrm{SIC}}\partial_t\boldsymbol E_\mathrm M.\label{SICetad}
\end{align}
The electrode field is given separately by Equation~\eqref{bulkElectrodeField}. Interfacial
overpotentials are evaluated from Equation~\eqref{eta}; they are not bulk fields. In the monovalent
electroneutral bulk, both occupied densities equal $\rho_\mathrm{salt,mac}/2$, so the first two
logarithmic terms combine with coefficient $t_{+,\mathrm V}^{\mathrm{SIC}}-t_{p,\mathrm
V}^{\mathrm{SIC}}/N$.

The remaining governing equations and boundary conditions to complete the system are enumerated in Table~\ref{tab:SICElectrolytetab}.

\begin{remark} From Equation~\eqref{SICtransference_number}, we see that removing free anions does not by itself give unit cation transference when the polymer is charged and mobile. Equation~\eqref{SICtransference_number} gives $t_{+,\mathrm V}^{\mathrm{SIC}}=1$ only when $D_p\rho_p=0$; for a nonzero polymer density ($\rho_p>0$) this requires $D_p=0$. In that case
$\rho_+=\rho_p$ cannot remain valid pointwise up to a Faradaic boundary: cations cross the interface
while the polyanion is immobile. The local-electroneutral SIC equations must therefore be
interpreted as an outer bulk approximation, with a thin interfacial layer in which Poisson's equation accounts for local charge separation. A uniformly valid electroneutral model would require effective boundary
conditions obtained by matching through that layer; the bulk electroneutral constraint is not
imposed directly on $\Sigma_\mathrm d$.\end{remark}{}

\section{Numerical Simulations}
\label{sec:numerical-simulations}
Both DIC and SIC simulations are performed in MOOSE using continuous linear
finite elements and adaptive second-order backward differentiation (BDF2).
All results shown in this section use the same 1920-element mesh, cosine-graded
within each material to resolve interfacial layers, with $320/1280/320$
elements in the left electrode, electrolyte, and right electrode, respectively.
The maximum time step is $5\times10^{-5}$, and a constrained Newton method
enforces the polymer lower bound. Additional solver settings, conservation
checks, and mesh and time-step comparisons are provided in Supplementary
Section~\ref{sec:numericalChecksSI}.

For both DIC and SIC, the initial conditions are chosen to be locally
electroneutral in each material and hence globally electroneutral over the
complete cell; local electroneutrality is not imposed during subsequent
evolution. Preservation of global electroneutrality is checked numerically
using the total-charge diagnostics reported in Supplementary
Sections~\ref{sec:dicNumericalChecksSI} and~\ref{sec:sicNumericalChecksSI}.

\subsection{One-dimensional DIC charge--discharge cycle}
\label{sec:utilization}
We solve the nondimensional DIC model on $[-0.25,1.25]$, with electrolyte
on $[0,1]$ and electrodes of thickness $\delta=0.25$. The implemented
interfacial kinetics use $f_\mathrm{inc}(x)=x$, the linear-affinity specialization
of the general reaction law. Potential is continuous at the interfaces,
fixed to zero at the left terminal, and has zero normal derivative at the
right terminal. This fixes the potential reference without prescribing voltage.

\paragraph{Parameters and protocol.}
The applied current is $\hat I_\mathrm{app}=-1.6$ for $0\leq\hat t<0.05$
and $+1.6$ for $0.05\leq\hat t\leq0.10$. Thus the imposed charge dose per
half-cycle is $|\hat I_\mathrm{app}|\hat t_{\rm rev}=0.08$, where $\hat t_{\rm rev}$ is the reversal time, or the time when the sign of the applied current changes.
Table~\ref{tab:dicNumericalParameters} gives the material and initial-state
parameters. These are dimensionless illustrative parameters; a specific
material and dimensional length, concentration, and diffusivity calibration
are not assigned.
\begin{table*}
\caption{Parameters for the DIC cycle at current magnitude $1.6$. 
}
\label{tab:dicNumericalParameters}
\begin{ruledtabular}\begin{tabular}{lll}
Quantity & Value & Meaning\\
\midrule
$(\phi_+,\phi_-,\phi_p)_0$ & $(0.14645,0.14645,0.20710)$ & electrolyte initial state\\
$(\phi_{e_\mathrm d},\phi_{i_\mathrm d})_0$ & $(1.5,0.6)$ & electrode initial state\\
$(\hat D_+,\hat D_-,\hat D_p)$ & $(1,1,0.01)$ & electrolyte diffusivities\\
$(\hat D_e,\hat D_i)$ & $(100,1)$ & electrode diffusivities\\
$(\hat\epsilon_\mathrm d^2,\hat\epsilon_\mathrm M^2)$ & $(10^{-5},10^{-6})$ & Poisson coefficients\\
$(z_+,z_-,z_p)$ & $(1,-1,0)$ & DIC valences\\
$\rho_p^o/\rho_+^o$ & $0.1183625465$ & polymer reference-density ratio\\
$N$ & $10$ & polymerization index\\
$\hat k_\mathrm L=\hat k_\mathrm R$ & $4000$ & interfacial reaction coefficient\\
$\phi_{p,\min}$ & $10^{-10}$ & enforced polymer lower bound\\
\end{tabular}\end{ruledtabular}
\end{table*}

\paragraph{Concentration profiles.}
Figures~\ref{fig:dicLithiumProfiles}--\ref{fig:dicAnionProfiles} show nine
saved states from the DIC cycle. Solid curves denote the first
half-cycle, including the state at reversal; dashed curves denote the
second half-cycle. Figure ~\ref{fig:dicLithiumProfiles} shows that lithium is transferred from the right electrode to the
left during the first half-cycle and returns during the second. The
small Poisson coefficients, $(\hat\epsilon_\mathrm d^2,\hat\epsilon_\mathrm M^2)$ , permit sharp interfacial concentration layers. Vacancies and polymer redistribute in response to the transfer of the lithium from one electrode to the other. For example, Figure \label{fig:dicVacancyProfiles} shows that the volume fractions of vacancies near each electrode change in response to the conversion of lithium metal to the lithium cation so that during the charge cycle, the vacancies in the electrolyte are depleted near the right electrode and the vacancies in the right electrode are being produced. During the dicharge cycle, this reverses and the vacancies in the electrolyte are localized near the right electrode. Furthermore, volume fraction of polymers, electron density, and volume fraction of anions are shown in 
Figures ~\ref{fig:dicPolymerProfiles}, ~\ref{fig:dicElectronProfiles}, and ~\ref{fig:dicAnionProfiles}, respectively. In particular, Figure ~\ref{fig:dicAnionProfiles} shows that the anions get localized on the right and depleted at the left electrode during the charge cycle, which is expected due to the presence of electric field active along the negative x-axis (see Figure ~\ref{fig:dicPotentialProfiles}). Figure ~\ref{fig:dicPolymerProfiles} shows that polymers get depleted from the right electrode and localized near the left electrode during the charge cycle, which is expected on the basis of the local incompressibility constraint used in the model. The localization and depletion of anions and polymers, respectively, near the right electrode turn into depletion and localization during the discharge cycle due to the reversal of the electric field, which can be inferred from Figure ~\ref{fig:dicPotentialProfiles}. 
\begin{figure*}
\includegraphics[width=0.85\textwidth]{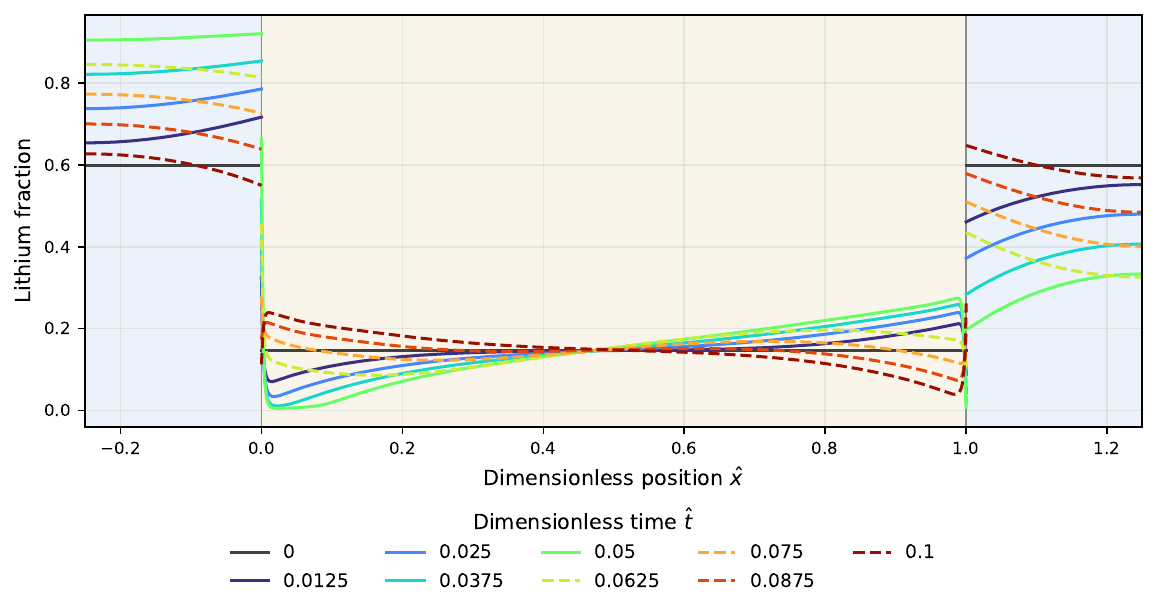}
\caption{Lithium-ion fraction $\phi_+$ in the electrolyte and intercalated-lithium fractions $\phi_{i,\mathrm d}$ in the electrodes. Current magnitude $1.6$, reversal at $\hat t=0.05$, and end time $0.10$.}
\label{fig:dicLithiumProfiles}
\end{figure*}
\begin{figure*}
\includegraphics[width=0.85\textwidth]{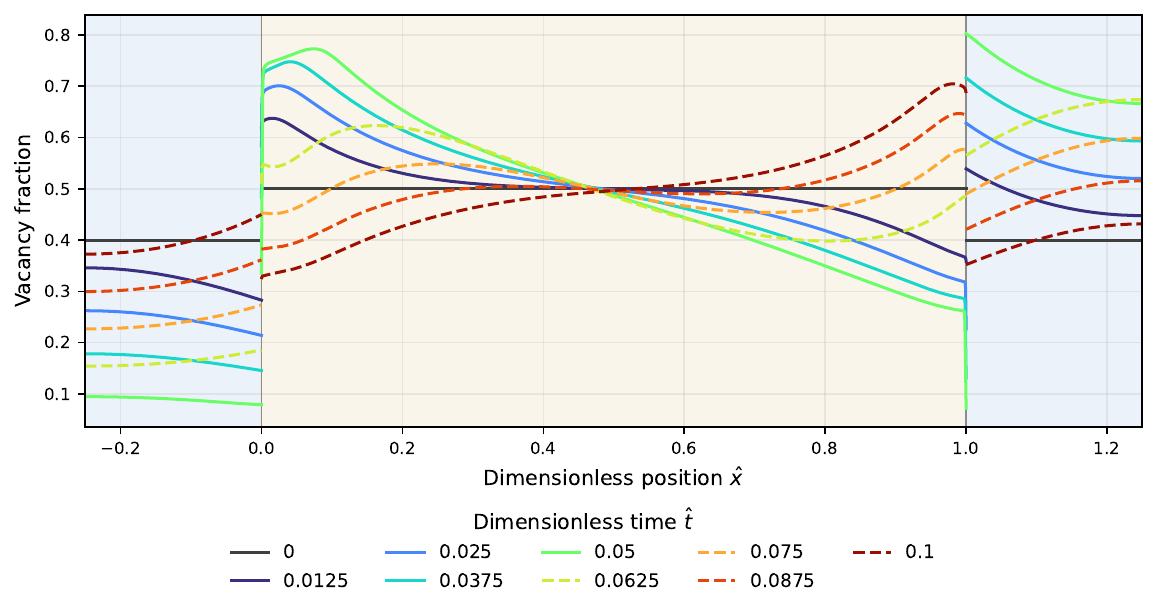}
\caption{Vacancy fractions in the electrodes and electrolyte. Current magnitude $1.6$, reversal at $\hat t=0.05$, and end time $0.10$.}
\label{fig:dicVacancyProfiles}
\end{figure*}
\begin{figure*}
\includegraphics[width=0.85\textwidth]{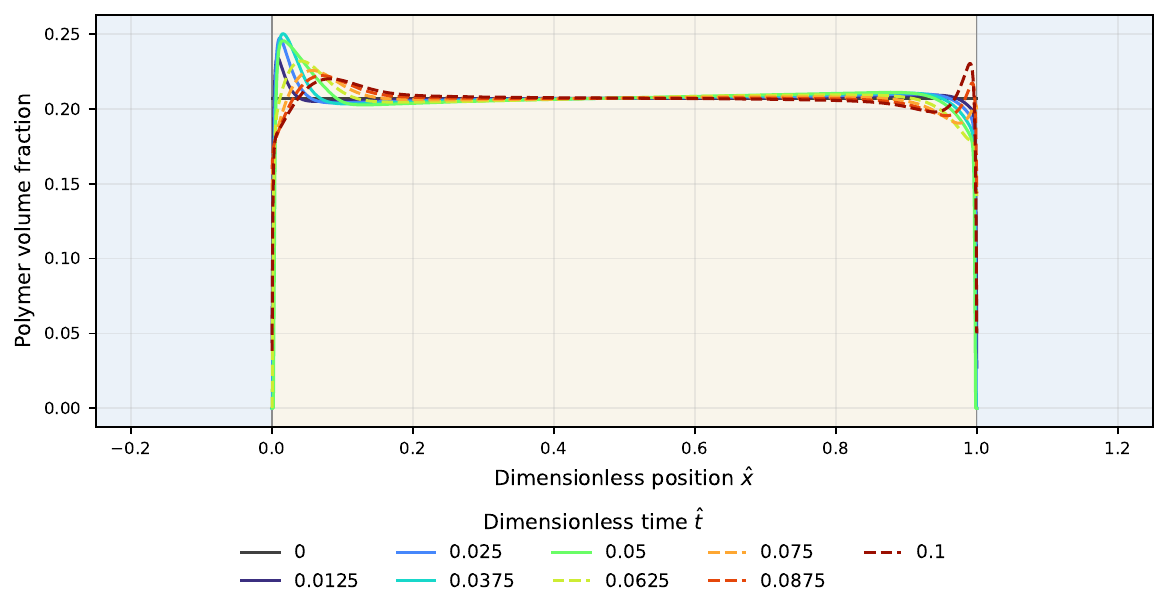}
\caption{Polymer volume fraction $\phi_p$ in the electrolyte. Current magnitude $1.6$, reversal at $\hat t=0.05$, and end time $0.10$.}
\label{fig:dicPolymerProfiles}
\end{figure*}
\begin{figure*}
\includegraphics[width=0.85\textwidth]{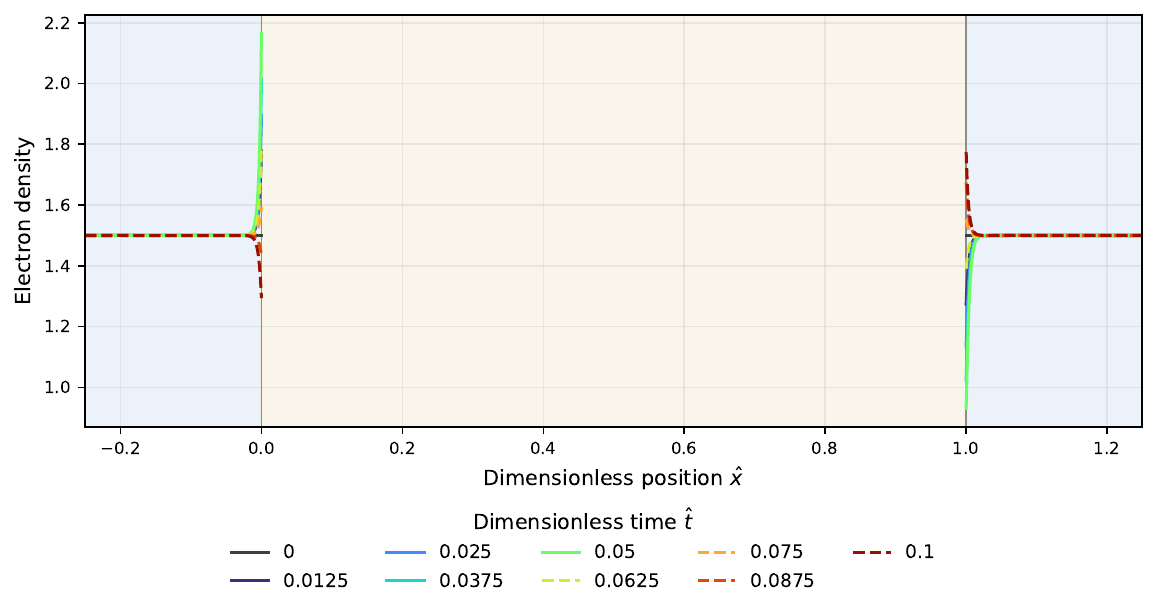}
\caption{Electron densities $\phi_{e,\mathrm d}$ in the electrodes. Current magnitude $1.6$, reversal at $\hat t=0.05$, and end time $0.10$.}
\label{fig:dicElectronProfiles}
\end{figure*}
\begin{figure*}
\includegraphics[width=0.85\textwidth]{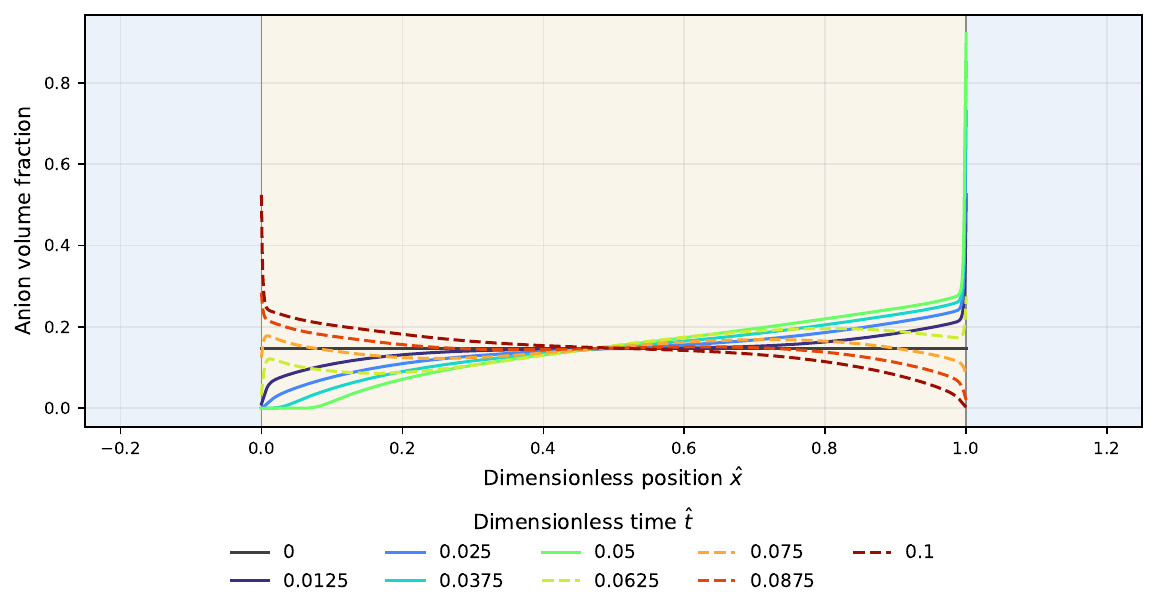}
\caption{Anion volume fraction $\phi_-$ in the electrolyte. Current magnitude $1.6$, reversal at $\hat t=0.05$, and end time $0.10$.}
\label{fig:dicAnionProfiles}
\end{figure*}
\begin{figure*}
\includegraphics[width=0.85\textwidth]{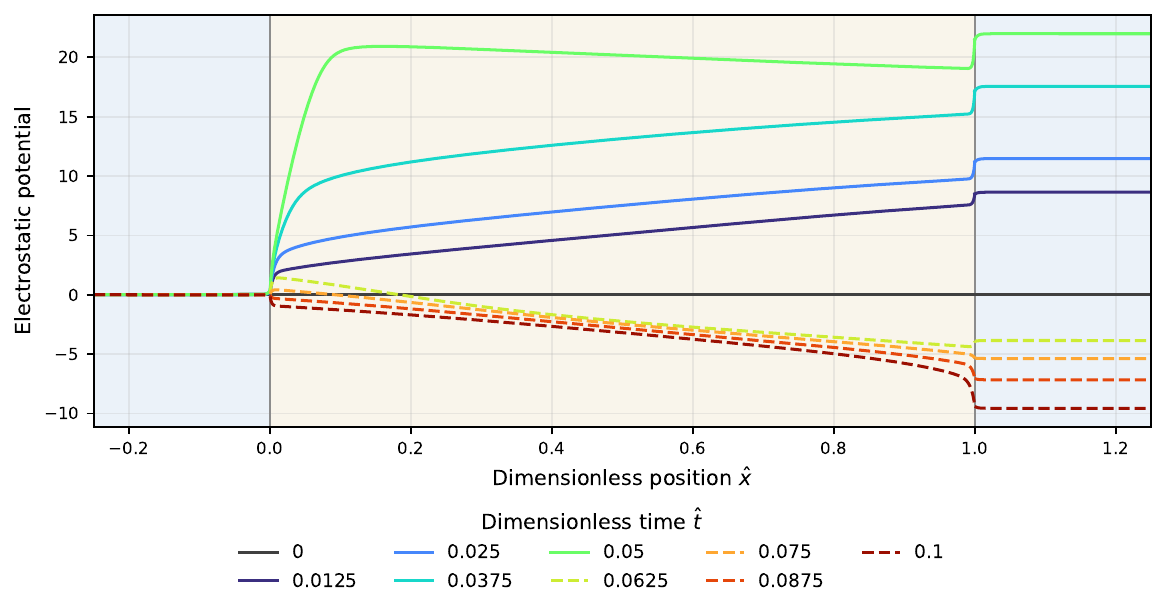}
\caption{Electrostatic potential relative to the left terminal. Current magnitude $1.6$, reversal at $\hat t=0.05$, and end time $0.10$.}
\label{fig:dicPotentialProfiles}
\end{figure*}

\paragraph{Potential and utilization.}
Define the mean utilization of electrode $\mathrm d$ by
\begin{equation}
u_\mathrm d(\hat t)=\frac{1}{\delta}\int_{\hat\Omega_\mathrm d}
\phi_{i_\mathrm d}(\hat x,\hat t)\,\mathrm d\hat x,
\qquad \mathrm d\in\{\mathrm L,\mathrm R\}.
\label{eq:numericalUtilization}
\end{equation}
Both electrodes initially have mean utilization of $0.6$. At reversal the left
and right values are $0.911208$ and $0.289917$; at the end they are
$0.603018$ and $0.596357$. Figure~\ref{fig:dicUtilizationHistories} shows
this nearly reversible transfer with a small residual redistribution.
The terminal electrochemical potential difference is evaluated from
Equation~\eqref{valueECBCNonDim} and shown in Figure~\ref{fig:dicTerminalPotential}. 
The largest saved electrochemical
potential difference strictly before reversal is $31.4216$; its final
value is $-9.57392$. The value sampled exactly at the abrupt current
reversal is sensitive to time resolution, as quantified in Supplementary Section~\ref{sec:dicNumericalChecksSI}.
\begin{figure*}
\includegraphics[width=0.85\textwidth]{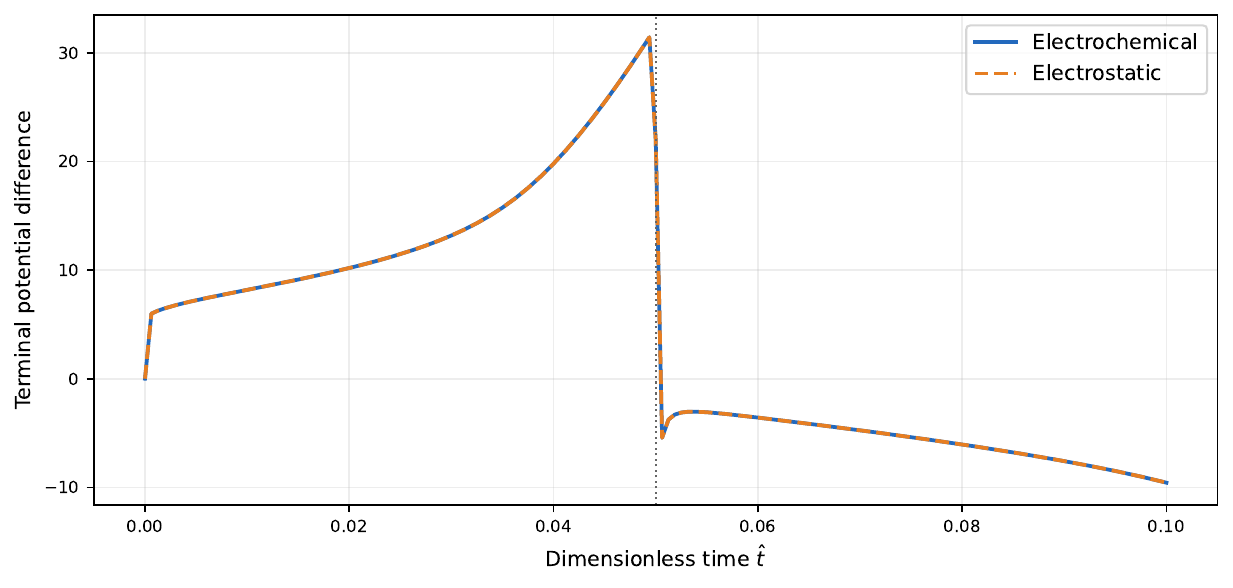}
\caption{Terminal electrochemical and electrostatic potential differences
for the current-$1.6$ DIC cycle. The vertical line marks current reversal.}
\label{fig:dicTerminalPotential}
\end{figure*}
\begin{figure*}
\includegraphics[width=0.85\textwidth]{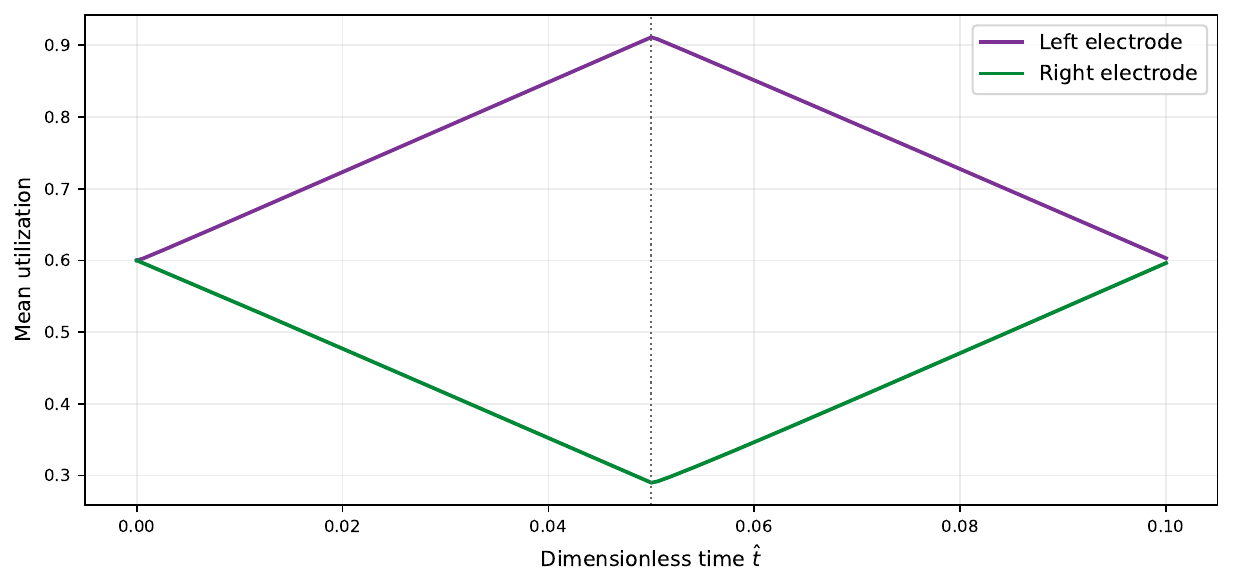}
\caption{Mean left- and right-electrode utilization during the current-$1.6$
DIC cycle.}
\label{fig:dicUtilizationHistories}
\end{figure*}

\clearpage

\subsection{One-dimensional SIC charge--discharge cycle}
\label{sec:sicNumerical}
We next solve the SIC model with charged polymer segments and no free
anions. The geometry, diffusion time scale, electrode diffusivities,
reaction coefficient, Poisson coefficients, boundary conditions, and numerical
methods are the same as for DIC. In particular, the applied current is
$-1.6$ until $\hat t=0.05$, then $+1.6$ until $\hat t=0.10$, giving the
same half-cycle charge dose $0.08$. Figures use the same snapshot times,
color sequence, and line conventions as the DIC figures.

\paragraph{SIC composition and comparison basis.}
The polymer segment valence is $z_p=-1$, and its diffusivity is
$\hat D_p=10^{-4}$, while $\hat D_+=1$. This smaller polymer mobility
represents a nearly cation-only conduction regime in the initial uniform
bulk: the conductivity-based cation transference number is
$1/(1+\hat D_p/\hat D_+)=0.99990001$. It need not remain constant during
the interfacial transient. With $\rho_p^o/\rho_+^o=0.1183625465$,
initial local neutrality requires $\phi_+=\frac{\rho_p^o}{\rho_+^o}\phi_p$.
Table~\ref{tab:sicNumericalParameters} lists the differences from DIC.
The initial electrode utilization is chosen for zero interfacial reaction
rate in this neutral uniform state.
\begin{table*}[!hbp]
\caption{Composition and mobility differences between the matched-current
DIC and SIC cycles. Other parameters and numerical settings are as in the
DIC study.}
\label{tab:sicNumericalParameters}
\begin{ruledtabular}\begin{tabular}{lcc}
Quantity & DIC & SIC\\
\midrule
Initial cation fraction & $0.14645$ & $0.052917789$\\
Initial free-anion fraction & $0.14645$ & absent\\
Initial polymer fraction & $0.20710$ & $0.447082211$\\
Initial electrolyte vacancy fraction & $0.5$ & $0.5$\\
Polymer segment valence & $0$ & $-1$\\
Polymer diffusivity & $0.01$ & $0.0001$\\
Initial electrode utilization & $0.6$ & $0.3014498233$\\
Initial right-electrode lithium inventory & $0.15$ & $0.075362456$\\
\end{tabular}\end{ruledtabular}
\end{table*}
The imposed half-cycle charge is approximately $53.3\%$ of the initial
right-electrode lithium inventory in DIC and $106.2\%$ in SIC. These ratios
compare terminal charge dose with initial inventory, rather than the actual
fraction of lithium removed: solving Poisson's equation allows local charge
accumulation within the cell while preserving global electroneutrality. Thus the calculations match applied current
and duration, but not initial utilization or fractional capacity demand.
Differences between DIC and SIC cannot be attributed solely to polymer charge.

\paragraph{Concentration profiles and polarization.}
Figures~\ref{fig:sicLithiumProfiles}--\ref{fig:sicPotentialProfiles} show
substantial cation depletion extending into the electrolyte from the right
side near reversal, together with polymer accumulation at the right
interface. The minimum saved cation and electrolyte vacancy fractions are
$5.0962\times10^{-5}$ and $7.1569\times10^{-4}$, respectively. The polymer
fraction ranges from $0.299566$ to $0.999233$, remaining above its imposed
lower bound. The minimum saved right-electrode electron density is
$0.0042583$, and the minimum local right-electrode utilization is
$0.00049822$. All saved fields are finite and admissible; these extrema
need not occur at the same position or time.
\begin{figure*}[!hbp]
\includegraphics[width=0.85\textwidth]{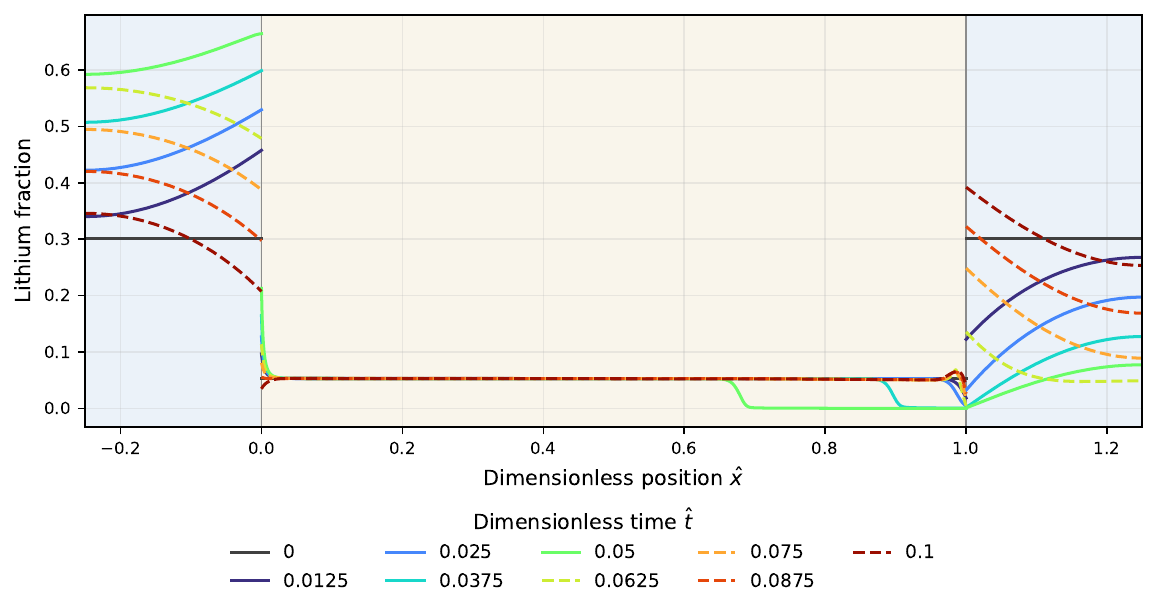}
\caption{Lithium-ion fraction in the SIC electrolyte and intercalated-lithium fractions in the electrodes. Current magnitude $1.6$, reversal at $\hat t=0.05$, and end time $0.10$. Solid curves include the reversal state; dashed curves follow reversal.}
\label{fig:sicLithiumProfiles}
\end{figure*}
\begin{figure*}[!hbp]
\includegraphics[width=0.85\textwidth]{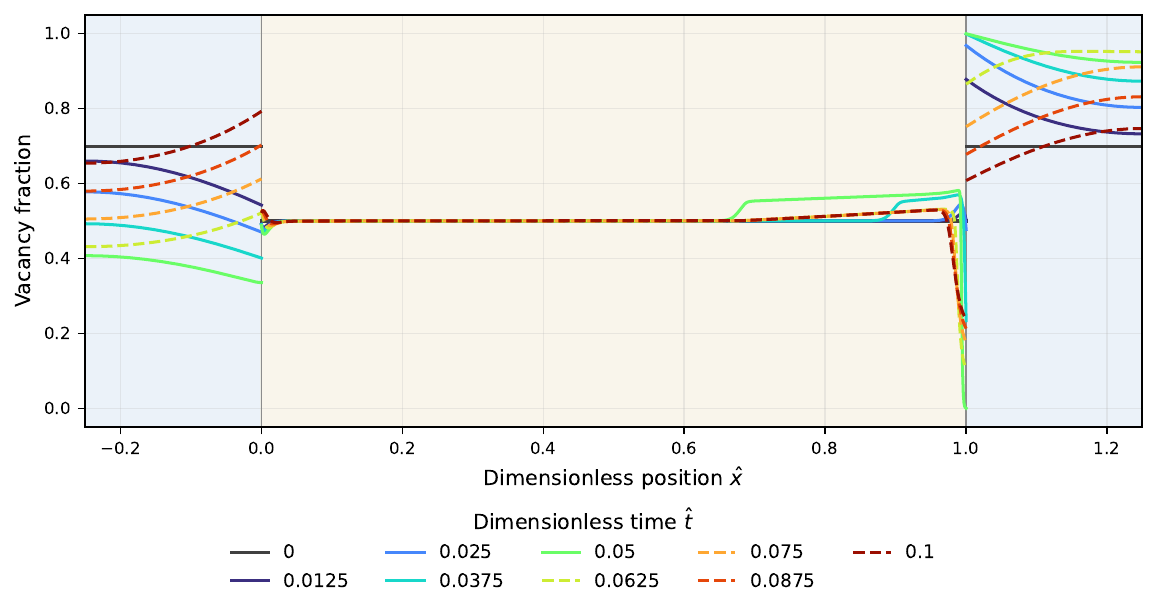}
\caption{Vacancy fractions in the SIC electrolyte and electrodes. Current magnitude $1.6$, reversal at $\hat t=0.05$, and end time $0.10$. Solid curves include the reversal state; dashed curves follow reversal.}
\label{fig:sicVacancyProfiles}
\end{figure*}
\begin{figure*}[!hbp]
\includegraphics[width=0.85\textwidth]{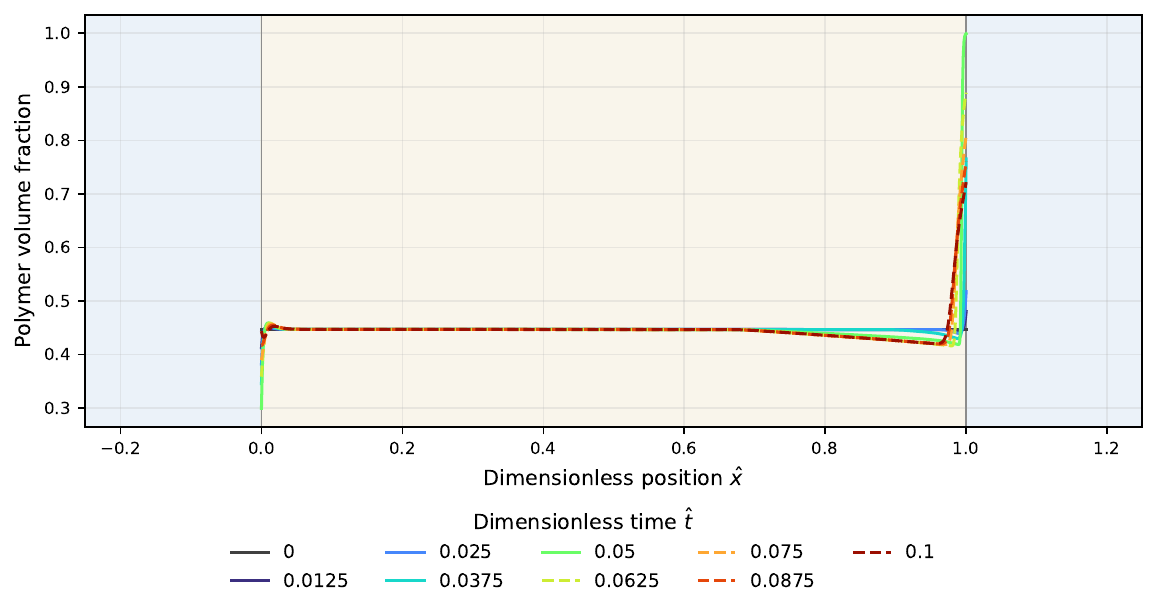}
\caption{Charged-polymer volume fraction $\phi_p$ in the SIC electrolyte. Current magnitude $1.6$, reversal at $\hat t=0.05$, and end time $0.10$. Solid curves include the reversal state; dashed curves follow reversal.}
\label{fig:sicPolymerProfiles}
\end{figure*}
\begin{figure*}[!hbp]
\includegraphics[width=0.85\textwidth]{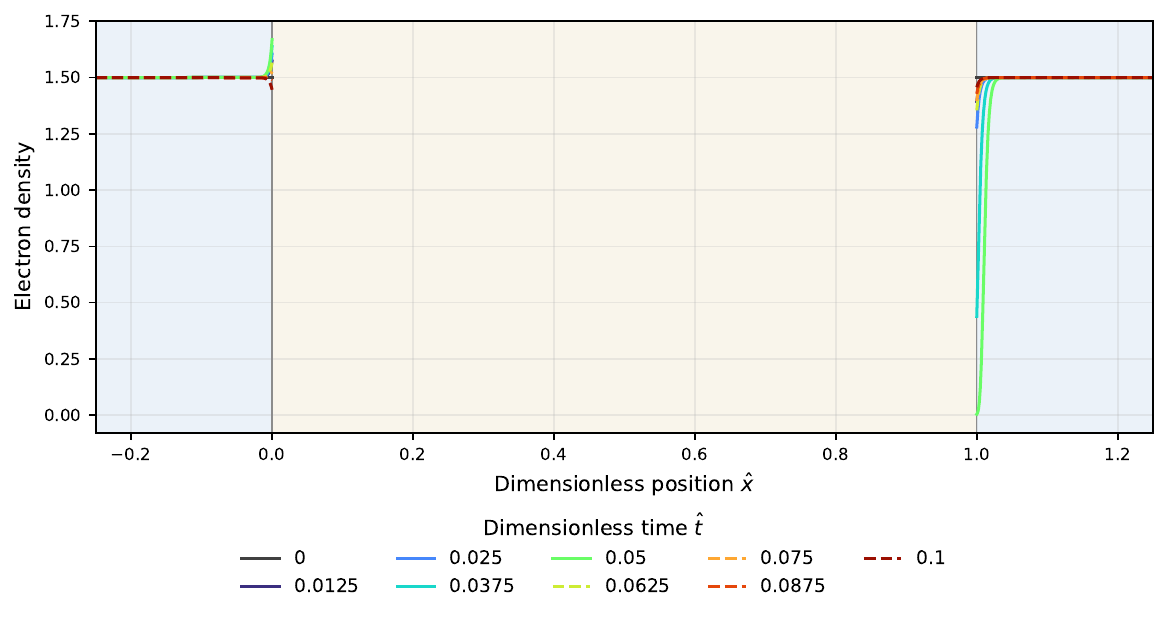}
\caption{Electron densities in the electrodes of the SIC cell. Current magnitude $1.6$, reversal at $\hat t=0.05$, and end time $0.10$. Solid curves include the reversal state; dashed curves follow reversal.}
\label{fig:sicElectronProfiles}
\end{figure*}
\begin{figure*}[!hbp]
\includegraphics[width=0.85\textwidth]{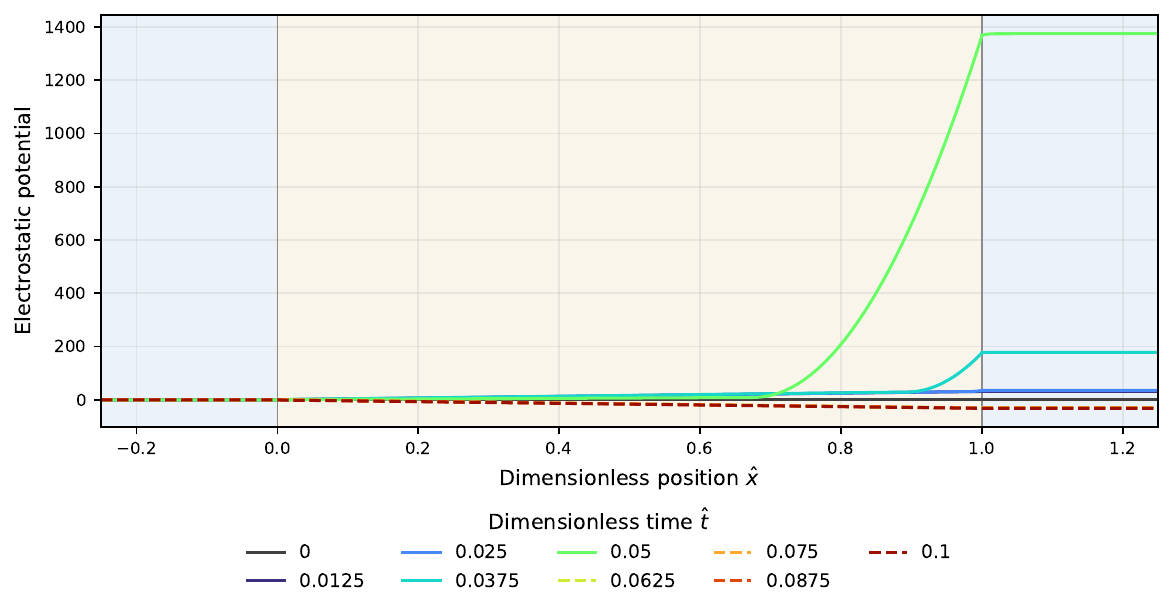}
\caption{Electrostatic potential in the SIC cell relative to the left terminal. Current magnitude $1.6$, reversal at $\hat t=0.05$, and end time $0.10$. Solid curves include the reversal state; dashed curves follow reversal.}
\label{fig:sicPotentialProfiles}
\end{figure*}

The terminal electrochemical potential difference reaches a maximum saved
magnitude of $1374.35$, compared with the much smaller polarization in DIC.
At the last saved state strictly before reversal, $\hat t=0.049375$, its
value is already $1291.26$. Thus the large polarization extends beyond the
exact switching sample. Its final value is $-31.39374$.
Figure~\ref{fig:sicTerminalPotential} distinguishes the electrochemical
potential difference from the electrostatic potential difference.
These are dimensionless model outputs, not calibrated operating voltages.
\begin{figure*}[!hbp]
\includegraphics[width=0.85\textwidth]{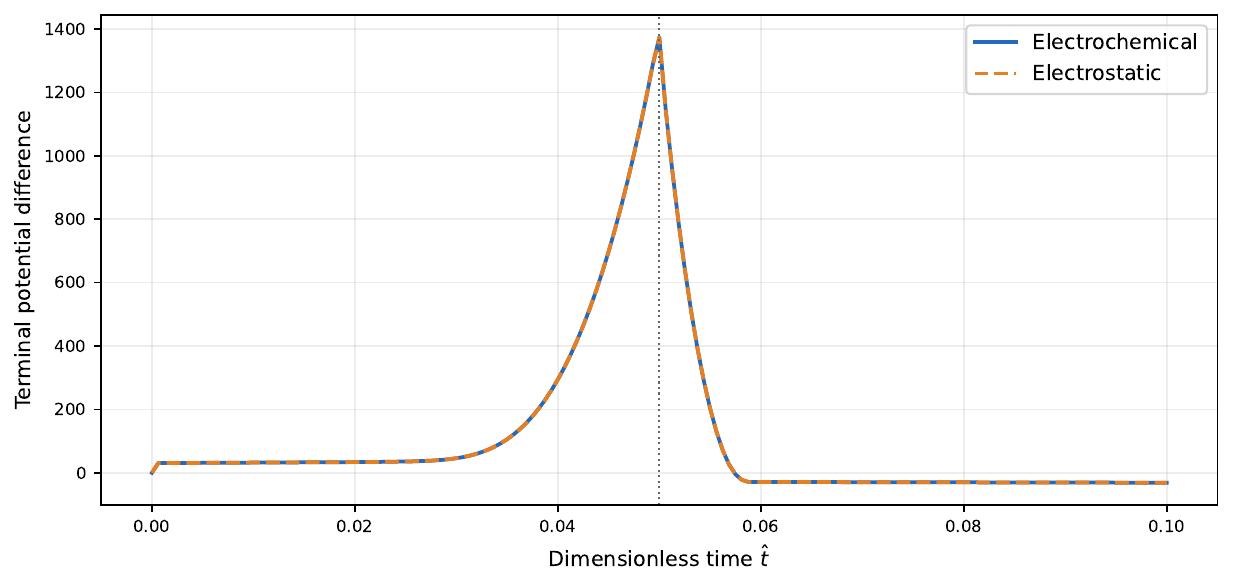}
\caption{Terminal electrochemical and electrostatic potential differences
for the current-$1.6$ SIC cycle. The vertical line marks current reversal.}
\label{fig:sicTerminalPotential}
\end{figure*}

\paragraph{Electrode utilization.}
Using Equation~\eqref{eq:numericalUtilization}, mean left- and right-electrode
utilizations at reversal are $0.6188325$ and $0.0499962$. Their final values
are $0.3021047$ and $0.3022262$, close to the common initial value
$0.3014498$. The right electrode continues losing lithium briefly after
the terminal current reverses (Figure~\ref{fig:sicUtilizationHistories}).
Reversal of imposed current does not require instantaneous reversal of
each interfacial reaction.
\begin{figure*}[!hbp]
\includegraphics[width=0.85\textwidth]{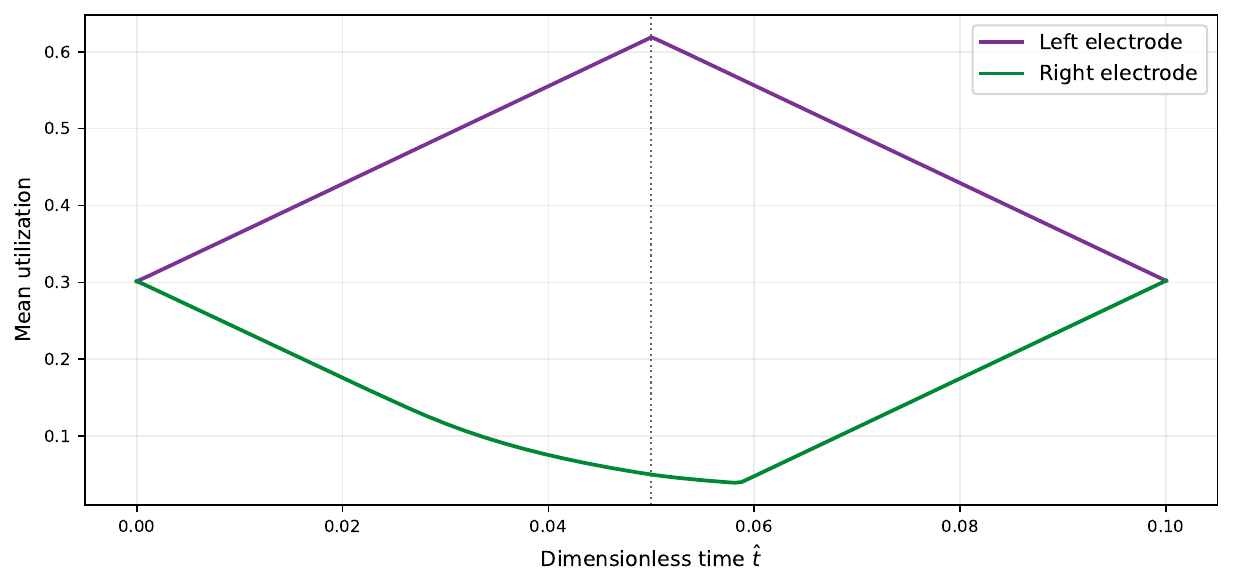}
\caption{Mean electrode utilization during the current-$1.6$ SIC cycle
with the same electrode colors as in DIC.}
\label{fig:sicUtilizationHistories}
\end{figure*}

Material calibration and assessment of the model assumptions in the depleted
regime are needed before quantitative physical predictions are inferred.
\clearpage

\section{Conclusions}
\label{sec:conclusion}
In this paper, an isothermal model of a salt-in-polyelectrolyte battery is presented. {The model
satisfies an isothermal open-system free-energy balance with nonnegative internal dissipation. It
conserves the total number of lithium nuclei (mobile and intercalated), the numbers of free anions
and polymer segments, the total number of vacancies across all three domains, and the total charge.
It also preserves global electroneutrality when the initial and boundary data satisfy the
corresponding compatibility condition. The system free energy itself need not decrease during driven
charging.} The one-dimensional DIC cycle at dimensionless current magnitude $1.6$
demonstrates lithium transfer between electrodes, sharp interfacial concentration
layers, and electrostatic polarization with the reduced Poisson coefficients.
Electrode utilization nearly returns to its initial state after current reversal.
Charge and lithium inventories are conserved to near machine precision, while
a small polymer inventory drift remains in the DIC calculation.
Integrated utilization is less sensitive to resolution than interfacial profiles
and the voltage transient at current reversal. The corresponding SIC cycle shows strong cation depletion and much larger
polarization, with delayed right-electrode recovery after current reversal.
Its charge, lithium, and polymer inventories are conserved to solver accuracy.
The same imposed charge is a larger demand relative to the initial electrode
lithium inventory in SIC; the composition and polymer mobility also differ.
The contrast therefore cannot be attributed solely to polymer charge. Both
cases constitute dimensionless illustrative studies; material calibration,
further interfacial and time resolution, and assessment of the model assumptions
in the strongly depleted regime remain future work.

While it is frequently employed in battery
modeling~\cite{kumar2024dynamic,munteanu2024single,doyle1993modeling,bazant2005current,chu2005electrochemical,danilov2010modeling,pang2019experimental,woeppel2020single,wolff2018model,landstorfer2011advanced,habscheid2025finite,ferguson2014lithium,hu2023three,jiang2022thermal,latz2011thermodynamic,bazant2013theory},
the isothermal assumption is a limitation of the presented model. Relaxing this assumption is an
important direction for future work. The representation of the dielectric responses in the
electrodes and electrolyte as field-independent constants is reasonable under the following
conditions: (i) electric fields are sufficiently small that nonlinear polarization effects do not
occur~\cite{rezaei2021interfacial},
(ii) densities vary smoothly over distances much larger than a molecule, so the material looks locally uniform~\cite{ballenegger2005dielectric},
{and (iii) retained bound-polarization modes relax rapidly relative to electrochemical time scales,
so their response may be treated as instantaneous. In this regime, the dielectric response may be
represented by a time-scale-appropriate effective background permittivity. Rapid modes contribute
their relaxed response; the high-frequency limit would instead apply to modes that are frozen on the
modeled time scale. Mobile-ion polarization is excluded because it is modeled
explicitly~\cite{kumar2019understanding}.}
Nevertheless, several studies have shown that spatial variations in the dielectric response can play
a significant role. In a foundational paper, it was demonstrated that an ion's solvation energy is
sensitive to its distance from the planar surface separating two dielectric media because the
boundary between dielectric media induces polarization effects, effectively changing the relative
permittivity~\cite{makov1994solvation}.
More recently, it was shown that the solvent reorganization energy is dominated by the medium's
dielectric polarization; hence, using a fixed solvation energy derived from a bulk permittivity can
lead to inaccurate predictions~\cite{fraggedakis2021theory,zhang2025lithium}. Relaxing the
constant-permittivity assumption to incorporate the dependency on the electric field, densities, and
time is another important direction for future work.

\begin{acknowledgments}
  This work is supported by the Center for Nanophase Materials Sciences (CNMS), which is a US
  Department of Energy, Office of Science User Facility at Oak Ridge National Laboratory (ORNL).
  CB was funded by a Graduate Advancement and Training Education (GATE) fellowship provided by The
  Science Alliance, which is a Tennessee Higher Education Commission (THEC) center of excellence
  administered by The University of Tennessee-Oak Ridge Innovation Institute (UT-ORII), and by
  NSF-DMS 2309547. RK acknowledges support from the Fast and Cooperative
  Ion Transport in Polymer-Based Materials (FaCT), an Energy
  Frontier Research Center funded by the U.S. Department of
  Energy, Office of Science, Office of Basic Energy Sciences. SMW was funded by NSF-DMS 2309547 and
  NSF-DMS 2607995. RK dedicates this work to Prof. M. Muthukumar for his foundational contributions to polyelectrolyte physics.  
\end{acknowledgments}

\section*{Statements and Declarations}
\noindent \textbf{Supplementary Information:}
Proofs, mathematical rearrangements used in Section~\ref{sec:recapturingKnownDynamics}, and the
calculations for Section~\ref{sec:compDFN}, and the conservation and numerical accuracy
checks for the DIC and SIC simulations are available in the Supplementary Information.

\begin{samepage}
\noindent \textbf{Use of Artificial Intelligence:}
All analyses were developed by the human authors. OpenAI GPT 6 Astra was used
to check and improve the calculations and proofs, identify and correct
calculation errors, and suggest grammatical corrections and improvements to
the manuscript. The initial MATLAB and MOOSE codes were written and tested
by the human authors. AI generated a second version of the MOOSE code
designed to avoid negative densities by incorporating numerical constraints.
This AI-generated MOOSE code was used to generate all simulation results
reported in this paper. AI was also used to generate the plots.\par
\end{samepage}

\noindent \textbf{Competing Interests:} The authors have no competing interests to declare that are
relevant to the content of this article.

\noindent \textbf{Data Availability:}
The numerical results generate simulation data. The data underlying the figures, together with
the run inputs and post-processing metadata, will be deposited in an archived repository; repository
and accession details will be added before publication.

\noindent \textbf{Materials Availability:} Not applicable.

\noindent \textbf{Code Availability:} The MOOSE application, run inputs, and figure-generation
scripts used for the  calculations will be archived with the numerical data. Repository and
version details will be added before publication.

\noindent \textbf{Author Contributions:}
Rajeev Kumar and Steven Wise devised early forms of the model, conceptualized the problem, and
managed the research. Charlotte Beckford refined the model, conducted the mathematical analyses,
developed and tested the codes, and created the first draft of the manuscript. All authors have
read, revised, and verified the manuscript.

\nocite{dreyer2016new}
\nocite{mongcopa2018relationship}
\bibliography{references}

@article{marcus1956electrostatic,
  title={Electrostatic free energy and other properties of states having nonequilibrium polarization. I},
  author={Marcus, Rudolph A},
  journal={Journal of Chemical Physics},
  volume={24},
  number={5},
  pages={979--989},
  year={1956},
  publisher={American Institute of Physics}
}

@book{newman2021electrochemical,
  title={Electrochemical systems},
  author={Newman, John and Balsara, Nitash P},
  year={2021},
  publisher={John Wiley \& Sons}
}

@article{Doi_2011,
doi = {10.1088/0953-8984/23/28/284118},
url = {https://dx.doi.org/10.1088/0953-8984/23/28/284118},
year = {2011},
month = {jun},
publisher = {},
volume = {23},
number = {28},
pages = {284118},
author = {Masao Doi},
title = {Onsager's variational principle in soft matter},
journal = {Journal of Physics: Condensed Matter},
}

@book{groot1962non,
  title={Non-equilibrium Thermodynamics},
  author={Groot, S.R. and Mazur, P.},
  isbn={9780444101730},
  lccn={lc62005264},
  series={Dover books on physics and chemistry},
  url={https://books.google.com/books?id=3b-wAAAAIAAJ},
  year={1962},
  publisher={North-Holland Publishing Company}
}

@article{li2025emergent,
  title={Emergent Nanostructure and Ion Transport in Polyzwitterion/Polyanion Blends},
  author={Li, Hongwei and Zhu, Qinyu and Shinohara, Yuya and Wang, Yangyang and Christakopoulos, Panagiotis and Kudlack, Autumn F and Huang, Zitan and Bonnesen, Peter V and Do, Changwoo and Rahman, Md Anisur and others},
  journal={Macromolecules},
  volume={58},
  number={16},
  pages={8658--8669},
  year={2025},
  publisher={ACS Publications}
}

@article{zhu2024using,
  title={Using Data-Science Approaches to Unravel Insights for Enhanced Transport of Lithium Ions in Single-Ion Conducting Polymer Electrolytes},
  author={Zhu, Qinyu and Liu, Yifan and Shepard, Lauren B and Bhattacharya, Debjyoti and Sinnott, Susan B and Reinhart, Wesley F and Cooper, Valentino R and Kumar, Rajeev},
  journal={Chemistry of Materials},
  volume={36},
  number={24},
  pages={11934--11946},
  year={2024},
  publisher={ACS Publications}
}

@article{singh2025revisiting,
  title={Revisiting the role of mobile ion size for the activation barrier of charge transport in single ion conducting polymers},
  author={Singh, Harmandeep and Popov, Ivan and Lehmann, Michelle L and Rahman, Md Anisur and Schweizer, Kenneth S and Kumar, Rajeev and Saito, Tomonori and Sokolov, Alexei P and Gainaru, Catalin P},
  journal={ACS Applied Polymer Materials},
  volume={7},
  number={15},
  pages={9661--9668},
  year={2025},
  publisher={ACS Publications}
}

@article{kumar2020capacitance,
  title={Capacitance of thin films containing polymerized ionic liquids},
  author={Kumar, Rajeev and Mahalik, Jyoti P and Silmore, Kevin S and Wojnarowska, Zaneta and Erwin, Andrew and Ankner, John F and Sokolov, Alexei P and Sumpter, Bobby G and Bocharova, Vera},
  journal={Science Advances},
  volume={6},
  number={26},
  pages={eaba7952},
  year={2020},
  publisher={American Association for the Advancement of Science}
}

@article{pesko2018comparing,
  title={Comparing cycling characteristics of symmetric lithium-polymer-lithium cells with theoretical predictions},
  author={Pesko, Danielle M and Feng, Zhange and Sawhney, Simar and Newman, John and Srinivasan, Venkat and Balsara, Nitash P},
  journal={Journal of the Electrochemical Society},
  volume={165},
  number={13},
  pages={A3186--A3194},
  year={2018},
  publisher={The Electrochemical Society}
}

@article{doyle1996comparison,
  title={Comparison of modeling predictions with experimental data from plastic lithium ion cells},
  author={Doyle, Marc and Newman, John and Gozdz, Antoni S and Schmutz, Caroline N and Tarascon, Jean-Marie},
  journal={Journal of the Electrochemical Society},
  volume={143},
  number={6},
  pages={1890--1903},
  year={1996},
  publisher={The Electrochemical Society, Inc.}
}

@article{fuller1994simulation,
  title={Simulation and optimization of the dual lithium ion insertion cell},
  author={Fuller, Thomas F and Doyle, Marc and Newman, John},
  journal={Journal of the Electrochemical Society},
  volume={141},
  number={1},
  pages={1--10},
  year={1994},
  publisher={The Electrochemical Society, Inc.}
}

@article{west1982modeling,
  title={Modeling of porous insertion electrodes with liquid electrolyte},
  author={West, Keld and Jacobsen, Torben and Atlung, Sven},
  journal={Journal of the Electrochemical Society},
  volume={129},
  number={7},
  pages={1480--1485},
  year={1982},
  publisher={The Electrochemical Society, Inc.}
}

@article{bazant2013theory,
  title={Theory of chemical kinetics and charge transfer based on nonequilibrium thermodynamics},
  author={Bazant, Martin Z},
  journal={Accounts of Chemical Research},
  volume={46},
  number={5},
  pages={1144--1160},
  year={2013},
  publisher={ACS Publications}
}

@article{fraggedakis2021theory,
  title={Theory of coupled ion-electron transfer kinetics},
  author={Fraggedakis, Dimitrios and McEldrew, Michael and Smith, Raymond B and Krishnan, Yamini and Zhang, Yirui and Bai, Peng and Chueh, William C and Shao-Horn, Yang and Bazant, Martin Z},
  journal={Electrochimica Acta},
  volume={367},
  pages={137432},
  year={2021},
  publisher={Elsevier}
}

@article{makov1994solvation,
  title={Solvation and ionization near a dielectric surface},
  author={Makov, G and Nitzan, A},
  journal={Journal of Physical Chemistry},
  volume={98},
  number={13},
  pages={3459--3466},
  year={1994},
  publisher={ACS Publications}
}

@article{landstorfer2011advanced,
  title={An advanced model framework for solid electrolyte intercalation batteries},
  author={Landstorfer, Manuel and Funken, Stefan and Jacob, Timo},
  journal={Physical Chemistry Chemical Physics},
  volume={13},
  number={28},
  pages={12817--12825},
  year={2011},
  publisher={Royal Society of Chemistry}
}

@article{wolff2018model,
  title={Model based assessment of performance of lithium-ion batteries using single ion conducting electrolytes},
  author={Wolff, Nicolas and R{\"o}der, Fridolin and Krewer, Ulrike},
  journal={Electrochimica Acta},
  volume={284},
  pages={639--646},
  year={2018},
  publisher={Elsevier}
}

@article{danilov2010modeling,
  title={Modeling all-solid-state Li-ion batteries},
  author={Danilov, D and Niessen, RAH and Notten, PHL},
  journal={Journal of the Electrochemical Society},
  volume={158},
  number={3},
  pages={A215},
  year={2011},
  publisher={IOP Publishing},
  doi={10.1149/1.3521414}
}

@article{pang2019experimental,
  title={Experimental and numerical analysis to identify the performance limiting mechanisms in solid-state lithium cells under pulse operating conditions},
  author={Pang, Mei-Chin and Hao, Yucang and Marinescu, Monica and Wang, Huizhi and Chen, Mu and Offer, Gregory J},
  journal={Physical Chemistry Chemical Physics},
  volume={21},
  number={41},
  pages={22740--22755},
  year={2019},
  publisher={Royal Society of Chemistry}
}

@article{chu2005electrochemical,
  title={Electrochemical thin films at and above the classical limiting current},
  author={Chu, Kevin T and Bazant, Martin Z},
  journal={SIAM Journal on Applied Mathematics},
  volume={65},
  number={5},
  pages={1485--1505},
  year={2005},
  publisher={SIAM}
}

@article{bazant2005current,
  title={Current-voltage relations for electrochemical thin films},
  author={Bazant, Martin Z and Chu, Kevin T and Bayly, Bruce J},
  journal={SIAM Journal on Applied Mathematics},
  volume={65},
  number={5},
  pages={1463--1484},
  year={2005},
  publisher={SIAM}
}

@article{chattopadhyay2023applications,
  title={Applications of polymer electrolytes in lithium-ion batteries: a review},
  author={Chattopadhyay, Jayeeta and Pathak, Tara Sankar and Santos, Diogo MF},
  journal={Polymers},
  volume={15},
  number={19},
  pages={3907},
  year={2023},
  publisher={MDPI}
}

@article{deguenon2023overcoming,
  title={Overcoming the challenges of integrating variable renewable energy to the grid: A comprehensive review of electrochemical battery storage systems},
  author={Deguenon, Lere and Yamegueu, Daniel and Kadri, Sani Moussa and Gomna, Aboubakar},
  journal={Journal of Power Sources},
  volume={580},
  pages={233343},
  year={2023},
  publisher={Elsevier}
}

@article{zhang2025lithium,
  title={Lithium-ion intercalation by coupled ion-electron transfer},
  author={Zhang, Yirui and Fraggedakis, Dimitrios and Gao, Tao and Pathak, Shakul and Zhuang, Debbie and Grosu, Cristina and Samantaray, Yash and Neto, Armando R. C. and Duggirala, Sravani R. and Huang, Botao and Zhu, Yun Guang and Giordano, Livia and Tatara, Ryoichi and Agarwal, Harsh and Stephens, Ryan M. and Bazant, Martin Z. and Shao-Horn, Yang},
  journal={Science},
  volume={390},
  number={6768},
  pages={eadq2541},
  year={2025},
  publisher={American Association for the Advancement of Science},
  doi={10.1126/science.adq2541}
}

@article{kumar2024dynamic,
  title={Dynamic density functional theory of polymers with salt in electric fields},
  author={Kumar, Rajeev and Zhu, Qinyu},
  journal={Journal of Chemical Physics},
  volume={161},
  number={10},
  year={2024},
  pages={104902},
  publisher={AIP Publishing}
}

@article{dickinson2019butler,
  title={The Butler-Volmer equation for polymer electrolyte membrane fuel cell (PEMFC) electrode kinetics: A critical discussion},
  author={Dickinson, Edmund JF and Hinds, Gareth},
  journal={Journal of the Electrochemical Society},
  volume={166},
  number={4},
  pages={F221--F231},
  year={2019},
  publisher={IOP Publishing},
  doi={10.1149/2.0361904jes}
}

@article{shah2019comparing,
  title={Comparing measurements of limiting current of electrolytes with theoretical predictions up to the solubility limit},
  author={Shah, Deep B and Kim, Hong Keun and Nguyen, Hien Q and Srinivasan, Venkat and Balsara, Nitash P},
  journal={Journal of Physical Chemistry C},
  volume={123},
  number={39},
  pages={23872--23881},
  year={2019},
  publisher={ACS Publications}
}

@article{dreyer2016new,
  title={A new perspective on the electron transfer: recovering the Butler--Volmer equation in non-equilibrium thermodynamics},
  author={Dreyer, Wolfgang and Guhlke, Clemens and M{\"u}ller, R{\"u}diger},
  journal={Physical Chemistry Chemical Physics},
  volume={18},
  number={36},
  pages={24966--24983},
  year={2016},
  publisher={Royal Society of Chemistry}
}

@article{doyle1993modeling,
  title={Modeling of galvanostatic charge and discharge of the lithium/polymer/insertion cell},
  author={Doyle, Marc and Fuller, Thomas F and Newman, John},
  journal={Journal of the Electrochemical Society},
  volume={140},
  number={6},
  pages={1526--1533},
  year={1993},
  publisher={IOP Publishing},
  doi={10.1149/1.2221597}
}

@article{latz2011thermodynamic,
  title={Thermodynamic consistent transport theory of Li-ion batteries},
  author={Latz, Arnulf and Zausch, Jochen},
  journal={Journal of Power Sources},
  volume={196},
  number={6},
  pages={3296--3302},
  year={2011},
  publisher={Elsevier}
}

@article{hu2023three,
  title={A Three-Dimensional, Thermodynamically and Variationally Consistent, Fully Coupled, Electro-Chemo-Thermo-Mechanical Model of Solid-State Batteries},
  author={Hu, Tianchen and Messner, Mark C and Barai, Pallab and Barua, Bipul},
  journal={Journal of the Electrochemical Society},
  volume={170},
  number={12},
  pages={123501},
  year={2023},
  publisher={IOP Publishing},
  doi={10.1149/1945-7111/ad09f5}
}

@article{yildiz2024comparative,
  title={A comparative review of models for all-solid-state li-ion batteries},
  author={Yildiz, Erkin and Serpelloni, Mattia and Salvadori, Alberto and Cabras, Luigi},
  journal={Batteries},
  volume={10},
  number={5},
  pages={150},
  year={2024},
  publisher={MDPI},
  doi={10.3390/batteries10050150}
}

@article{munteanu2024single,
  title={Single-Particle Model of Li-ion battery--Model calibration and validation against real data in an electric vehicular application},
  author={Munteanu, Iulian and Bratcu, Antoneta Iuliana and Thivel, Pierre-Xavier and Bultel, Yann and Georges, Didier and Decaux, C{\'e}line},
  journal={IFAC-PapersOnLine},
  volume={58},
  number={13},
  pages={23--30},
  year={2024},
  publisher={Elsevier}
}

@phdthesis{ferguson2014lithium,
  title={Lithium-ion battery modeling using non-equilibrium thermodynamics},
  author={Ferguson, Todd Richard},
  year={2014},
  school={Massachusetts Institute of Technology}
}

@misc{habscheid2025finite,
  title={A finite element solver for a thermodynamically consistent electrolyte model},
  author={Habscheid, Jan and Singh, Satyvir and Theisen, Lambert and Braun, Stefanie and Torrilhon, Manuel},
  year={2025},
  eprint={2505.16296},
  archivePrefix={arXiv}
}

@article{brosa2022continuum,
  title={A continuum of physics-based lithium-ion battery models reviewed},
  author={Brosa Planella, Ferran and Ai, Weilong and Boyce, Adam M and Ghosh, Abir and Korotkin, Ivan and Sahu, Smita and Sulzer, Valentin and Timms, Robert and Tranter, Thomas G and Zyskin, Maxim and others},
  journal={Progress in Energy},
  volume={4},
  number={4},
  pages={042003},
  year={2022},
  publisher={IOP Publishing}
}

@article{jiang2022thermal,
  title={A thermal-electrochemical-mechanical coupled model based on non-equilibrium thermodynamics of Li-ion batteries},
  author={Jiang, Wenjuan and Zhou, Qi and Lu, Feng and Chen, Yufang and Ma, Zengsheng},
  journal={Journal of Energy Storage},
  volume={55},
  pages={105655},
  year={2022},
  publisher={Elsevier}
}

@article{alkhedher2024electrochemical,
  title={Electrochemical and thermal modeling of lithium-ion batteries: A review of coupled approaches for improved thermal performance and safety lithium-ion batteries},
  author={Alkhedher, Mohammad and Al Tahhan, Aghyad B and Yousaf, Jawad and Ghazal, Mohammed and Shahbazian-Yassar, Reza and Ramadan, Mohamad},
  journal={Journal of Energy Storage},
  volume={86},
  pages={111172},
  year={2024},
  publisher={Elsevier}
}

@article{mehta2024mathematical,
  title={Mathematical modelling of electrochemical, thermal and degradation processes in lithium-ion cells—A comprehensive review},
  author={Mehta, Rohit and Gupta, Amit},
  journal={Renewable and Sustainable Energy Reviews},
  volume={192},
  pages={114264},
  year={2024},
  publisher={Elsevier}
}

@article{kumar2017rayleighian,
  title={A Rayleighian approach for modeling kinetics of ionic transport in polymeric media},
  author={Kumar, Rajeev and Mahalik, Jyoti P and Bocharova, Vera and Stacy, Eric W and Gainaru, Catalin and Saito, Tomonori and Gobet, Mallory P and Greenbaum, Steve and Sumpter, Bobby G and Sokolov, Alexei P},
  journal={Journal of Chemical Physics},
  volume={146},
  number={6},
  pages={064902},
  year={2017},
  publisher={AIP Publishing},
  doi={10.1063/1.4975309}
}

@article{kumar2025ferro,
  title = {Modeling kinetic effects of charged vacancies on electromechanical responses of ferroelectrics: Rayleighian approach},
  author = {Kumar, Rajeev and Zhang, Shuaifang and Ganesh, P.},
  journal = {Physical Review Research},
  volume = {7},
  number = {1},
  pages = {013059},
  numpages = {12},
  year = {2025},
  month = {Jan},
  publisher = {American Physical Society},
  doi = {10.1103/PhysRevResearch.7.013059},
  url = {https://link.aps.org/doi/10.1103/PhysRevResearch.7.013059}
}

@article{woeppel2020single,
  title={Single-versus dual-ion conductors for electric double layer gating: finite element modeling and hall-effect measurements},
  author={Woeppel, Aaron and Xu, Ke and Kozhakhmetov, Azimkhan and Awate, Shubham and Robinson, Joshua A and Fullerton-Shirey, Susan K},
  journal={ACS Applied Materials \& Interfaces},
  volume={12},
  number={36},
  pages={40850--40858},
  year={2020},
  publisher={ACS Publications}
}

@article{boettcher2020potentially,
  title={Potentially confusing: potentials in electrochemistry},
  author={Boettcher, Shannon W and Oener, Sebastian Z and Lonergan, Mark C and Surendranath, Yogesh and Ardo, Shane and Brozek, Carl and Kempler, Paul A},
  journal={ACS Energy Letters},
  volume={6},
  number={1},
  pages={261--266},
  year={2020},
  publisher={ACS Publications}
}

@article{kumar2019understanding,
  title={Understanding the effects of dipolar interactions on the thermodynamics of diblock copolymer melts},
  author={Kumar, Rajeev and Li, Wei and Sumpter, Bobby G and Muthukumar, Murugappan},
  journal={Journal of Chemical Physics},
  volume={151},
  number={5},
  pages={054902},
  year={2019},
  publisher={AIP Publishing},
  doi={10.1063/1.5114799}
}

@article{rezaei2021interfacial,
  title={Interfacial, electroviscous, and nonlinear dielectric effects on electrokinetics at highly charged surfaces},
  author={Rezaei, Majid and Mitterwallner, Bernhard G and Loche, Philip and Uematsu, Yuki and Netz, Roland R and Bonthuis, Douwe Jan},
  journal={Journal of Physical Chemistry B},
  volume={125},
  number={18},
  pages={4767--4778},
  year={2021},
  publisher={ACS Publications}
}

@article{ballenegger2005dielectric,
  title={Dielectric permittivity profiles of confined polar fluids},
  author={Ballenegger, V and Hansen, J-P},
  journal={Journal of Chemical Physics},
  volume={122},
  number={11},
  pages={114711},
  year={2005},
  publisher={AIP Publishing},
  doi={10.1063/1.1845431}
}

@article{prigogine1948affinity,
  title={Affinity and reaction rate close to equilibrium},
  author={Prigogine, Ilya and Outer, Paul and Herbo, CL},
  journal={Journal of Physical and Colloid Chemistry},
  volume={52},
  number={2},
  pages={321--331},
  year={1948},
  publisher={ACS Publications},
  doi={10.1021/j150458a004}
}

@article{mongcopa2018relationship,
  title={Relationship between segmental dynamics measured by quasi-elastic neutron scattering and conductivity in polymer electrolytes},
  author={Mongcopa, Katrina Irene S and Tyagi, Madhusudan and Mailoa, Jonathan P and Samsonidze, Georgy and Kozinsky, Boris and Mullin, Scott A and Gribble, Daniel A and Watanabe, Hiroshi and Balsara, Nitash P},
  journal={ACS Macro Letters},
  volume={7},
  number={4},
  pages={504--508},
  year={2018},
  publisher={ACS Publications}
}

\clearpage

\section*{Supplementary Information}
\addcontentsline{toc}{section}{Supplementary Information}

\renewcommand{\thefigure}{S\arabic{figure}}
\setcounter{figure}{0}
\renewcommand{\thetable}{S\arabic{table}}
\setcounter{table}{0}
\renewcommand{\theequation}{S\arabic{equation}}
\setcounter{equation}{0}

\makeatletter
\renewcommand{\thesubsection}{S\arabic{subsection}}
\renewcommand{\p@subsection}{} 
\makeatother
\setcounter{subsection}{0}

\subsection{Motivating Example for Equation~\eqref{sk}}
\label{sec:exProp}
Consider a symmetric lithium--polymer--lithium cell with LiTFSI salt added to the polymer electrolyte. The
number densities of $\mathrm{Li}^+$ and $\mathrm{TFSI}^-$ in the electrolyte are represented by
$\rho_+$ and $\rho_-$, respectively. The number densities of intercalated lithium in the left and
right electrodes are represented by $\rho_{i_\mathrm{d}}$, $\mathrm{d} \in \{\mathrm{L},
\mathrm{R}\}$, respectively. During a discharge cycle, the reactions occurring at the left and right
electrodes are represented by Equations~\eqref{leftLiReaction} and~\eqref{rightLiReaction} (depicted
in Figure~\ref{fig:reactionScheme}). During charging the written directions and stoichiometric
coefficients remain fixed, while the net rates change sign. The arrows in
Figure~\ref{fig:reactionScheme} show the discharge directions.

Suppose that we wish to represent the following forward reaction on $\Sigma_\mathrm{d}$
\begin{equation}
  \mathrm{Li} + \mathrm{V}_\mathrm{M} \rightarrow \mathrm{Li}^+ + e^- + \mathrm{V}_\mathrm{d}. \label{deintercalation}
\end{equation}
This implies
\begin{equation}
  s_{{n},\mathrm{d}}
  = \begin{cases}
    \phantom{-}1, & {n} \in \left\{ e_\mathrm{d},\mathrm{V}_\mathrm{d},+\right\}, \\
    -1, & {n} \in \left\{i_\mathrm{d},\mathrm{V}_\mathrm{M}\right\}, \\
    \phantom{-}0, & {n} \in \left\{e_{\mathrm{d}_0}, i_{\mathrm{d}_0}, \mathrm{V}_{\mathrm{d}_0}, -, p \right\},
  \end{cases} \label{deintercalationsk}
\end{equation} where ${\mathrm{d}, \mathrm{d}_0} \in \left\{\mathrm{L},\mathrm{R}\right\}, {\mathrm{d}}\neq \mathrm{d}_0$.

If instead we wish to represent
the following backward reaction on $\Sigma_\mathrm{d}$
\begin{equation}
  \mathrm{Li} + \mathrm{V}_\mathrm{M} \leftarrow \mathrm{Li}^+ + e^- + \mathrm{V}_\mathrm{d},
\end{equation}
then the stoichiometric coefficients in~\eqref{deintercalationsk} would be negated.

More generally, for every cation produced (consumed) by the reaction occurring on
$\Sigma_\mathrm{d}$, an electrolyte vacancy and an intercalated metal is consumed (produced) and an
electron and an electrode vacancy are produced (consumed). No other species are involved in the
reaction. The stoichiometric coefficients are therefore given by~\eqref{sk}.

\subsection{List of Symbols}
\label{sec:listofSymbols}
\begin{table*}
  \caption{\label{tab:addnlSymbols}Symbols used in the paper which are not model parameters or
  species.}
  \begin{ruledtabular}
    \begin{tabular}{ccc}
      Symbol & Definition & Units\\ \midrule
      $\Omega_\mathrm{L}$ & {Left/negative electrode} & -- \\
      $\Omega_\mathrm{M}$ & Electrolyte & -- \\
      $\Omega_\mathrm{R}$ & {Right/positive electrode} & -- \\
      $\Sigma_\mathrm{d}$ & The boundary between $\Omega_\mathrm{M}$ and $\Omega_\mathrm{d}$ & -- \\
      $\Gamma_\mathrm{d}$ & The far boundary of $\Omega_\mathrm{d}$ & -- \\
      $\partial$ & Used to indicate the boundary of a domain & -- \\
      $\boldsymbol{n}_\mathrm{d}$ & Outward-facing normal with respect to $\Omega_\mathrm{d}$ & -- \\
      $\boldsymbol{n}_\mathrm{M}$ & Outward facing normal with respect to $\Omega_\mathrm{M}$ & --\\
      $\tilde{\mu}_n$ & Electrochemical potential of species $n$ & $\mathrm{Joules}$ \\
      & (Defined in Equation~\eqref{chem_potentials}) & \\
      $\tilde{\mu}^*_n$ & Exchange electrochemical potential of species $n$ & $\mathrm{Joules}$ \\
      & (Defined in Equation~\eqref{exchange_chem_potentials}) & \\
      $M_\ell$ & Mobility of species $\ell$ & $\frac{1}{\left(\mathrm{Length}\right)\cdot\mathrm{Time}}$\\
      $\nu_\mathrm{d}$ & Reaction rate on $\Sigma_\mathrm{d}$ & $\frac{\mathrm{Moles}}{\left(\mathrm{Length}\right)^2\cdot\mathrm{Time}}$\\
      $\boldsymbol J_\ell$ & Number flux of species $\ell$, $-M_\ell\nabla(\tilde\mu_\ell^*/k_{\mathrm{B}}T)$ & $\frac{1}{\left(\mathrm{Length}\right)^2\cdot\mathrm{Time}}$\\
      $\boldsymbol j_{\rho_c}$ & Conduction-current density & $\frac{\mathrm{Amperes}}{\left(\mathrm{Length}\right)^2}$\\
      $\boldsymbol I_\mathrm{tot}$ & Total conduction-plus-displacement current density & $\frac{\mathrm{Amperes}}{\left(\mathrm{Length}\right)^2}$\\
      $\rho_c$ & Local electric-charge density & $\frac{\mathrm{Coulombs}}{\left(\mathrm{Length}\right)^3}$\\
      $Q(t)$ & Total electric charge in the complete cell & $\mathrm{Coulombs}$\\
      $\tilde{\mu}_{\ddagger, \mathrm{d}}$&  Excess electrochemical potential of the transition & $\mathrm{Joules}$ \\
      & state of the chemical reaction occurring on $\Sigma_\mathrm{d}$ & \\
    \end{tabular}
  \end{ruledtabular}
\end{table*}
The electron number flux written $\boldsymbol j_{e,\mathrm d}$ in the main text is $\boldsymbol
J_{e_\mathrm d}$ in the species-flux notation of this SI; these denote the same quantity.

\subsection{Proof of Theorem~\ref{thmThermoConsist}}
\label{sec:proofThmThermoConsist}
We introduce the notation $\dot{g} = \frac{\partial g}{\partial t}$ for $g$, a function.
\begin{enumerate}
  \item
  From the definition of the total free energy of the system, using integration by parts,
  Equations~\eqref{poissonElectrodes} and~\eqref{poissonElectrolyte},
  definition~\eqref{chem_potentials}, and the electrostatic potential's external circuit-electrode boundary
  conditions, the terminal electrostatic terms vanish: $\dot\psi_\mathrm L=0$ on $\Gamma_\mathrm L$
  by the fixed reference condition~\eqref{psiDirichletBCLeft}, and $\nabla\psi_\mathrm R\cdot\bm
  n_\mathrm R=0$ on $\Gamma_\mathrm R$ by Equation~\eqref{psiNeumannBCRight}. Therefore,
  \begin{eqnarray}
    \frac{\dot{F}}{k_{\mathrm{B}}T}
    = \sum_{\mathrm{d}\in\left\{\mathrm{L},\mathrm{R}\right\}} \int_{\Omega_{\mathrm{d}}}\sum_{{\ell}\in \{e_\mathrm{d}, i_\mathrm{d},\mathrm{V}_\mathrm{d}\}} \dot{\rho}_{\ell}\frac{\tilde{\mu}_{{\ell}}}{k_{\mathrm{B}}T}\,\mathrm{d}\bm r
    + \int_{\Omega_\mathrm{M}} \sum_{\ell \in \{+, -, p, \mathrm{V}_\mathrm{M}\}} \dot{\rho}_\ell\frac{\tilde{\mu}_\ell}{k_{\mathrm{B}}T}\mathrm{d}\bm r\nonumber\\
    -\sum_{\mathrm{d}\in\left\{\mathrm{L},\mathrm{R}\right\}}\int_{\Sigma_\mathrm{d}} \left\{ \frac{e\dot{\psi}_\mathrm{M}}{k_{\mathrm{B}}T} \frac{\epsilon_\mathrm{M}}{4 \pi l_\mathrm{B}} \nabla \frac{e\psi_\mathrm{M}}{k_{\mathrm{B}}T} \!\cdot\! \bm n_\mathrm{M}+ \frac{e\dot{\psi}_\mathrm{d}}{k_{\mathrm{B}}T} \frac{\epsilon_\mathrm{d}}{4 \pi l_\mathrm{B}} \nabla \frac{e\psi_\mathrm{d}}{k_{\mathrm{B}}T} \!\cdot\! \bm n_\mathrm{d}\right\}\, \mathrm{d}\bm S
    \label{FDeriv1}.
  \end{eqnarray}

  Boundary conditions~\eqref{transmission_cont}
  and~\eqref{transmission_jump} make the surface integral in~\eqref{FDeriv1} vanish. Using
  definition~\eqref{exchange_chem_potentials} and the no voids
  assumptions~\eqref{localincompresselectrodes}
  and~\eqref{localincompresselectrolyte},~\eqref{FDeriv1} simplifies to
  \begin{eqnarray}
    \frac{\dot{F}}{k_{\mathrm{B}}T} = \sum_{\mathrm{d}\in\left\{\mathrm{L},\mathrm{R}\right\}}\int_{\Omega_{\mathrm{d}}}\sum_{{\ell} \in \left\{e_\mathrm{d},i_\mathrm{d}\right\}}\dot{\rho}_{{\ell}} \frac{\tilde{\mu}_{{\ell}}^*}{k_{\mathrm{B}}T} \,\mathrm{d}\bm r
    + \int_{\Omega_\mathrm{M}}\sum_{\ell \in \left\{+,-,p\right\}}\dot{\rho}_\ell \frac{\tilde{\mu}_\ell^*}{k_{\mathrm{B}}T}  \, \mathrm{d}\bm r.\label{FDeriv2}
  \end{eqnarray}

  Using Equations~\eqref{ellHatGovEqn} and~\eqref{ellGovEqn}, integration by parts, the periodic
  boundary conditions, and {the physical boundary
  conditions~\eqref{noFluxI},~\eqref{fluxECBC},~\eqref{ellHat_BC}, and~\eqref{ell_BC}},
  without yet specifying the form of the reaction rate $\nu_\mathrm{d}$, Equation~\eqref{FDeriv2}
  becomes
  \begingroup
  \footnotesize
  \begin{eqnarray}
    \frac{\dot{F}}{k_{\mathrm{B}}T} &=& -\sum_{\mathrm{d} \in \{\mathrm{L},\mathrm{R}\}} \int_{\Omega_\mathrm{d}} \left\{\sum_{{\ell} \in \{ e_\mathrm{d},i_\mathrm{d} \}} M_{{\ell}}\left|\nabla \left( \frac{\tilde{\mu}_{{\ell}}^*}{k_{\mathrm{B}}T}\right) \right|^2  \right\} \, \mathrm{d}\bm r - \int_{\Omega_\mathrm{M}} \left\{\sum_{\ell \in \{+, -, p\}} M_{\ell}\left|\nabla \left( \frac{\tilde{\mu}_{\ell}^*}{k_{\mathrm{B}}T} \right)\right|^2\right\} \, \mathrm{d}\bm r\nonumber\\
    &+& \int_{\Gamma_\mathrm{L}} M_{e_\mathrm{L}}\nabla\left(\frac{\tilde{\mu}_{e_\mathrm{L}}^*}{k_{\mathrm{B}}T}\right)\!\cdot\!\bm n_\mathrm{L}\left(\frac{\tilde{\mu}_{e_\mathrm{L}}^*}{k_{\mathrm{B}}T}\right)\,\mathrm{d}\bm S\nonumber\\
    &+& \int_{\Gamma_\mathrm{R}} M_{e_\mathrm{R}}\nabla\left(\frac{\tilde{\mu}_{e_\mathrm{R}}^*}{k_{\mathrm{B}}T}\right)\!\cdot\!\bm n_\mathrm{R}\left(\frac{\tilde{\mu}_{e_\mathrm{R}}^*}{k_{\mathrm{B}}T}\right)\,\mathrm{d}\bm S\nonumber\\
    &-& \sum_{\mathrm{d} \in \{\mathrm{L},\mathrm{R}\}} \int_{\Sigma_\mathrm{d}} N_A\left\{\frac{A^\mathrm{r}_\mathrm{d}}{k_{\mathrm{B}}T}-\frac{A^\mathrm{p}_\mathrm{d}}{k_{\mathrm{B}}T}\right\}\nu_\mathrm{d}\, \mathrm{d}\bm S.\label{reactionRateEmerges}
  \end{eqnarray}
  \endgroup
  In the last term of Equation~\eqref{reactionRateEmerges}, we used~\cite{prigogine1948affinity},
  $\sum_{\ell\in\left\{e_\mathrm{d},i_\mathrm{d},+,-,p\right\}}s_{\ell,\mathrm{d}}\left(\frac{\tilde{\mu}_\ell^*}{k_{\mathrm{B}}T}\right)
  = -\frac{A_\mathrm{d}}{k_{\mathrm{B}}T} =
  -\frac{A^\mathrm{r}_\mathrm{d}}{k_{\mathrm{B}}T}+\frac{A^\mathrm{p}_\mathrm{d}}{k_{\mathrm{B}}T}$.
  Thus its integrand is
  \begin{equation}
    -\frac{A_\mathrm{d}}{k_{\mathrm{B}}T}\left(N_A \nu_\mathrm{d}\right) = -N_A\left\{\frac{A^\mathrm{r}_\mathrm{d}}{k_{\mathrm{B}}T}-\frac{A^\mathrm{p}_\mathrm{d}}{k_{\mathrm{B}}T}\right\}\nu_\mathrm{d}.\label{secondSurfaceIntegrand}
  \end{equation}
  Since $f_\mathrm{inc}(x)$ is a nondecreasing function, $\forall \, x, y, z \in \mathbb{R}$,
  \begin{eqnarray*}
    0 \leq (\left[x-z\right]-\left[y-z\right]) \left(f_\mathrm{inc}(x-z)-f_\mathrm{inc}(y-z)\right)\\
    = \left(x-y\right)\left(f_\mathrm{inc}(x-z)-f_\mathrm{inc}(y-z)\right)
  \end{eqnarray*}
  Letting $x:= \frac{A^\mathrm{r}_\mathrm{d}}{k_{\mathrm{B}}T}$, $y:=
  \frac{A^\mathrm{p}_\mathrm{d}}{k_{\mathrm{B}}T}$, and $z =
  \frac{\tilde{\mu}_{\ddagger,\mathrm{d}}}{k_{\mathrm{B}}T}$ we have
  \begin{eqnarray}
    0 \leq \left\{ \frac{A^\mathrm{r}_\mathrm{d}}{k_{\mathrm{B}}T}-\frac{A^\mathrm{p}_\mathrm{d}}{k_{\mathrm{B}}T}\right\} \left[ f_\mathrm{inc} \left( \frac{A^\mathrm{r}_\mathrm{d}-\tilde{\mu}_{\ddagger,\mathrm{d}}}{k_{\mathrm{B}}T}\right) - f_\mathrm{inc} \left( \frac{A^\mathrm{p}_\mathrm{d}-\tilde{\mu}_{\ddagger,\mathrm{d}}}{k_{\mathrm{B}}T}\right)\right] \nonumber\\
    =\left\{ \frac{A^\mathrm{r}_\mathrm{d}}{k_{\mathrm{B}}T}-\frac{A^\mathrm{p}_\mathrm{d}}{k_{\mathrm{B}}T}\right\} \frac{\nu_\mathrm{d}}{k_\mathrm d}.
  \end{eqnarray}

  For $k_\mathrm d>0$, monotonicity of $f_\mathrm{inc}$ shows that the product multiplying the
  explicit minus sign in the interfacial term of~\eqref{reactionRateEmerges} is nonnegative; when
  $k_\mathrm d=0$, $\nu_\mathrm d=0$ and that term vanishes. Because $N_A>0$, the two bulk terms and
  the interfacial term in~\eqref{reactionRateEmerges} are therefore all nonpositive contributions to
  the free-energy rate.

  {We now consider the two external-terminal surface integrals in~\eqref{reactionRateEmerges}.}

  {At the external terminals, the one-dimensional constant-area current boundary conditions give
  \[
  M_{e_\mathrm{L}}\nabla\left(\frac{\tilde{\mu}_{e_\mathrm{L}}^*}{k_{\mathrm{B}}T}\right)\!\cdot\!\bm n_\mathrm{L}=-\frac{I_\mathrm{app}}{e},
  \qquad
  M_{e_\mathrm{R}}\nabla\left(\frac{\tilde{\mu}_{e_\mathrm{R}}^*}{k_{\mathrm{B}}T}\right)\!\cdot\!\bm n_\mathrm{R}=\frac{I_\mathrm{app}}{e},
  \]
  and the ideal-current-collector assumption makes each terminal electron electrochemical potential
  uniform, with $\tilde{\mu}_{e_\mathrm{L}|\Gamma_L}^*-\tilde{\mu}_{e_\mathrm{R}|\Gamma_R}^*=eV$. In
  general geometry the boundary power is retained as the sum of the two
  electrochemical-potential-weighted terminal-current integrals. With balanced integrated currents
  it factors as $I_\mathrm{app}^{\mathrm{tot}}V$; for congruent constant-area terminals with matched
  current density, the sum in~\eqref{reactionRateEmerges} becomes}
  \begin{equation}
    \int_{\Gamma_\mathrm{R}}\left\{\frac{I_\mathrm{app}}{e} \left(\frac{-eV}{k_{\mathrm{B}}T}\right)\right\} \, \mathrm{d}\bm S = - \int_{\Gamma_\mathrm{R}} I_\mathrm{app} \frac{V}{k_{\mathrm{B}}T} \, \mathrm{d}\bm S.
  \end{equation}
  {Let $\mathcal D_\mathrm{int}\geq0$ denote $k_{\mathrm{B}}T$ times the sum of the three
  nonnegative quantities preceded by explicit minus signs in Equation~\eqref{reactionRateEmerges}:
  the electrode bulk integral, the electrolyte bulk integral, and the interfacial reaction integral.
  Thus $\mathcal D_\mathrm{int}$ has units of energy per unit time. The complete balance is
  therefore
  \begin{equation}
    \dot F=-\mathcal D_\mathrm{int}-\int_{\Gamma_\mathrm R}I_\mathrm{app}V\,\mathrm dS.
  \end{equation}
  Equivalently,
  \begin{equation}
    \dot F+\int_{\Gamma_\mathrm R}I_\mathrm{app}V\,\mathrm dS=-\mathcal D_\mathrm{int}\leq0.
  \end{equation}
  The terminal term is external electrical power, and its sign depends on whether the cell receives
  or delivers energy. Thus the internal free energy can increase during charging. The substitution
  $V=I_\mathrm{app}R$ is not made: $V$ is the full cell voltage, including chemical electromotive
  force and polarization, so $V/I_\mathrm{app}$ is not a constitutive contact resistance. When the
  net electrical power supplied through the terminals is zero, the free energy satisfies $\dot
  F=-\mathcal D_\mathrm{int}\leq0$. Under applied electrical power, the full balance accounts for
  both internal dissipation and energy exchanged with the external circuit.}

  \item {Vacancies are massless site variables and the electron mass is neglected, so the physically
  relevant conservation statements are number-inventory laws. Define
  \begin{align*}
    N_\ell(t)&:=\int_{\Omega_\mathrm M}\rho_\ell\,\mathrm d\boldsymbol r,
    &&\ell\in\{+,-,p,\mathrm{V}_\mathrm{M}\},\\
    N_{\ell_\mathrm d}(t)&:=\int_{\Omega_\mathrm d}\rho_{\ell_\mathrm d}\,\mathrm d\boldsymbol r,
    &&\ell\in\{e,i,\mathrm V\},\quad \mathrm d\in\{\mathrm L,\mathrm R\}.
  \end{align*}
  Since $s_{i_\mathrm{d},\mathrm d}+s_{+,\mathrm d}=0$ and neither intercalated lithium nor lithium
  ions cross the exterior boundary,
  \begin{equation}
    N_\mathrm{Li}:=N_{i_\mathrm L}+N_{i_\mathrm R}+N_+,
    \qquad \frac{\mathrm dN_\mathrm{Li}}{\mathrm dt}=0.
  \end{equation}
  Because $s_{-,\mathrm d}=s_{p,\mathrm d}=0$ and their exterior fluxes vanish, $\dot N_-=\dot
  N_p=0$. The total number of vacancies across the electrolyte and both electrodes satisfies
  \begin{equation}
    \frac{\mathrm d}{\mathrm dt}(N_{\mathrm{V}_\mathrm L}+N_{\mathrm{V}_\mathrm R}+N_{\mathrm{V}_\mathrm{M}})=0
  \end{equation}
  because $s_{\mathrm{V}_\mathrm{d},\mathrm d}+s_{\mathrm{V}_\mathrm{M},\mathrm d}=0$. In addition,
  $N_{i_\mathrm{d}}+N_{\mathrm{V}_\mathrm{d}}$ is fixed for each electrode by the lattice-site
  constraint. 
  }

  \item Interfacial reactions conserve charge. From~\cite{dreyer2016new}, at each interface, the sum
  of the products of the valencies and stoichiometric coefficients should be zero. Using the
  assumption that the electrode vacancies and intercalated species are neutrally charged, this
  implies that on $\Sigma_\mathrm{d}$, for $\mathrm{d} \in \{\mathrm{L}, \mathrm{R}\}$,
  \begin{align}
    0 &=  - s_{e_\mathrm{d},\mathrm{d}} +
    \sum_{\ell \in \{+, -, p\}}z_\ell s_{\ell,\mathrm{d}}.\label{charge_conservation}
  \end{align}
  Charge conservation of the complete cell follows from this interfacial condition together with the
  balanced integrated terminal currents~\eqref{fluxECBC}. As the electrode vacancies and
  intercalated species are neutral, the charge density at time $t$ is
  \begin{equation}
    \rho_c\left(\boldsymbol{r},t\right) = \begin{cases}
      e\left(\mathrm{q}_\mathrm{d}-\rho_{e_\mathrm{d}}\right), & \boldsymbol{r} \in \Omega_\mathrm{d},\\
      \sum_{\ell \in \left\{+,-, p\right\}} ez_\ell \rho_\ell, & \boldsymbol{r} \in \Omega_\mathrm{M},
    \end{cases}
  \end{equation}

  so the total charge in the system at time $t$ is
  \begin{align}
    Q\left(t\right)&= \sum_{\mathrm{d}\in\left\{\mathrm{L},\mathrm{R}\right\}}\int_{\Omega_\mathrm{d}}\rho_c\,\mathrm{d}\boldsymbol{r} +   \int_{\Omega_\mathrm{M}}\rho_c\,\mathrm{d}\boldsymbol{r}\nonumber\\
    &= \sum_{\mathrm{d} \in \{\mathrm{L}, \mathrm{R}\}} \int_{\Omega_{\mathrm{d}}} e\left(\mathrm{q}_\mathrm{d}-\rho_{e_\mathrm{d}}\right) \, \mathrm{d}\boldsymbol{r} + \int_{\Omega_\mathrm{M}} \sum_{\ell \in \{+, -, p\}} ez_\ell \rho_\ell \,\mathrm{d} \boldsymbol{r}.\label{totalSystemCharge}
  \end{align}
  Computing the time derivative,
  \begin{align}
    \dot Q &= \sum_{\mathrm{d}\in\left\{\mathrm{L},\mathrm{R}\right\}} \int_{\Omega_\mathrm{d}}-e\dot{\rho}_{e_\mathrm{d}} \mathrm{d}\boldsymbol{r} + \int_{\Omega_\mathrm{M}} \left\{\sum_{\ell \in \{+, -, p\}} ez_\ell \dot{\rho}_\ell \right\}\,\mathrm{d}\boldsymbol{r}.
  \end{align}
  Using Equations~\eqref{ellHatGovEqn} and~\eqref{ellGovEqn}, the divergence theorem, boundary
  conditions~\eqref{ellHat_BC},~\eqref{ell_BC}, and~\eqref{fluxECBC}, and the periodic boundary
  conditions of the species,
  \begin{align}
    \dot Q
    &= \sum_{\mathrm{d}\in\left\{\mathrm{L},\mathrm{R}\right\}} \int_{\Sigma_\mathrm{d}} \left[ -s_{e_\mathrm{d},\mathrm{d}} +\sum_{\ell \in \left\{+, -, p\right\}} z_\ell s_{\ell,\mathrm{d}}\right]
    \, \left(eN_A \nu_\mathrm{d}\right) \mathrm{d}\boldsymbol{S} .\label{dotRhoC}
  \end{align}
  The external-terminal contributions cancel by~\eqref{fluxECBC}, as used in
  obtaining~\eqref{dotRhoC}, and the remaining interfacial contributions vanish
  by~\eqref{charge_conservation}. We therefore conclude $\dot Q = 0$.

  \item We prove that $Q\left(0\right) = 0$; global electroneutrality for all time then follows
  immediately from charge conservation in item 3.
  We observe,
  \begin{align}
    Q\left(0\right) &= \sum_{\mathrm{d} \in \{\mathrm{L}, \mathrm{R}\}} \int_{\Omega_{\mathrm{d}}} e\left(\mathrm{q}_\mathrm{d}-\rho_{e_\mathrm{d}}\left(\boldsymbol{r},0\right)\right) \, \mathrm{d}\boldsymbol{r} + \int_{\Omega_\mathrm{M}} \sum_{\ell \in \{+, -, p\}} ez_\ell \rho_\ell\left(\boldsymbol{r},0\right) \,\mathrm{d} \boldsymbol{r}.\label{rhoC0}
  \end{align}
  The second term in~\eqref{rhoC0} is the initial total charge in the electrolyte. By assumption,
  this is zero. From~\eqref{varrho}, the first term in~\eqref{rhoC0} is zero as well. We conclude
  $Q\left(0\right) = 0$.
\end{enumerate}

\subsection{Proof of Theorem~\ref{thmdivFreeI}}
\label{sec:proofThmDivFreeI}
Subtracting the conservation equation with flux~\eqref{jrhoC} from Equation~\eqref{dotRhoCAlt}, two
expressions for the same quantity $\dot{\rho}_c$, we find
\begin{equation}
  0
  = \begin{cases}
    \nabla \cdot \left(\frac{\epsilon_\mathrm{d}}{4 \pi l_\mathrm{B}}\frac{e^2}{k_{\mathrm{B}}T} \frac{\partial\bm E_\mathrm{d}}{\partial t}+ eM_{e_\mathrm{d}}\nabla \left( \frac{\tilde{\mu}_{e_\mathrm{d}}^*}{k_{\mathrm{B}}T}\right) \right), & \bm r \in\Omega_\mathrm{d},\\
    \nabla \cdot \left(\frac{\epsilon_\mathrm{M}}{4 \pi l_\mathrm{B}}\frac{e^2}{k_{\mathrm{B}}T} \frac{\partial \bm E_\mathrm{M}}{\partial t}- e\sum_{\ell \in \left\{+, -, p\right\}}z_\ell M_{\ell}\nabla \left( \frac{\tilde{\mu}_{\ell}^*}{k_{\mathrm{B}}T}\right) \right), & \bm r \in\Omega_\mathrm{M}.
  \end{cases}
\end{equation}
{The bulk part of Theorem~\ref{thmdivFreeI} follows immediately from
definition~\eqref{totalElectricCurrentDensity}. At $\Sigma_\mathrm d$, differentiating
normal-displacement continuity gives cancellation of the two displacement-current normal components.
The reaction flux conditions give
\begin{equation}
  \boldsymbol j_{\rho_c,\mathrm d}\!\cdot\!\boldsymbol n_\mathrm d
  =eN_A s_{e_\mathrm{d},\mathrm d}\nu_\mathrm d,
  \qquad
  \boldsymbol j_{\rho_c,\mathrm M}\!\cdot\!\boldsymbol n_\mathrm M
  =-eN_A\!\sum_{\ell\in\{+,-,p\}}z_\ell s_{\ell,\mathrm d}\nu_\mathrm d.
\end{equation}
Reaction charge conservation, $s_{e_\mathrm{d},\mathrm d}=\sum_\ell z_\ell s_{\ell,\mathrm d}$,
therefore yields
\begin{equation}
  \boldsymbol I_{\mathrm{tot},\mathrm d}\!\cdot\!\boldsymbol n_\mathrm d
  +\boldsymbol I_{\mathrm{tot},\mathrm M}\!\cdot\!\boldsymbol n_\mathrm M=0.
\end{equation}
Thus the total current is divergence-free across the composite cell in the distributional sense.
Spatial uniformity follows only in one dimension.}

\subsection{Mathematical Rearrangements}
\subsubsection{Equations~\eqref{electronGovEqn}-~\eqref{p_gov_eqn_non_dilute}}
\label{FickianDiffSI}
For $\mathrm{d} \in \left\{\mathrm{L},\mathrm{R}\right\}$, in $\Omega_\mathrm{d}$
\begin{align}
  \frac{ \partial \rho_{e_\mathrm{d}}}{\partial t}
  &= \nabla \!\cdot\!  \left[ D_{e_\mathrm{d}} \rho_{e_\mathrm{d}}\nabla\left(\ln \phi_{e_\mathrm d} + 1 - \frac{e\psi_\mathrm{d}}{k_{\mathrm{B}}T}\right)\right]\nonumber\\
  &= \nabla \!\cdot\!\left[D_{e_\mathrm{d}}\left( \nabla \rho_{e_\mathrm{d}} - \rho_{e_\mathrm{d}} \nabla \left( \frac{e\psi_\mathrm{d}}{k_{\mathrm{B}}T}\right)\right) \right]\\
  \frac{\partial \rho_{i_\mathrm{d}}}{\partial t}
  &=
  \nabla \!\cdot\! \left[ D_{i_\mathrm{d}} \rho_{i_\mathrm{d}}\nabla\left(\ln\left(\frac{\rho_{i_\mathrm{d}}}{\rho_{i_\mathrm{d}}^o}\right) + 1 -\frac{\rho_{\mathrm{V}_\mathrm{d}}^o}{\rho_{i_\mathrm{d}}^o}\left[ \ln \left(\frac{\rho_{\mathrm{V}_\mathrm{d}}}{\rho_{\mathrm{V}_\mathrm{d}}^o}\right) + 1 \right]\right)\right]\nonumber\\
  &=
  \nabla \!\cdot\! \left[ D_{i_\mathrm{d}} \rho_{i_\mathrm{d}}\nabla\left(\ln\left(\frac{\rho_{i_\mathrm{d}} }{\rho_{\mathrm{V}_\mathrm{d}}}\right) \right)\right]\nonumber\\
  &= \nabla \!\cdot\! \left[ \frac{D_{i_\mathrm{d}}}{1-\frac{\rho_{i_\mathrm{d}}}{\rho_{i_\mathrm{d}}^o}} \nabla \rho_{i_\mathrm{d}}\right]\label{rhoI}.
\end{align}
In $\Omega_\mathrm{M}$, for $\ell \in \{+, -\}$,
\begin{align}
  \frac{ \partial \rho_\ell}{\partial t}
  &= \nabla \!\cdot\! \left[ D_\ell \rho_\ell\nabla\left(\ln\left(\frac{\rho_\ell}{\rho_\ell^o}\right)  + 1 + z_\ell \frac{e\psi_\mathrm{M}}{k_{\mathrm{B}}T}-\frac{\rho_{\mathrm{V}_\mathrm{M}}^o}{\rho_{\ell}^o}\left[ \ln \left(\frac{\rho_{\mathrm{V}_\mathrm{M}}}{\rho_{\mathrm{V}_\mathrm{M}}^o}\right) + 1 \right]\right)\right] \nonumber \\
  &= \nabla \!\cdot\! \left[ D_\ell \left( \nabla \rho_\ell + z_\ell \rho_\ell\nabla \left( \frac{e\psi_\mathrm{M}}{k_{\mathrm{B}}T}\right)-\frac{\rho_{\mathrm{V}_\mathrm{M}}^o}{\rho_\ell^o}\frac{\rho_\ell}{ \rho_{\mathrm{V}_\mathrm{M}}}\nabla \rho_{\mathrm{V}_\mathrm{M}}\right)\right] \nonumber\\
  &= \nabla \!\cdot\! \left[ D_\ell \left( \nabla \rho_\ell + z_\ell \rho_\ell\nabla \left( \frac{e\psi_\mathrm{M}}{k_{\mathrm{B}}T}\right)-\frac{\rho_+^o}{\rho_\ell^o}\frac{\rho_\ell}{\rho_{\mathrm{V}_\mathrm{M}}}\nabla \rho_{\mathrm{V}_\mathrm{M}}\right)\right] \\
  \frac{\partial \rho_p}{\partial t} &=  \nabla \!\cdot\! \left[ D_p\left( \frac{1}{N}\nabla \rho_p + z_p \rho_p\nabla \left( \frac{e\psi_\mathrm{M}}{k_{\mathrm{B}}T}\right)-\frac{\rho_+^o}{\rho_p^o}\frac{\rho_p}{\rho_{\mathrm{V}_\mathrm{M}}}\nabla  \rho_{\mathrm{V}_\mathrm{M}}\right)\right].
\end{align}
\begin{remark}
  The second equality in Equation~\eqref{rhoI} uses
  $\rho_{i_\mathrm{d}}^o=\rho_{\mathrm{V}_\mathrm{d}}^o$, and the third uses
  Equation~\eqref{localincompresselectrodes}.
\end{remark}

\subsection{Derivation of the Salt Chemical Potential}
In~\ref{sec:transportCoeffs}, we use~\cite{kumar2024dynamic} to find Equations~\ref{conductivity}-~\ref{ratio_salt_diff_thermodynamic_factor}. 
In~\cite{kumar2024dynamic}, they consider 
\begin{align}
    \mu_\mathrm{salt}^* = \frac{\frac{\delta F}{\delta \phi_+}-\frac{\delta F}{\delta \phi_\mathrm{M}}}{1+\frac{\rho_+^o}{\rho_-^o}}+\frac{\frac{\delta F}{\delta \phi_-}-\frac{\delta F}{\delta \phi_\mathrm{M}}}{1+\frac{\rho_-^o}{\rho_+^o}}
\end{align}
where $\phi_\mathrm{M} = 1-\phi_+-\phi_- = \phi_p$, so
\begin{align}
   \frac{\delta F}{\delta \phi_+} - \frac{\delta F}{\delta \phi_p} 
    &=k_BT\left(\rho_+^o\left[ \ln \phi_+ + 1 + \frac{e\psi}{k_BT} \right] - \frac{\rho_p^o}{N}\left[\ln \phi_p + 1 \right]\right) \\
    \frac{\delta F}{\delta \phi_-} - \frac{\delta F}{\delta \phi_p} 
    &=  k_BT\left(\rho_-^o\left[ \ln \phi_- + 1 - \frac{e\psi}{k_BT} \right] - \frac{\rho_p^o}{N}\left[\ln \phi_p + 1 \right]\right)\\
    \frac{\mu_\mathrm{salt}^*}{k_BT} &= \frac{\rho_+^o\rho_-^o}{\rho_+^o+\rho_-^o}\left[\ln \phi_+ + \ln \phi_- + 2\right] - 
    \frac{\rho_p}{N}\left[ \ln \phi_p + 1\right].
\end{align}
As the system described in~\cite{kumar2024dynamic} has no vacancies, $\phi_p=1-\phi_+-\phi_-$. Substituting,
\begin{align}
      \frac{\mu_\mathrm{salt}^*}{k_BT} &= \frac{\rho_+^o\rho_-^o}{\rho_+^o+\rho_-^o}\left[\ln \phi_+ + \ln \phi_- + 2\right] -
    \frac{\rho_p}{N}\left[ \ln\left(1-\phi_+-\phi_-\right)+ 1\right].
\end{align}
From~\cite{kumar2024dynamic}, $\phi_\mathrm{salt} = \phi_++\phi_-$. Using $\rho_c = e\left(\rho_+ - \rho_-\right)$,
\begin{equation}
    \phi_+ = \frac{\phi_\mathrm{salt}+\frac{\rho_c}{e\rho_-^o}}{1+\frac{\rho_+^o}{\rho_-^o}}, \qquad \phi_-=\frac{\phi_\mathrm{salt}-\frac{\rho_c}{e\rho_+^o}}{1+\frac{\rho_-^o}{\rho_+^o}}. \label{chargedSpeciesPhiSaltRelations}
\end{equation}
Substituting,
\begin{align}
   \frac{\mu_\mathrm{salt}^*}{k_BT}  &= \frac{\rho_+^o\rho_-^o}{\rho_+^o+\rho_-^o}\left[\ln \left( \frac{\phi_\mathrm{salt}+\frac{\rho_c}{e\rho_-^o}}{1+\frac{\rho_+^o}{\rho_-^o}}\right)+\ln \left( \frac{\phi_\mathrm{salt}-\frac{\rho_c}{e\rho_+^o}}{1+\frac{\rho_-^o}{\rho_+^o}}\right)+2 \right] \nonumber\\
    &\quad- \frac{\rho_p^o}{N}\left[ \ln\left(1-\frac{\phi_\mathrm{salt}+\frac{\rho_c}{e\rho_-^o}}{1+\frac{\rho_+^o}{\rho_-^o}}-\frac{\phi_\mathrm{salt}-\frac{\rho_c}{e\rho_+^o}}{1+\frac{\rho_-^o}{\rho_+^o}}\right) + 1\right].
\end{align}
We find
\begin{align}
    \frac{\partial \mu_\mathrm{salt}^*}{\partial \phi_\mathrm{salt}} \frac{1}{k_BT} &= \frac{\rho_+^o\rho_-^o}{\rho_+^o+\rho_-^o}\left[\frac{1}{\phi_\mathrm{salt}+\frac{\rho_c}{e\rho_-^o}}+\frac{1}{\phi_\mathrm{salt}-\frac{\rho_c}{e\rho_+^o}} \right] + \frac{\rho_p^o}{N}\frac{1}{1-\frac{\phi_\mathrm{salt}+\frac{\rho_c}{e\rho_-^o}}{1+\frac
    {\rho_+^o}{\rho_-^o}}-\frac{\phi_\mathrm{salt}-\frac{\rho_c}{e\rho_+^o}}{1+\frac{\rho_-^o}{\rho_+^o}}}.\label{saltchempotKumarSystem}
\end{align}

In our system, we analogously consider
\begin{align}
    \mu_\mathrm{salt}^* &= \frac{\frac{\delta F}{\delta \phi_+}-\frac{\delta F}{\delta \phi_{\mathrm{V}_\mathrm{M}}}}{1+\frac{\rho_+^o}{\rho_-^o}}+\frac{\frac{\delta F}{\delta \phi_-}-\frac{\delta F}{\delta \phi_{\mathrm{V}_\mathrm{M}}}}{1+\frac{\rho_-^o}{\rho_+^o}}\nonumber\\
    &= 
    k_BT\left(\frac{\rho_+^o\rho_-^o}{\rho_+^o+\rho_-^o}\left[\ln \phi_+ +\ln \phi_- +2 \right]- \rho_+^o\left[ \ln\left(\phi_{v_\mathrm{M}}\right) + 1\right]\right).
\end{align}
When $\phi_p=0$, $\phi_{v_\mathrm{M}} = 1-\phi_+-\phi_-$ and we find
\begin{align}
     \frac{\mu_\mathrm{salt}^*}{k_BT} &= 
    \frac{\rho_+^o\rho_-^o}{\rho_+^o+\rho_-^o}\left[\ln \phi_+ +\ln \phi_- +2 \right]- \rho_+^o\left[ \ln\left(1-\phi_+-\phi_-\right) + 1\right].
\end{align}
Using~\eqref{chargedSpeciesPhiSaltRelations},
\begin{align}
     \frac{\mu_\mathrm{salt}^*}{k_BT}  &= \frac{\rho_+^o\rho_-^o}{\rho_+^o+\rho_-^o}\left[\ln \left( \frac{\phi_\mathrm{salt}+\frac{\rho_c}{e\rho_-^o}}{1+\frac{\rho_+^o}{\rho_-^o}}\right)+\ln \left( \frac{\phi_\mathrm{salt}-\frac{\rho_c}{e\rho_+^o}}{1+\frac{\rho_-^o}{\rho_+^o}}\right)+2 \right] \nonumber\\
    &\quad-\rho_+^o\left[ \ln\left(1-\frac{\phi_\mathrm{salt}+\frac{\rho_c}{e\rho_-^o}}{1+\frac{\rho_+^o}{\rho_-^o}}-\frac{\phi_\mathrm{salt}-\frac{\rho_c}{e\rho_+^o}}{1+\frac{\rho_-^o}{\rho_+^o}}\right) + 1\right].
\end{align}
We find
\begin{align}
    \frac{\partial \mu_\mathrm{salt}^*}{\partial \phi_\mathrm{salt}} \frac{1}{k_BT} &= \frac{\rho_+^o\rho_-^o}{\rho_+^o+\rho_-^o}\left[\frac{1}{\phi_\mathrm{salt}+\frac{\rho_c}{e\rho_-^o}}+\frac{1}{\phi_\mathrm{salt}-\frac{\rho_c}{e\rho_+^o}} \right] + \rho_+^o\frac{1}{1-\frac{\phi_\mathrm{salt}+\frac{\rho_c}{e\rho_-^o}}{1+\frac
    {\rho_+^o}{\rho_-^o}}-\frac{\phi_\mathrm{salt}-\frac{\rho_c}{e\rho_+^o}}{1+\frac{\rho_-^o}{\rho_+^o}}}.\label{saltchempotOurSystem}
\end{align}
If we let $\rho_\mathrm{M}^o = \frac{\rho_p^o}{N}$ for the system in~\cite{kumar2024dynamic}, and $\rho_\mathrm{M}^o=\rho_+^o$ for our system,~\eqref{saltchempotKumarSystem} and~\eqref{saltchempotOurSystem} match; the common value is
\begin{align}
      \frac{\partial \mu_\mathrm{salt}^*}{\partial \phi_\mathrm{salt}} \frac{1}{k_BT} &= \frac{\rho_+^o\rho_-^o}{\rho_+^o+\rho_-^o}\left[\frac{1}{\phi_\mathrm{salt}+\frac{\rho_c}{e\rho_-^o}}+\frac{1}{\phi_\mathrm{salt}-\frac{\rho_c}{e\rho_+^o}} \right] + \rho_\mathrm{M}^o\frac{1}{1-\frac{\phi_\mathrm{salt}+\frac{\rho_c}{e\rho_-^o}}{1+\frac
    {\rho_+^o}{\rho_-^o}}-\frac{\phi_\mathrm{salt}-\frac{\rho_c}{e\rho_+^o}}{1+\frac{\rho_-^o}{\rho_+^o}}} \nonumber\\
    &= \frac{\rho_+^o\rho_-^o}{\rho_+^o+\rho_-^o}\left[ \frac{1}{\left(1+\frac{\rho_+^o}{\rho_-^o}\right)\phi_+}+\frac{1}{\left(1+\frac{\rho_-^o}{\rho_+^o}\right)\phi_-}\right] + \frac{\rho_\mathrm{M}^o}{1-\phi_+-\phi_-}\nonumber \\
    &= \frac{\rho_+^o\rho_-^o}{\rho_+^o+\rho_-^o}\left[ \frac{\rho_-^o}{\left(\rho_-^o+\rho_+^o\right)\phi_+}+\frac{\rho_+^o}{\left(\rho_+^o+\rho_-^o\right)\phi_-}\right] + \frac{\rho_\mathrm{M}^o}{\phi_\mathrm{M}}\nonumber \\
     &= \frac{\rho_+^o\rho_-^o}{\rho_+^o+\rho_-^o}\left[ \frac{\rho_-^o}{\left(\rho_-^o+\rho_+^o\right)\frac{\rho_+}{\rho_+^o}}+\frac{\rho_+^o}{\left(\rho_+^o+\rho_-^o\right)\frac{\rho_-}{\rho_-^o}}\right] + \frac{\rho_\mathrm{M}^o}{\phi_\mathrm{M}}\nonumber \\
    &= \left( \frac{\rho_+^o\rho_-^o}{\rho_+^o+\rho_-^o}\right)^2 \left[\frac{1}{\rho_+}+\frac{1}{\rho_-} \right] + \frac{\rho_\mathrm{M}^o}{\phi_\mathrm{M}}
\end{align}

Then this single corresponding expression for the salt chemical potential is substituted into the equations for physical transport coefficients in~\cite{kumar2024dynamic} to find Equations~\eqref{conductivity}-~\eqref{ratio_salt_diff_thermodynamic_factor}.

\subsection{Comparison with Doyle--Fuller--Newman model}
\label{DFNComp}

We compare our DIC limit with Reference~\cite{doyle1993modeling}, using a separate local notation
for its equations. Here $F_A=eN_A$ is Faraday's constant, $R_\mathrm g=N_Ak_\mathrm B$ is the gas
constant, $c$ is salt molar concentration, $D$ is salt diffusivity, $\kappa$ is electrolyte
conductivity, and $t_+^0$ is the polymer-rest-frame cation transference number. The applied current
density is $I$, electrolyte current density is $\boldsymbol i_2$, and cation molar flux is
$\boldsymbol N_+$. The salt stoichiometry is $\nu_+$; $s_+,s_1$ are reaction stoichiometric
coefficients in the source's convention, and $n$ is the electron-transfer number.

The composite-electrode electrolyte fraction is $\varepsilon_\mathrm F$, solid conductivity is
$\sigma_\mathrm s$, interfacial area per bulk volume is $a$, and interfacial molar flux is $j_n$. We
use $a$ consistently in both source terms. Thicknesses $\delta_s,\delta_c$ describe separator and
cathode. Particle radius is $R_s$, radial coordinate is $r$, solid lithium concentration is $c_s$,
and solid diffusivity is $D_s$; $\bar c_s$ is the normalized unit-step diffusion response and $t'$
is integration time.

In the kinetic law, $\boldsymbol i$ is interfacial current density, $k_2$ a rate coefficient,
$c_{\max}$ the electrolyte site capacity, and $c_T$ the solid lithium capacity. The $\alpha$'s are
charge-transfer exponents. The phase potentials are $\Phi_1,\Phi_2$, with $\eta_\mathrm
F:=\Phi_1-\Phi_2$; the kinetic driving voltage is $\eta_\mathrm F-U'$, not our interfacial
overpotential $\eta_\mathrm d$. The standard potentials are $U_2^\theta,U_{\mathrm{ref}}^\theta$,
and $\beta,b_U$ parameterize the concentration-dependent potential correction. Thus $F_A$,
$\sigma_\mathrm s$, and $\varepsilon_\mathrm F$ are distinct from our free energy, electrolyte
conductivity, and permittivities. The source's empirical offset is renamed $b_U$ to reserve $\zeta$
for polymer friction. All concentrations in this comparison are molar, rather than number densities. The
comparison equations are
\begin{table}
  \caption{Equations from~\cite{doyle1993modeling}.}
  \label{tab:Neumanntab1}
  \begin{tabular}{@{}lll@{}}
    \toprule
    {Variable} & {Equation} &{Boundary condition}\\
    \midrule
    & Separator & \\
    $c$ & $\frac{\partial c}{\partial t} = \nabla \cdot D\nabla c - \frac{\boldsymbol{i}_2\cdot \nabla t_+^0}{z_+ \nu_+ F_A}$ & $\boldsymbol{N}_+=I/F_A$ at $x=0$ \\
    $\eta_\mathrm F$ & $\nabla \eta_\mathrm F = \frac{I}{\kappa} + \frac{R_\mathrm g T}{F_A}\left(\frac{s_+}{n\nu_+} +\frac{t_+^0}{z_+\nu_+}\right)\nabla \ln c$ & --\\
    & Composite cathode & \\
    $c$ & $\varepsilon_\mathrm F \frac{\partial c}{\partial t} = \nabla \cdot \left( \varepsilon_\mathrm F D \nabla c\right) - \frac{\boldsymbol{i}_2 \cdot \nabla t_+^0}{z_+ \nu_+ F_A} + \frac{a j_n\left(1-t_+^0\right)}{\nu_+}$ & $\nabla c = 0$ at $x=\delta_s + \delta_c$\\
    $\eta_\mathrm F$ & $\nabla \eta_\mathrm F =\frac{-\left(I-\boldsymbol{i}_2\right)}{\sigma_\mathrm s} + \frac{\boldsymbol{i}_2}{\kappa} + \frac{R_\mathrm g T}{F_A}\left(\frac{s_+}{n\nu_+}+\frac{t_+^0}{z_+\nu_+} \right) \nabla \ln c$ & $\nabla \eta_\mathrm F = -I/\sigma_\mathrm s$ at $x = \delta_s+\delta_c$ \\
    $\boldsymbol{i}_2$ & $aj_n = \frac{-s_1}{nF_A} \nabla \cdot \boldsymbol{i}_2$& $\boldsymbol{i}_2 = I$ at $x = \delta_s$ \\
    $c_s$ & $j_n = -D_s \int_0^t \frac{\partial c_s}{\partial t} \left(R_s, t'\right) \frac{\partial \bar{c}_s}{\partial r} \left(R_s, t-t'\right)\,\mathrm d t' $& --\\
    $j_n$ & Equation~\eqref{Eqn17} & --\\
    \bottomrule
  \end{tabular}
\end{table}
\begin{align}
  \boldsymbol{i} = F_A k_2 \left(c_\mathrm{max}-c\right)^{\alpha_C}c^{\alpha_A}&\left[c_s\exp\left(\frac{\alpha_a F_A}{R_\mathrm g T}\left(\eta_\mathrm F-U'\right)\right)\right.\nonumber\\
  &\left.-\left(c_T-c_s\right)\exp\left(-\frac{\alpha_c F_A}{R_\mathrm g T}\left(\eta_\mathrm F-U'\right)\right)  \right] \label{Eqn17}\\
  U' &= U_2^\theta - U_{\mathrm{ref}}^\theta + \frac{R_\mathrm g T}{F_A}\left(\beta c_s+b_U\right) \\
  \eta_\mathrm F &= \Phi_1-\Phi_2.
\end{align}
{In our DIC model, the following quantity is the total mobile-ion number density:}
\begin{align}
\rho_\mathrm{salt}:=\rho_++\rho_=.\label{rhoSalt}
\end{align}
{Thus, under local electroneutrality for a 1:1 salt, the salt formula-unit density is
$\rho_\mathrm{salt}^{\mathrm{fu}}=\rho_\mathrm{ion}/2$ and
$c_\mathrm{salt}=\rho_\mathrm{ion}/(2N_A)$.}
The ionic charge density is
\begin{align}
  \rho_c = e\left(\rho_+ - \rho_-\right).\label{rhoC}
\end{align}
In the most general DIC case, for $\ell \in \left\{+, -\right\}$,
\begin{align}
  \frac{\partial \phi_\ell}{\partial t} &= \nabla \cdot \left[ D_\ell \left(\nabla \phi_\ell + z_\ell \phi_\ell \nabla  \left(\frac{e\psi_\mathrm{M}}{k_{\mathrm{B}}T}\right)- \frac{\rho_{\mathrm{V}_\mathrm{M}}^o}{\rho_\ell^o} \frac{\phi_\ell}{\phi_{\mathrm{V}_\mathrm{M}}} \nabla \phi_{\mathrm{V}_\mathrm{M}}\right)\right] \\
  \frac{\partial \rho_\ell}{\partial t}&= \nabla \cdot \left[ D_\ell  \left(\nabla \rho_\ell + z_\ell \rho_\ell \nabla  \left(\frac{e\psi_\mathrm{M}}{k_{\mathrm{B}}T}\right) - \frac{\rho_+^o}{\rho_\ell^o} \frac{\rho_\ell}{\phi_{\mathrm{V}_\mathrm{M}}} \nabla \phi_{\mathrm{V}_\mathrm{M}}\right)\right],\label{phiEll}
\end{align}
where $z_+ = 1, z_- = -1$, and where we used the fact that $\rho_{\mathrm{V}_\mathrm{M}}^o =
\rho_+^o$. As such, we have
\begin{align}
  \frac{\partial \rho_\mathrm{salt}}{\partial t} &= \nabla \cdot \left[D_+\nabla \rho_++D_-\nabla \rho_- + \left(D_+\rho_+-D_-\rho_- \right) \nabla  \left(\frac{e\psi_\mathrm{M}}{k_{\mathrm{B}}T}\right)\right.\nonumber\\
  &\left.\qquad-\left(D_+\frac{\rho_+^o}{\rho_+^o}\rho_++D_-\frac{\rho_+^o}{\rho_-^o}\rho_-\right)\frac{\nabla \phi_{\mathrm{V}_\mathrm{M}}}{\phi_{\mathrm{V}_\mathrm{M}}}\right] \nonumber\\
  &= \nabla \cdot \left[ \left( D_++\frac{1}{\rho_+^o\phi_{\mathrm{V}_\mathrm{M}}}\left(D_+\frac{\rho_+^o}{\rho_+^o}\rho_++D_-\frac
  {\rho_+^o}{\rho_-^o}\rho_-\right) \right)\nabla \rho_+\right.\nonumber\\
  &\qquad+\left. \left(D_-+\frac{1}{\rho_-^o\phi_{\mathrm{V}_\mathrm{M}}}\left(D_+\frac{\rho_+^o}{\rho_+^o}\rho_++D_-\frac
  {\rho_+^o}{\rho_-^o}\rho_-\right) \right)\nabla \rho_- \right.\nonumber\\
  &\qquad+\left. \left(\frac{1}{\rho_p^o\phi_{\mathrm{V}_\mathrm{M}}}\left(D_+\frac{\rho_+^o}{\rho_+^o}\rho_++D_-\frac
  {\rho_+^o}{\rho_-^o}\rho_-\right) \right)\nabla \rho_p \right.\nonumber \\
  &\left.\qquad+ \left(D_+\rho_+-D_-\rho_- \right) \nabla \left( \frac
  {e\psi_\mathrm{M}}{k_{\mathrm{B}}T}\right) \right]\label{rhoSaltGeneralCase} \\
  \frac{\partial \rho_c}{\partial t} &= \nabla \cdot \left[ e\left( D_+\nabla \rho_+ - D_-\nabla \rho_- + \left( D_+\rho_++D_-\rho_-\right)\nabla \left( \frac{e\psi_\mathrm{M} }{k_{\mathrm{B}}T}\right)\right.\right.\nonumber\\
  &\left.\left.\qquad-\left( D_+\frac{\rho_+^o}{\rho_+^o}\rho_+-D_-\frac{\rho_+^o}{\rho_-^o}\rho_-\right)\frac{\nabla \phi_{\mathrm{V}_\mathrm{M}}}{\phi_{\mathrm{V}_\mathrm{M}}}\right)\right] \nonumber\\
  &= \nabla \cdot \left[
  e\left( D_++\frac{1}{\rho_+^o\phi_{\mathrm{V}_\mathrm{M}}}\left( D_+\frac{\rho_+^o}{\rho_+^o}\rho_+-D_-\frac{\rho_+^o}{\rho_-^o}\rho_-\right)\right)\nabla \rho_+
  \right.\nonumber\\
  &\left.\qquad+e\left( -D_-+\frac{1}{\rho_-^o\phi_{\mathrm{V}_\mathrm{M}}}\left( D_+\frac{\rho_+^o}{\rho_+^o}\rho_+-D_-\frac{\rho_+^o}{\rho_-^o}\rho_-\right)\right)\nabla \rho_-\right. \nonumber\\
  &\left.\qquad+e\left(\frac{1}{\rho_p^o\phi_{\mathrm{V}_\mathrm{M}}}\left(D_+\frac{\rho_+^o}{\rho_+^o}\rho_+-D_-\frac{\rho_+^o}{\rho_-^o}\rho_-\right)\right)\nabla \rho_p \right. \nonumber \\
  &\left.\qquad+e\left(D_+\rho_++D_-\rho_- \right) \nabla\left( \frac{e\psi_\mathrm{M}}{k_{\mathrm{B}}T}\right) \right].
\end{align}

From the above governing equation for $\frac{\partial \rho_c}{\partial t}$, as $\frac{\partial
\rho_c}{\partial t} = -\nabla \cdot \boldsymbol{j}_{\rho_c}$, where $\boldsymbol{j}_{\rho_c}$ is the
ionic current, we have that, in the electrolyte,
\begin{align}
  -\boldsymbol{j}_{\rho_c} &=  e\left[ \left( D_++\frac{1}{\rho_+^o\phi_{\mathrm{V}_\mathrm{M}}}\left( D_+\frac{\rho_+^o}{\rho_+^o}\rho_+-D_-\frac{\rho_+^o}{\rho_-^o}\rho_-\right)\right)\nabla \rho_+
  \right.\nonumber\\
  &\qquad\left.+ \left(-D_-+\frac{1}{\rho_-^o\phi_{\mathrm{V}_\mathrm{M}}}\left( D_+\frac{\rho_+^o}{\rho_+^o}\rho_+-D_-\frac{\rho_+^o}{\rho_-^o}\rho_-\right)\right)\nabla \rho_-\right. \nonumber\\
  &\left.\qquad+\frac{1}{\rho_p^o\phi_{\mathrm{V}_\mathrm{M}}}\left(D_+\frac{\rho_+^o}{\rho_+^o}\rho_+-D_-\frac{\rho_+^o}{\rho_-^o}\rho_-\right)\nabla \rho_p \right. \nonumber \\
  &\left.\qquad+\left(D_+\rho_++D_-\rho_- \right) \nabla\left( \frac{e\psi_\mathrm{M}}{k_{\mathrm{B}}T}\right) \right].
\end{align}
This implies that in the electrolyte,
\begin{align}
  \nabla  \left(\frac{e\psi_\mathrm{M}}{k_{\mathrm{B}}T}\right)
  &= \frac{-\frac{\boldsymbol{j}_{\rho_c}}{e}}{\left(D_+\rho_++D_-\rho_-\right) } \nonumber\\
  &-\frac{\left[ D_++ \frac{1}{\rho_+^o\phi_{\mathrm{V}_\mathrm{M}}}\left( D_+\frac{\rho_+^o}{\rho_+^o}\rho_+-D_-\frac{\rho_+^o}{\rho_-^o}\rho_-\right)\right]\nabla \rho_+}{\left(D_+\rho_++D_-\rho_-\right) } \nonumber\\
  &-\frac{\left[-D_-+\frac{1}{\rho_-^o\phi_{\mathrm{V}_\mathrm{M}}}\left( D_+\frac{\rho_+^o}{\rho_+^o}\rho_+-D_-\frac{\rho_+^o}{\rho_-^o}\rho_-\right)\right]\nabla \rho_- }{\left(D_+\rho_++D_-\rho_-\right) }\nonumber\\
  &-\frac{\left[\frac{1}{\rho_p^o\phi_{\mathrm{V}_\mathrm{M}}}\left( D_+\frac{\rho_+^o}{\rho_+^o}\rho_+-D_-\frac{\rho_+^o}{\rho_-^o}\rho_-\right)\right]\nabla \rho_p }{\left(D_+\rho_++D_-\rho_-\right) }. \label{nablaPsi}
\end{align}
Substituting~\eqref{nablaPsi} into Equation~\eqref{rhoSaltGeneralCase},
\begin{align}
  \frac{\partial \rho_\mathrm{salt}}{\partial t}
  &= \nabla \cdot \left[ \left( D_++\frac{1}{\rho_+^o\phi_{\mathrm{V}_\mathrm{M}}}\left(D_+\frac{\rho_+^o}{\rho_+^o}\rho_++D_-\frac
  {\rho_+^o}{\rho_-^o}\rho_-\right) \right.\right. \nonumber\\
  &\quad\qquad\left.\left.-\frac{\left(D_+\rho_+-D_-\rho_- \right) }{\left(D_+\rho_++D_-\rho_-\right)}\left[ D_++\frac{1}{\rho_+^o\phi_{\mathrm{V}_\mathrm{M}}}\left( D_+\frac{\rho_+^o}{\rho_+^o}\rho_+-D_-\frac{\rho_+^o}{\rho_-^o}\rho_-\right)\right]\right)\nabla \rho_+\right.\nonumber\\
  &\qquad+\left. \left(D_-+\frac{1}{\rho_-^o\phi_{\mathrm{V}_\mathrm{M}}}\left(D_+\frac{\rho_+^o}{\rho_+^o}\rho_++D_-\frac
  {\rho_+^o}{\rho_-^o}\rho_-\right)\right.\right. \nonumber\\
  &\quad\qquad\left.\left.-\frac{\left(D_+\rho_+-D_-\rho_- \right) }{\left(D_+\rho_++D_-\rho_-\right)}\left[ -D_-+\frac{1}{\rho_-^o\phi_{\mathrm{V}_\mathrm{M}}}\left( D_+\frac{\rho_+^o}{\rho_+^o}\rho_+-D_-\frac{\rho_+^o}{\rho_-^o}\rho_-\right)\right] \right)\nabla \rho_- \right.\nonumber\\
  &\qquad+\left. \left(\frac{1}{\rho_p^o\phi_{\mathrm{V}_\mathrm{M}}}\left(D_+\frac{\rho_+^o}{\rho_+^o}\rho_++D_-\frac
  {\rho_+^o}{\rho_-^o}\rho_-\right) \right.\right. \nonumber\\
  &\quad\qquad\left.\left.-\frac{\left(D_+\rho_+-D_-\rho_- \right) }{\left(D_+\rho_++D_-\rho_-\right)}\left[ \frac{1}{\rho_p^o\phi_{\mathrm{V}_\mathrm{M}}}\left( D_+\frac{\rho_+^o}{\rho_+^o}\rho_+-D_-\frac{\rho_+^o}{\rho_-^o}\rho_-\right)\right]\right)\nabla \rho_p \right.\nonumber \\
  &\left.\qquad+ \frac{\left(D_+\rho_+-D_-\rho_- \right) }{\left(D_+\rho_++D_-\rho_-\right)} \left( -\frac{\boldsymbol{j}_{\rho_c}}{e}\right) \right] \nonumber
  \\
  &= \nabla \cdot \left[ \frac{2D_+D_-\rho_-}{\left(D_+\rho_++D_-\rho_-\right)}\left(\frac{\rho_+}{\rho_+^o}\frac{\rho_+^o}{\phi_{\mathrm{V}_\mathrm{M}}}\left(\frac{1}{\rho_-^o}+\frac{1}{\rho_+^o}\right)+1\right) \nabla \rho_+\right.\nonumber\\
  &\left.\qquad+\frac{2D_+D_-\rho_+}{\left(D_+\rho_++D_-\rho_-\right)}\left( \frac{\rho_-}{\rho_-^o}\frac{\rho_+^o}{\phi_{\mathrm{V}_\mathrm{M}}}\left(\frac{1}{\rho_-^o}+\frac{1}{\rho_+^o}\right) + 1\right) \nabla \rho_- \right.\nonumber\\
  &\left.\qquad+\frac{2D_+D_-\rho_+\rho_-}{\left(D_+\rho_++D_-\rho_-\right)}\frac{\rho_+^o}{\rho_p^o}\frac{1}{\phi_{\mathrm{V}_\mathrm{M}}}\left(\frac{1}{\rho_-^o}+\frac{1}{\rho_+^o}\right)\nabla \rho_p\right. \nonumber \\
  &\left.\qquad- \frac{D_+\rho_+-D_-\rho_-}{D_+\rho_++D_-\rho_-} \left( \frac{\boldsymbol{j}_{\rho_c}}{e}\right)\right].\label{saltAbsenceIonicCurrent}
\end{align}
Furthermore,
\begin{align}
  \frac{\partial \rho_c}{\partial t}
  &= -\nabla \cdot \left[{\boldsymbol{j}_{\rho_c}} \right] \\
  \frac{\partial \rho_{p}}{\partial t}
  &=\nabla \cdot \left[ D_p\left( \frac{1}{N}+\frac{\rho_+^o}{\rho_p^o}\frac{\phi_p}{\phi_{\mathrm{V}_\mathrm{M}}}\right)\nabla \rho_p+D_p\frac{\phi_p}{\phi_{\mathrm{V}_\mathrm{M}}}\nabla \rho_+ + D_p\frac{\rho_+^o}{\rho_-^o}\frac{\phi_p}{\phi_{\mathrm{V}_\mathrm{M}}}\nabla \rho_-\right]. \label{33p}
\end{align}
From Equations~\eqref{rhoSalt} and~\eqref{rhoC},
\begin{align}
  \rho_\mathrm{salt} +\frac{\rho_c}{e}&= 2 \rho_+ \\
  \rho_\mathrm{salt} - \frac{\rho_c}{e} &= 2 \rho_-.
\end{align}
Continuing from Equation~\eqref{saltAbsenceIonicCurrent},
\begin{align}
  \frac{\partial \rho_\mathrm{salt}}{\partial t} &= \nabla \cdot \left[ \frac{D_+D_-\rho_+\rho_-}{D_+\rho_++D_-\rho_-}\left(\frac{1}{\rho_+}+\frac{1}{\rho_-}+\frac{\rho_+^o}{\phi_{\mathrm{V}_\mathrm{M}}}\left(\frac{1}{\rho_+^o}+\frac{1}{\rho_-^o}\right)^2\right)\nabla \rho_\mathrm{salt}\right.\nonumber\\
  &\quad+\left.\frac{D_+D_-\rho_+\rho_-}{D_+\rho_++D_-\rho_-}\left(\frac{1}{\rho_+}-\frac{1}{\rho_-}+\frac{\rho_+^o}{\phi_{\mathrm{V}_\mathrm{M}}}\left(\frac{1}{\rho_+^o}+\frac{1}{\rho_-^o}\right)\left(\frac{1}{\rho_+^o}-\frac{1}{\rho_-^o}\right)\right)\nabla \left(\frac
  {\rho_c}{e}\right)\right.\nonumber\\
  &\quad+\frac{2D_+D_-\rho_+\rho_-}{D_+\rho_++D_-\rho_-}\frac{\rho_+^o}{\rho_p^o}\frac{1}{\phi_{\mathrm{V}_\mathrm{M}}}\left( \frac{1}{\rho_+^o}+\frac{1}{\rho_-^o}\right) \nabla\rho_p\nonumber\\
  &\quad\left.- \frac{D_+\rho_+-D_-\rho_-}{D_+\rho_++D_-\rho_-}\frac{\boldsymbol{j}_{\rho_c}}{e}\right] \nonumber\\
  &= \nabla \cdot \left[
  \frac{D_+D_-\rho_+\rho_-}{D_+\rho_++D_-\rho_-}\left(\frac{1}{\rho_+}+\frac{1}{\rho_-}+\frac{\rho_+^o}{\phi_{\mathrm{V}_\mathrm{M}}}\left(\frac{1}{\rho_+^o}+\frac{1}{\rho_-^o}\right)^2\right)\nabla \rho_\mathrm{salt}
  \right.\nonumber\\
  &\quad+t_{+,\mathrm{V}}\left(1-t_{+,\mathrm{V}}\right)\frac{k_BT}{e^2}\sigma\left(\frac{1}{\rho_+}-\frac{1}{\rho_-}+\frac{\left(\rho_+^o\right)^2}{\rho_{\mathrm{V}_\mathrm{M}}}\left(\frac{1}{\left(\rho_+^o\right)^2}-\frac{1}{\left(\rho_-^o\right)^2}\right)\right)\nabla \left( \frac{\rho_c}{e}\right)
  \nonumber\\
  &\quad+t_{+,\mathrm{V}}\left(1-t_{+,\mathrm{V}}\right) \frac{k_BT}{e^2}\sigma\left[2 \frac{\rho_+^o}{\rho_p^o}\frac{1}{\phi_{v_\mathrm{M}}}\left(\frac{1}{\rho_+^o}+\frac{1}{\rho_-^o}\right)\right]\nabla \rho_p\nonumber\\
  &\quad- \left(t_{+,\mathrm{V}}-\left(1-t_{+,\mathrm{V}}\right)\right) \frac{\boldsymbol{j}_{\rho_c}}{e} \nonumber\\
  &= \nabla \cdot \left[
 \frac{D_+D_-\rho_+\rho_-}{D_+\rho_++D_-\rho_-}\left(\frac{1}{\rho_+}+\frac{1}{\rho_-}+\frac{\rho_+^o}{\phi_{\mathrm{V}_\mathrm{M}}}\left(\frac{1}{\rho_+^o}+\frac{1}{\rho_-^o}\right)^2\right)\nabla \rho_\mathrm{salt}
  \right.\nonumber\\
  &\quad+ t_{+,\mathrm{V}}t_{-,\mathrm{V}}\frac{k_BT}{e^2}\sigma\left(\frac{1}{\rho_+}-\frac{1}{\rho_-}+\frac{1}{\rho_{\mathrm{V}_\mathrm{M}}}\left[1-\left(\frac{\rho_+^o}{\rho_-^o}\right)^2\right] \right)\nabla \left( \frac{\rho_c}{e}\right)\nonumber\\
  &\quad+t_{+,\mathrm{V}}\left(1-t_{+,\mathrm{V}}\right) \frac{k_BT}{e^2}\sigma \left[2 \frac{\rho_+^o}{\rho_p^o}\frac{1}{\rho_{v_\mathrm{M}}}\left(1+\frac{\rho_+^o}{\rho_-^o}\right)\right]\nabla \rho_p- \left(t_{+,\mathrm{V}}-t_{-,\mathrm{V}}\right) \frac{\boldsymbol{j}_{\rho_c}}{e} \nonumber\\
  &= \nabla \cdot \left[ D_\mathrm{salt}^{\mathrm{eff}}\nabla \rho_{\mathrm{salt}}+t_{+,\mathrm{V}}t_{-,\mathrm{V}}\frac{k_BT}{e^2}\sigma\xi_\mathrm{DIC}\nabla \left(\frac{\rho_c}{e}\right) + t_{+,\mathrm{V}}t_{-,\mathrm{V}}\frac{k_BT}{e^2}\sigma\chi^*_{\mathrm{DIC}}\nabla \rho_p-\left(t_{+,\mathrm{V}}-t_{-,\mathrm{V}} \right) \frac{\boldsymbol j_{\rho_c}}{e}\right],
\end{align}
where we used $t_{+,\mathrm{V}}$ from Equation~\eqref{transference_number}, $t_{-,\mathrm{V}} =
1-t_{+,\mathrm{V}}$, and where we defined
\begin{align}
  D_\mathrm{salt}^{\mathrm{eff}} := \frac{D_+D_-\rho_+\rho_-}{D_+\rho_++D_-\rho_-}
  \left[\frac{1}{\rho_+}+\frac{1}{\rho_-}
  +\frac{\rho_+^o}{\phi_{\mathrm{V}_\mathrm{M}}}
  \left(\frac{1}{\rho_+^o}+\frac{1}{\rho_-^o}\right)^2\right]\\
  \xi_\mathrm{DIC} :=  \frac{1}{\rho_+}-\frac{1}{\rho_-}+\frac{1}{\rho_{\mathrm{V}_\mathrm{M}}}\left[1-\left(\frac{\rho_+^o}{\rho_-^o}\right)^2\right] \\
  \chi^*_{\mathrm{DIC}} := \frac{2D_+D_-\rho_+\rho_-}{D_+\rho_++D_-\rho_-}\frac{\rho_+^o}{\rho_p^o}\frac{1}{\phi_{\mathrm{V}_\mathrm{M}}}\left(\frac{1}{\rho_+^o}+\frac{1}{\rho_-^o}\right).
\end{align}
In particular, to obtain the second inequality we used that
\begin{align}
  D_+\rho_+ + D_-\rho_- &= \frac{k_{\mathrm{B}}T}{e^2}\sigma \\
  D_+ \rho_+ &= t_{+,\mathrm{V}}\frac{k_{\mathrm{B}}T}{e^2} \sigma \\
  D_-\rho_- &= \left(1-t_{+,\mathrm{V}}\right)\frac{k_{\mathrm{B}}T}{e^2}\sigma = t_{-,\mathrm{V}}\frac{k_{\mathrm{B}}T}{e^2}\sigma,
\end{align}
so
\begin{align}
\frac{D_+D_-\rho_+\rho_-}{D_+\rho_++D_-\rho_-} &= t_{+,\mathrm{V}} t_{-,\mathrm{V}} \frac{k_BT}{e^2}\sigma.
\end{align}
Furthermore, from~\eqref{jrhoC} and~\eqref{totalElectricCurrentDensity}, $\boldsymbol{j}_{\rho_c} =
\boldsymbol{I}_\mathrm{tot}-\frac{\epsilon_\mathrm{M}}{4\pi l_\mathrm{B}}\frac{e^2}{k_{\mathrm{B}}T}
\frac{\partial \boldsymbol{E}_\mathrm{M}}{\partial t}$.
We conclude
\begin{align}
  \frac{\partial \rho_\mathrm{salt}}{\partial t} &= \nabla \cdot \left[ {D_\mathrm{salt}^{\mathrm{eff}}} \nabla \rho_\mathrm{salt} + \xi_\mathrm{DIC} t_{+,\mathrm{V}} t_{-,\mathrm{V}} \frac{k_{\mathrm{B}}T}{e^2}\sigma \nabla \left(\frac{\rho_c}{e}\right) + \chi_\mathrm{DIC}^{*} t_{+,\mathrm{V}} t_{-,\mathrm{V}} \frac{k_{\mathrm{B}}T}{e^2}\sigma \nabla \rho_p \right.\nonumber\\
  &\quad\left.- \left( t_{+,\mathrm{V}}-t_{-,\mathrm{V}}\right)\left( \frac{\boldsymbol{I}_\mathrm{tot}}{e}-\frac{\epsilon_\mathrm{M}}{4\pi l_\mathrm{B}}\frac{e}{k_{\mathrm{B}}T}\frac{\partial \boldsymbol{E}_\mathrm{M}}{\partial t}\right)\right].
\end{align}
{The coefficient $D_\mathrm{salt}^{\mathrm{eff}}$ is taken directly from the ionic-flux reduction.
It avoids assigning a single reference density to the aggregate fraction
$\phi_p+\phi_{\mathrm{V}_\mathrm{M}}$, whose constituents generally have different reference
volumes. The salt diffusivity formula imported from the earlier single-neutral-constituent theory is
used only for comparison or in the polymer-free specialization.}
{All quantities $\rho_\ell^{\mathrm{eq}}>0$ used below are spatially uniform bulk reference
densities. Hence $\nabla\rho_\ell^{\mathrm{eq}}=0$ and
$\nabla\ln(\rho_\ell/\rho_\ell^{\mathrm{eq}})=\nabla\ln\rho_\ell$. They provide dimensionless
logarithmic ratios and are not spatially varying equilibrium profiles obtained by solving Poisson's equation
together with the species equilibrium conditions.}
We work to complete the system by finding governing equations for the polymer density, charge
density, and electric field. From Equation~\eqref{33p} and~\eqref{poissonElectrolyte},
\begin{align}
  \frac{\partial \rho_p}{\partial t} &= \nabla \cdot \left[ \frac{D_p\phi_p}{2\phi_{\mathrm{V}_\mathrm{M}}}\left(1+\frac{\rho_+^o}{\rho_-^o}\right)
  \nabla \rho_\mathrm{salt}+ \frac{D_p\phi_p}{2\phi_{\mathrm{V}_\mathrm{M}}}\left(1-\frac{\rho_+^o}{\rho_-^o}\right)\nabla \left( \frac{\rho_c}{e}\right)\right.\nonumber\\
  &\quad\left. +D_p \left( \frac{1}{N}+\frac{\rho_+^o}{\rho_p^o}\frac{\phi_p}{\phi_{\mathrm{V}_\mathrm{M}}}\right) \nabla \rho_p\right]\label{rhoVMEqn}\\
  \nabla \cdot \left[ \frac{\epsilon_\mathrm{M}}{4\pi l_\mathrm{B}}\frac{e^2}{k_{\mathrm{B}}T}\boldsymbol{E}_\mathrm{M}\right] &= \rho_c.
\end{align}
From Equation~\eqref{nablaPsi},
\begin{align}
  -\frac{e}{k_{\mathrm{B}}T}\boldsymbol{E}_\mathrm{M} &= - \frac{\boldsymbol{j}_{\rho_c}}{e}\frac{e^2}{k_{\mathrm{B}}T} \frac{1}{\sigma} \nonumber\\
  &-\frac{\nabla \rho_\mathrm{salt}}{2}\left[ \frac{D_+-D_-+\frac{1}{\phi_{\mathrm{V}_\mathrm{M}}}\left(\frac{1}{\rho_+^o}+\frac{1}{\rho_-^o}\right)\left(D_+\rho_+-D_-\frac{\rho_+^o}{\rho_-^o}\rho_-\right)}{D_+\rho_++D_-\rho_-}\right] \nonumber\\
  &-\frac{\nabla (\rho_c/e)}{2}\left[ \frac{D_++D_-+\frac{1}{\phi_{\mathrm{V}_\mathrm{M}}}\left(\frac{1}{\rho_+^o}-\frac{1}{\rho_-^o}\right)\left(D_+\rho_+-D_-\frac{\rho_+^o}{\rho_-^o}\rho_-\right)}{D_+\rho_++D_-\rho_-}\right] \nonumber \\
  &-\frac{1}{\rho_p^o\phi_{\mathrm{V}_\mathrm{M}}}\frac{D_+\rho_+-D_-\frac{\rho_+^o}{\rho_-^o}\rho_-}{D_+\rho_++D_-\rho_-}\nabla \rho_p
  \nonumber \\
  &= -\frac{e}{k_{\mathrm{B}}T}\frac{1}{\sigma}\left[ \boldsymbol{I}_\mathrm{tot} - \frac{\epsilon_\mathrm{M}}{4\pi l_\mathrm{B}}\frac{e^2}{k_{\mathrm{B}}T}\frac{\partial \boldsymbol{E}_\mathrm{M}}{\partial t}\right] \nonumber\\
  &-\frac{\nabla \rho_\mathrm{salt}}{2}\left[ \frac{t_{+,\mathrm{V}}}{\rho_+}-\frac{t_{-,\mathrm{V}}}{\rho_-}+\frac{1}{\phi_{\mathrm{V}_\mathrm{M}}}\left( \frac{1}{\rho_+^o}+\frac{1}{\rho_-^o}\right)\left(t_{+,\mathrm{V}}-\frac{\rho_+^o}{\rho_-^o}t_{-,\mathrm{V}}\right)\right] \nonumber\\
  &-\frac{\nabla (\rho_c/e)}{2}\left[ \frac{t_{+,\mathrm{V}}}{\rho_+}+\frac{t_{-,\mathrm{V}}}{\rho_-}+\frac{1}{\phi_{\mathrm{V}_\mathrm{M}}}\left( \frac{1}{\rho_+^o}-\frac{1}{\rho_-^o}\right)\left(t_{+,\mathrm{V}}-\frac{\rho_+^o}{\rho_-^o}t_{-,\mathrm{V}}\right)\right] \nonumber\\
  &-\frac{\nabla \rho_p}{\rho_p^o \phi_{\mathrm{V}_\mathrm{M}}}\left( t_{+,\mathrm{V}}-\frac{\rho_+^o}{\rho_-^o}t_{-,\mathrm{V}}\right)\\
  \frac{\epsilon_\mathrm{M}}{4\pi l_\mathrm{B}}\frac{e^2}{k_{\mathrm{B}}T} \frac{\partial \boldsymbol{E}_\mathrm{M}}{\partial t} &= \boldsymbol{I}_\mathrm{tot} -\sigma \boldsymbol{E}_\mathrm{M} + \sigma \frac{k_{\mathrm{B}}T}{e} \times \nonumber \\
  &\left[ \frac{\nabla \rho_\mathrm{salt}}{2}\left(
  \frac{t_{+,\mathrm{V}}}{\rho_+}-\frac{t_{-,\mathrm{V}}}{\rho_-}+\frac{1}{\phi_{\mathrm{V}_\mathrm{M}}}\left( \frac{1}{\rho_+^o}+\frac{1}{\rho_-^o}\right) \left(t_{+,\mathrm{V}}-\frac{\rho_+^o}{\rho_-^o}t_{-,\mathrm{V}} \right)\right)\right. \nonumber\\
  &\left.
  +\frac{\nabla (\rho_c/e)}{2}\left(
  \frac{t_{+,\mathrm{V}}}{\rho_+}+\frac{t_{-,\mathrm{V}}}{\rho_-}+\frac{1}{\phi_{\mathrm{V}_\mathrm{M}}}\left( \frac{1}{\rho_+^o}-\frac{1}{\rho_-^o}\right) \left(t_{+,\mathrm{V}}-\frac{\rho_+^o}{\rho_-^o}t_{-,\mathrm{V}} \right)
  \right)
  \right. \nonumber\\
  &\left.+\frac{\nabla \rho_p}{\rho_p^o\phi_{\mathrm{V}_\mathrm{M}}} \left( t_{+,\mathrm{V}}-\frac{\rho_+^o}{\rho_-^o}t_{-,\mathrm{V}}\right) \right].
\end{align}
From the relationship between $\boldsymbol{E}_\mathrm{M}, \rho_c$,
\begin{align}
  \frac{\partial \rho_c}{\partial t} &= \nabla \cdot \left[
  \boldsymbol{I}_\mathrm{tot} -\sigma \boldsymbol{E}_\mathrm{M} + \sigma \frac{k_{\mathrm{B}}T}{e} \times \right. \nonumber\\
  &\qquad\left.\left[ \frac{\nabla \rho_\mathrm{salt}}{2}\left(
  \frac{t_{+,\mathrm{V}}}{\rho_+}-\frac{t_{-,\mathrm{V}}}{\rho_-}+\frac{1}{\phi_{\mathrm{V}_\mathrm{M}}}\left( \frac{1}{\rho_+^o}+\frac{1}{\rho_-^o}\right) \left(t_{+,\mathrm{V}}-\frac{\rho_+^o}{\rho_-^o}t_{-,\mathrm{V}} \right)\right)\right. \right.\nonumber\\
  &\qquad\left.\left.
  +\frac{\nabla (\rho_c/e)}{2}\left(
  \frac{t_{+,\mathrm{V}}}{\rho_+}+\frac{t_{-,\mathrm{V}}}{\rho_-}+\frac{1}{\phi_{\mathrm{V}_\mathrm{M}}}\left( \frac{1}{\rho_+^o}-\frac{1}{\rho_-^o}\right) \left(t_{+,\mathrm{V}}-\frac{\rho_+^o}{\rho_-^o}t_{-,\mathrm{V}} \right)
  \right)
  \right. \right.\nonumber\\
  &\qquad\left.\left.+\frac{\nabla \rho_p}{\rho_p^o\phi_{\mathrm{V}_\mathrm{M}}} \left( t_{+,\mathrm{V}}-\frac{\rho_+^o}{\rho_-^o}t_{-,\mathrm{V}}\right) \right]
  \right]. \label{rhoCDIC}
\end{align}

We derive bulk electric-field relations directly from the current laws. Every reference density
$\rho_\ell^{\mathrm{eq}}$ below is a positive spatially uniform normalization; no equilibrium
profile or equilibrium potential is introduced. For a domain label $a\in\{\mathrm L,\mathrm
M,\mathrm R\}$, write $\boldsymbol j_{\rho_c}=\boldsymbol
I_\mathrm{tot}-\varepsilon_a\partial_t\boldsymbol E_a$ in $\Omega_a$. In the DIC electrolyte,
\begin{align}
  \frac{\boldsymbol j_{\rho_c}}e&=-M_+\nabla(\tilde\mu_+^*/k_{\mathrm B}T)+M_-\nabla(\tilde\mu_-^*/k_{\mathrm B}T),\label{I}\\
  M_+&=D_+\rho_+,\label{mPlus}\\
  M_-&=D_-\rho_-.\label{mMinus}
\end{align}
Expanding the chemical-potential gradients and solving for $\boldsymbol E_\mathrm M$ gives
Equation~\eqref{etaDIC}, entirely within $\Omega_\mathrm M$. Equivalently,
\begin{align}
  \boldsymbol E_\mathrm M&=\frac{\boldsymbol j_{\rho_c}}{\sigma}+\frac{k_{\mathrm B}T}{e}\left[t_{+,\mathrm V}\nabla\ln\rho_+-t_{-,\mathrm V}\nabla\ln\rho_--\left(t_{+,\mathrm V}-\frac{\rho_+^o}{\rho_-^o}t_{-,\mathrm V}\right)\nabla\ln\rho_{\mathrm V_M}\right] \nonumber\\
  &= \frac{\boldsymbol j_{\rho_c}}{\sigma}+\frac{k_{\mathrm B}T}{e}\Bigg[\left( \frac{t_{+,\mathrm{V}}}{\rho_\mathrm{salt}+\frac{\rho_c}{e}}-\frac{t_{-,\mathrm{V}}}{\rho_\mathrm{salt}-\frac{\rho_c}{e}}+\frac{1}{2}\frac{\left(t_{+,\mathrm{V}}-\frac{\rho_+^o}{\rho_-^o}t_{-,\mathrm{V}}\right)\left(\frac{1}{\rho_+^o}+\frac{1}{\rho_-^o}\right)}{1-\frac{\rho_\mathrm{salt}+\frac{\rho_c}{e}}{2\rho_+^o}-\frac{\rho_\mathrm{salt}-\frac{\rho_c}{e}}{2\rho_-^o}-\frac{\rho_p}{\rho_p^o}}\right)\nabla\rho_\mathrm{salt} \nonumber\\
  &+ \left( \frac{t_{+,\mathrm{V}}}{\rho_\mathrm{salt}+\frac{\rho_c}{e}}+\frac{t_{-,\mathrm{V}}}{\rho_\mathrm{salt}-\frac{\rho_c}{e}}+\frac{1}{2}\frac{\left(t_{+,\mathrm{V}}-\frac{\rho_+^o}{\rho_-^o}t_{-,\mathrm{V}}\right)\left(\frac{1}{\rho_+^o}-\frac{1}{\rho_-^o}\right)}{1-\frac{\rho_\mathrm{salt}+\frac{\rho_c}{e}}{2\rho_+^o}-\frac{\rho_\mathrm{salt}-\frac{\rho_c}{e}}{2\rho_-^o}-\frac{\rho_p}{\rho_p^o}}\right)\nabla\left(\frac{\rho_c}{e}\right)\nonumber\\
  &+\frac{\left(t_{+,\mathrm{V}}-\frac{\rho_+^o}{\rho_-^o}t_{-,\mathrm{V}}\right)\frac{1}{\rho_p^o}}{1-\frac{\rho_\mathrm{salt}+\frac{\rho_c}{e}}{2\rho_+^o}-\frac{\rho_\mathrm{salt}-\frac{\rho_c}{e}}{2\rho_-^o}-\frac{\rho_p}{\rho_p^o}}\nabla \rho_p\Bigg]\label{etaGovEqnBeforeSimplification}
\end{align}
Furthermore, from~\eqref{eq:eleceq},
\begin{align}
     -\Delta \psi_\mathrm{d}^\mathrm{eq} = -\left(\widehat E_\mathrm d^\circ+\frac{k_{\mathrm{B}} T}{e}\ln \left[\left(
  \frac{\frac{\rho_\mathrm{salt}+\frac{\rho_c}{e}}{2}}{\rho_+^o-
\frac{\rho_\mathrm{salt}+\frac{\rho_c}{e}}{2}
  -\frac{\rho_+^o}{\rho_-^o}\frac{\rho_\mathrm{salt}-\frac{\rho_c}{e}}{2}-\frac{\rho_+^o}{\rho_p^o}\rho_p}\right)\frac{ q_\mathrm{d}-\frac{\rho_c}{e}}{\rho_+^o}\frac{\rho_{i_\mathrm{d}}^o-\rho_{i_\mathrm{d}}}{\rho_{i_\mathrm{d}}}
\right]\right).\label{eq:eleceqSI}
\end{align} As observed in Section~\ref{sec:recapturingKnownDynamics}, $\eta_\mathrm{d} = -\Delta \psi_\mathrm{d}^\mathrm{eq}$.

The logarithmic gradients here mean gradients of the ratios to their uniform reference densities. In
each electrode, $\rho_c = e\left(q_\mathrm{d}-\rho_e\right)$. Hence, 
\begin{align}
    -\boldsymbol{j}_{\rho_c} &= e \left[ -D_{e_\mathrm{d}}\nabla \rho_{e_\mathrm{d}} +D_{e_\mathrm{d}}\rho_{e_\mathrm{d}}\nabla \left( \frac{e\psi_\mathrm{d}}{k_BT}\right)\right].
\end{align}
This implies that in the electrode,
\begin{align}
    \nabla \left( \frac{e\psi_\mathrm{d}}{k_BT}\right) &= \frac{-\frac{\boldsymbol{j}_{\rho_c}}{e}}{D_{e_\mathrm{d}}\rho_{e_\mathrm{d}}}+ \frac{\nabla \rho_{e_\mathrm{d}}}{\rho_{e_\mathrm{d}}}.
\end{align}
Using $\boldsymbol{E}_\mathrm{d} = -\nabla \psi_\mathrm{d}$, $\rho_{e_\mathrm{d}} = q_\mathrm{d}-\frac{\rho_c}{e}$ and defining
\begin{equation}
  \sigma_\mathrm d:=\frac{e^2}{k_{\mathrm B}T}D_{e_\mathrm d}\rho_{e_\mathrm d},\label{electrodeConductivity}
\end{equation}
we have
\begin{align}
    -\frac{e}{k_BT}\boldsymbol{E}_\mathrm{d}&= -\frac{e}{k_BT} \frac{\boldsymbol{j}_{\rho_c}}{\sigma_\mathrm{d}} - \frac{\nabla \left(\frac{\rho_c}{e}\right)}{q_\mathrm{d}-\frac{\rho_c}{e}} \label{EdOrig}\\
    &= -\frac{e}{k_BT}\frac{1}{\sigma_\mathrm{d}}\left[\boldsymbol{I}_\mathrm{tot} - \frac{\epsilon_\mathrm{d}}{4\pi l_\mathrm{B}}\frac{e^2}{k_{\mathrm{B}}T}\frac{\partial \boldsymbol{E}_\mathrm{d}}{\partial t}\right]- \frac{\nabla \left(\frac{\rho_c}{e}\right)}{q_\mathrm{d}-\frac{\rho_c}{e}} \\
    \frac{\epsilon_\mathrm{d}}{4\pi l_\mathrm{B}}\frac{e^2}{k_BT} \frac{\partial \boldsymbol{E}_\mathrm{d}}{\partial t} &= \boldsymbol{I}_\mathrm{tot}-\sigma_\mathrm{d} \boldsymbol{E}_\mathrm{d}+\sigma_\mathrm{d} \frac{k_BT}{e} \frac{\nabla \left(\rho_c/e\right)}{q_\mathrm{d}-\rho_c/e}
\end{align}
From the relationship between $\boldsymbol{E}_\mathrm{d}, \rho_c$,
\begin{align}
    \frac{\partial \rho_c}{\partial t} &= \nabla \cdot \left[ \boldsymbol{I}_\mathrm{tot}-\sigma_\mathrm{d} \boldsymbol{E}_\mathrm{d}+\sigma_\mathrm{d} \frac{k_BT}{e} \frac{\nabla \left(\rho_c/e\right)}{q_\mathrm{d}-\rho_c/e}\right].\label{rhoCElectrode}
\end{align}
From~\eqref{EdOrig}
\begin{align}
    \boldsymbol{E}_\mathrm{d} = \frac{\boldsymbol{j}_{\rho_c}}{\sigma_\mathrm{d}} + \frac{k_BT}{e}\left[ \frac{\nabla \left( \frac{\rho_c}{e}\right)}{q_\mathrm{d}-\frac{\rho_c}{e}}\right].\label{Ed}
\end{align}
We use the interfacial reaction boundary conditions~\eqref{ellHat_BC} and~\eqref{ell_BC} along with~\eqref{sk} to enumerate the boundary conditions for $\rho_\mathrm{salt}, \rho_c, \rho_p, \rho_{i_\mathrm{d}}$ on $\Sigma_\mathrm{d}$. We use boundary condition~\eqref{transmission_jump} to enumerate the boundary conditions for $\boldsymbol{E}_\mathrm{M}$ on $\Sigma_\mathrm{d}$. We use ~\eqref{currentBCL} and~\eqref{currentBCR} to enumerate the boundary conditions for $\rho_c$ and~\eqref{noFluxI} no enumerate the boundary conditions for $\rho_{i_\mathrm{d}}$ on $\Gamma_\mathrm{d}$. Finally, we use~\eqref{psiNeumannBCRight} and global neutrality and the integrated Poisson equation (as mentioned in Section~\ref{sec:uniformI}) to enumerate the boundary conditions for $\boldsymbol{E}_\mathrm{d}$ on $\Gamma_\mathrm{d}$.

\begin{table}
  \caption{Governing equations and boundary conditions in the DIC model.}
  \label{tab:DICElectrolytetab}
  \resizebox{\textwidth}{!}{%
  \begin{tabular}{@{}lll@{}}
    \toprule
    {Variable} & {Equation} &{Boundary condition}\\
    \midrule
    & Electrolyte & \\
    $\rho_{\mathrm{salt}}$ & $\frac{\partial \rho_\mathrm{salt}}{\partial t} = \nabla \cdot \left[ {D_\mathrm{salt}^{\mathrm{eff}}}\nabla \rho_{\mathrm{salt}}+ \xi_\mathrm{DIC} t_{+,\mathrm{V}} t_{-,\mathrm{V}} \frac{k_{\mathrm{B}}T}{e^2}\sigma \nabla \left(\frac{\rho_c}{e}\right) \right.$ & $\left[ {D_\mathrm{salt}^{\mathrm{eff}}}\nabla \rho_{\mathrm{salt}}+\xi_\mathrm{DIC} t_{+,\mathrm{V}} t_{-,\mathrm{V}} \frac{k_{\mathrm{B}}T}{e^2}\sigma \nabla \left(\frac{\rho_c}{e}\right)\right.$\\
    & $\left.\quad+ \chi_\mathrm{DIC}^{*} t_{+,\mathrm{V}} t_{-,\mathrm{V}} \frac{k_{\mathrm{B}}T}{e^2}\sigma \nabla \rho_{p}- \left(t_{+,\mathrm{V}}- t_{-,\mathrm{V}}\right) \frac{\boldsymbol{j}_{\rho_c}}{e}\right] $ &
    $\left.\quad+ \chi_\mathrm{DIC}^{*} t_{+,\mathrm{V}} t_{-,\mathrm{V}} \frac{k_{\mathrm{B}}T}{e^2}\sigma \nabla \rho_{p} - \left(t_{+,\mathrm{V}}- t_{-,\mathrm{V}}\right)\frac{\boldsymbol{j}_{\rho_c}}{e}\right]
    \cdot \boldsymbol{n}_\mathrm{M}$\\
    & & $=-\frac{I_{ex,\mathrm{d}}}{e}\left[ \exp\left(-\frac{\alpha_\mathrm{d}e}{k_BT}\eta_\mathrm{d}\right)-\exp\left(\frac{\left(1-\alpha_\mathrm{d}\right)e}{k_BT}\eta_\mathrm{d}\right)\right]
    $ at $\Sigma_\mathrm{d}$\\
    $\rho_c$ & Equation~\eqref{rhoCDIC}& {$\boldsymbol{j}_{\rho_c}\cdot \boldsymbol{n}_\mathrm{M} = -{I_{ex,\mathrm{d}}}\left[ \exp\left(-\frac{\alpha_\mathrm{d}e}{k_BT}\eta_\mathrm{d}\right)-\exp\left(\frac{\left(1-\alpha_\mathrm{d}\right)e}{k_BT}\eta_\mathrm{d}\right)\right]$} at $\Sigma_\mathrm
    d$\\
    $\rho_{p}$ &  Equation~\eqref{rhoVMEqn}& $
    \boldsymbol{J}_{p}
    \cdot \boldsymbol{n}_\mathrm{M} =0
    $ at $\Sigma_\mathrm{d}$\\
    $\boldsymbol E_\mathrm M$ & Equation~\eqref{etaGovEqnBeforeSimplification} & $\epsilon_\mathrm{M} \boldsymbol{E}_\mathrm{M} \cdot \boldsymbol{n}_\mathrm{M} + \epsilon_\mathrm{d} \boldsymbol{E}_\mathrm{d} \cdot \boldsymbol{n}_\mathrm{d} = 0$ at $\Sigma_\mathrm{d}$\\
        $\eta_\mathrm{d}$ &Equation~\eqref{eq:eleceqSI}& --\\
    & Electrode & \\
$\rho_{i_\mathrm{d}}$ & Equation~\eqref{dimrhoI}& $\boldsymbol{J}_{i_\mathrm{d}} \cdot \boldsymbol{n}_\mathrm{d} = 0$ at $\Gamma_\mathrm{d}$
\\
& &$\boldsymbol{J}_{i_\mathrm{d}} \cdot \boldsymbol{n}_\mathrm{d} = \frac{I_{ex,\mathrm{d}}}{e}\left[ \exp\left(-\frac{\alpha_\mathrm{d}e}{k_BT}\eta_\mathrm{d}\right)-\exp\left(\frac{\left(1-\alpha_\mathrm{d}\right)e}{k_BT}\eta_\mathrm{d}\right)\right]$ at $\Sigma_\mathrm{d}$
\\
    $\rho_c$ & Equation~\eqref{rhoCElectrode}&
    $\boldsymbol{j}_{\rho_c} \cdot \boldsymbol{n}_\mathrm{L} = -{I_\mathrm{app}} $ at $\Gamma_\mathrm{L}$
    \\
    & & $\boldsymbol{j}_{\rho_c} \cdot \boldsymbol{n}_\mathrm{R} = \phantom{-}{I_\mathrm{app}} $ at $\Gamma_\mathrm{R}$
    \\
    $\boldsymbol{E}_\mathrm{d}$ & Equation~\eqref{Ed}&$\boldsymbol{E}_\mathrm{d}\cdot \boldsymbol{n}_\mathrm{d} = 0$ at $\Gamma_\mathrm{d}$\\
    \bottomrule
  \end{tabular}
  }
\end{table}

\clearpage
\subsubsection{Monovalent SIC flux reduction}\label{SISIC}
Set $\rho_-=0$, $z_+=1$, and $z_p=-1$, and write $c:=\rho_+$, $p:=\rho_p$, $v:=\phi_{\mathrm V_M}$,
$r:=\rho_+^o/\rho_p^o$, and $\widehat\psi_\mathrm M:=e\psi_\mathrm M/(k_{\mathrm B}T)$. The
conserved total charged-constituent density and charge-number density are
$S:=c+p=\rho_\mathrm{salt,mac}$ and $q_\mathrm n:=c-p=\rho_c/e$. This notation counts polymer
segments, not independently moving salt pairs. The original species equations are
\begin{align}
  \partial_t c&=\nabla\cdot D_+c(\nabla\ln c-\nabla\ln v+\nabla \widehat\psi_\mathrm M),\\
  \partial_t p&=\nabla\cdot D_pp(N^{-1}\nabla\ln p-r\nabla\ln v-\nabla \widehat\psi_\mathrm M).\label{SICP1}
\end{align}
All logarithms in gradient expressions are understood as ratios to uniform positive reference
values. Define $a:=D_+c$, $b:=D_pp$, $M:=a+b$, and $\boldsymbol j:=\boldsymbol j_{\rho_c}$. The
current law gives
\begin{equation}
  \frac{\boldsymbol j}e=-a(\nabla\ln c-\nabla\ln v+\nabla \widehat\psi_\mathrm M)+b(N^{-1}\nabla\ln p-r\nabla\ln v-\nabla \widehat\psi_\mathrm M).\label{ISIC}
\end{equation}
Consequently,
\begin{equation}
  \nabla \widehat\psi_\mathrm M=-\frac{\boldsymbol j}{eM}-\frac{a(\nabla\ln c-\nabla\ln v)-b(N^{-1}\nabla\ln p-r\nabla\ln v)}M.\label{SICNablaPsi}
\end{equation}
Since $\nabla\ln v=-(\nabla c+r\nabla p)/(\rho_+^o v)$, the sum of the chemical driving gradients is
\[
\nabla\ln c+N^{-1}\nabla\ln p-(1+r)\nabla\ln v
=\mathcal A\nabla c+\mathcal B\nabla p,
\]
with $\mathcal A,\mathcal B,\mathcal K=ab/M$ defined in the main text. Substitution into the species
equations gives
\begin{align}
  \partial_t c&=\nabla\cdot\left[\mathcal K(\mathcal A\nabla c+\mathcal B\nabla p)-t_{+,\mathrm V}^{\mathrm{SIC}}\frac{\boldsymbol j}e\right],\\
  \partial_t p&=\nabla\cdot\left[\mathcal K(\mathcal A\nabla c+\mathcal B\nabla p)+t_{p,\mathrm V}^{\mathrm{SIC}}\frac{\boldsymbol j}e\right],\\
  t_{+,\mathrm V}^{\mathrm{SIC}}&=a/M,\qquad t_{p,\mathrm V}^{\mathrm{SIC}}=b/M.\label{transferenceSIC}
\end{align}
Using $c=(S+q_\mathrm n)/2$ and $p=(S-q_\mathrm n)/2$ therefore yields
\begin{align}
  \partial_t S&=\nabla\cdot\left[\mathcal K(\mathcal A+\mathcal B)\nabla S+\mathcal K(\mathcal A-\mathcal B)\nabla q_\mathrm n-(t_{+,\mathrm V}^{\mathrm{SIC}}-t_{p,\mathrm V}^{\mathrm{SIC}})\frac{\boldsymbol j}e\right],\\
  \partial_t p&=\nabla\cdot\left[\frac{\mathcal K}2(\mathcal A+\mathcal B)\nabla S+\frac{\mathcal K}2(\mathcal A-\mathcal B)\nabla q_\mathrm n+t_{p,\mathrm V}^{\mathrm{SIC}}\frac{\boldsymbol j}e\right],\\
  \partial_t\rho_c&=-\nabla\cdot\boldsymbol j.
\end{align}
These expressions give $D_\mathrm{salt,mac}=\mathcal K(\mathcal A+\mathcal B)$ and
$\xi_\mathrm{SIC}=(\mathcal A-\mathcal B)/(\mathcal A+\mathcal B)$. The conductivity is
\begin{equation}
  \sigma_\mathrm{SIC}=e^2M/(k_{\mathrm B}T).\label{SICConductivity}
\end{equation}
Solving Equation~\eqref{ISIC} for the electrolyte electric field gives Equation~\eqref{SICetad}; the
electrode relation remains Equation~\eqref{bulkElectrodeField}. From Equation~\eqref{SICetad},
\begin{align}
     \boldsymbol E_\mathrm M&=\frac{\boldsymbol I_{\mathrm{tot},\mathrm M}}{\sigma_\mathrm{SIC}}-\frac{\epsilon_0\epsilon_\mathrm M}{\sigma_\mathrm{SIC}}\partial_t\boldsymbol E_\mathrm M \nonumber
 \\
  &+\frac{k_BT}{e}\left[\left(\frac{t_{+,\mathrm{V}}^\mathrm{SIC}}{S+q_n}-\frac{t_{p,\mathrm{V}}^\mathrm{SIC}}{N\left(S-q_n\right)}+\frac{t_{+,\mathrm{V}}^\mathrm{SIC}-rt_{p,\mathrm{V}}^\mathrm{SIC}}{1-\frac{1}{2}\left(\frac{1}{\rho_+^o}+\frac{1}{\rho_p^o}\right)S-\frac{1}{2}\left(\frac{1}{\rho_+^o}-\frac{1}{\rho_p^o}\right)q_n}\frac{1}{2}\left(\frac{1}{\rho_+^o}+\frac{1}{\rho_p^o}\right) \right)\nabla S \right.\nonumber\\
    &\quad+\left.\left(\frac{t_{+,\mathrm{V}}^\mathrm{SIC}}{S+q_n}+\frac{t_{p,\mathrm{V}}^\mathrm{SIC}}{N\left(S-q_n\right)}+\frac{t_{+,\mathrm{V}}^\mathrm{SIC}-rt_{p,\mathrm{V}}^\mathrm{SIC}}{1-\frac{1}{2}\left(\frac{1}{\rho_+^o}+\frac{1}{\rho_p^o}\right)S-\frac{1}{2}\left(\frac{1}{\rho_+^o}-\frac{1}{\rho_p^o}\right)q_n}\frac{1}{2}\left(\frac{1}{\rho_+^o}-\frac{1}{\rho_p^o}\right) \right)\nabla q_n \right].\label{EMSISIC}
\end{align}
Then,
\begin{align}
    \epsilon_0 \epsilon_\mathrm{M} \partial_t \boldsymbol{E}_\mathrm{M} &= \boldsymbol{I}_{\mathrm{tot,M}}-\sigma_\mathrm{SIC}\boldsymbol{E}_\mathrm{M} \nonumber\\
    &+\sigma_\mathrm{SIC}\frac{k_BT}{e}\left[\left(\frac{t_{+,\mathrm{V}}^\mathrm{SIC}}{S+q_n}-\frac{t_{p,\mathrm{V}}^\mathrm{SIC}}{N\left(S-q_n\right)}+\frac{t_{+,\mathrm{V}}^\mathrm{SIC}-rt_{p,\mathrm{V}}^\mathrm{SIC}}{1-\frac{1}{2}\left(\frac{1}{\rho_+^o}+\frac{1}{\rho_p^o}\right)S-\frac{1}{2}\left(\frac{1}{\rho_+^o}-\frac{1}{\rho_p^o}\right)q_n}\frac{1}{2}\left(\frac{1}{\rho_+^o}+\frac{1}{\rho_p^o}\right) \right)\nabla S \right.\nonumber\\
    &\quad+\left.\left(\frac{t_{+,\mathrm{V}}^\mathrm{SIC}}{S+q_n}+\frac{t_{p,\mathrm{V}}^\mathrm{SIC}}{N\left(S-q_n\right)}+\frac{t_{+,\mathrm{V}}^\mathrm{SIC}-rt_{p,\mathrm{V}}^\mathrm{SIC}}{1-\frac{1}{2}\left(\frac{1}{\rho_+^o}+\frac{1}{\rho_p^o}\right)S-\frac{1}{2}\left(\frac{1}{\rho_+^o}-\frac{1}{\rho_p^o}\right)q_n}\frac{1}{2}\left(\frac{1}{\rho_+^o}-\frac{1}{\rho_p^o}\right) \right)\nabla q_n \right].
\end{align}
From the relationship between $\boldsymbol{E}_\mathrm{M}, \rho_c$, 
\begin{align}
    \frac{\partial \rho_c}{\partial t} &= \nabla \cdot \left[\boldsymbol{I}_{\mathrm{tot,M}}-\sigma_\mathrm{SIC}\boldsymbol{E}_\mathrm{M} \right.\nonumber\\
    &+\left.\sigma_\mathrm{SIC}\frac{k_BT}{e}\left[\left(\frac{t_{+,\mathrm{V}}^\mathrm{SIC}}{S+q_n}-\frac{t_{p,\mathrm{V}}^\mathrm{SIC}}{N\left(S-q_n\right)}+\frac{t_{+,\mathrm{V}}^\mathrm{SIC}-rt_{p,\mathrm{V}}^\mathrm{SIC}}{1-\frac{1}{2}\left(\frac{1}{\rho_+^o}+\frac{1}{\rho_p^o}\right)S-\frac{1}{2}\left(\frac{1}{\rho_+^o}-\frac{1}{\rho_p^o}\right)q_n}\frac{1}{2}\left(\frac{1}{\rho_+^o}+\frac{1}{\rho_p^o}\right) \right)\nabla S \right.\right.\nonumber\\
    &\quad+\left.\left.\left(\frac{t_{+,\mathrm{V}}^\mathrm{SIC}}{S+q_n}+\frac{t_{p,\mathrm{V}}^\mathrm{SIC}}{N\left(S-q_n\right)}+\frac{t_{+,\mathrm{V}}^\mathrm{SIC}-rt_{p,\mathrm{V}}^\mathrm{SIC}}{1-\frac{1}{2}\left(\frac{1}{\rho_+^o}+\frac{1}{\rho_p^o}\right)S-\frac{1}{2}\left(\frac{1}{\rho_+^o}-\frac{1}{\rho_p^o}\right)q_n}\frac{1}{2}\left(\frac{1}{\rho_+^o}-\frac{1}{\rho_p^o}\right) \right)\nabla q_n \right] \right]\label{rhoCSISIC}.
\end{align}

We use these separate bulk
relations, with interfacial $\eta_\mathrm d$ determined only by Equation~\eqref{eta}.
In particular, from~\eqref{eq:eleceq},
\begin{align}
     -\Delta \psi_\mathrm{d}^\mathrm{eq} = -\left(\widehat E_\mathrm d^\circ+\frac{k_{\mathrm{B}} T}{e}\ln \left[\left(
  \frac{\frac{\rho_\mathrm{salt, mac}+\frac{\rho_c}{e}}{2}}{\rho_+^o-
\frac{\rho_\mathrm{salt, mac}+\frac{\rho_c}{e}}{2}
  -\frac{\rho_+^o}{\rho_p^o}
  \frac{\rho_\mathrm{salt, mac}-\frac{\rho_c}{e}}{2}}\right)\frac{ q_\mathrm{d}-\frac{\rho_c}{e}}{\rho_+^o}\frac{\rho_{i_\mathrm{d}}^o-\rho_{i_\mathrm{d}}}{\rho_{i_\mathrm{d}}}
\right]\right),\label{eq:eleceqSISIC}
\end{align} and, as observed in Section~\ref{sec:recapturingKnownDynamics}, $\eta_\mathrm{d} = -\Delta \psi_\mathrm{d}^\mathrm{eq}$.

High electrode conductivity and rapid electrode dielectric relaxation can make the electrode field
higher order, subject to the same checks as for the DIC model.
Define $S^{\mathrm{eq}}:=\rho_\mathrm{salt,mac}^{\mathrm{eq}}$ and $v^{\mathrm{eq}}:=\rho_{\mathrm
V_M}^{\mathrm{eq}}/\rho_+^o$, both positive spatially uniform reference values. For the formal
electroneutral bulk reduction, $c=p=S/2$, and
\begin{align}
  \boldsymbol E_\mathrm M&=\frac{\boldsymbol I_{\mathrm{tot},\mathrm M}}{\sigma_\mathrm{SIC}}+\frac{k_{\mathrm B}T}{e}\left[\left(t_{+,\mathrm V}^{\mathrm{SIC}}-\frac{t_{p,\mathrm V}^{\mathrm{SIC}}}N\right)\nabla\ln\frac{S}{S^{\mathrm{eq}}}-(t_{+,\mathrm V}^{\mathrm{SIC}}-r t_{p,\mathrm V}^{\mathrm{SIC}})\nabla\ln\frac{1-\frac{S}{2}\left(\frac{1}{\rho_+^o}+\frac{1}{\rho_p^o}\right)}{1-\frac{S^{\mathrm{eq}}}{2}\left(\frac{1}{\rho_+^o}+\frac{1}{\rho_p^o}\right)}\right]\nonumber\\
  &\quad-\frac{\epsilon_0\epsilon_\mathrm M}{\sigma_\mathrm{SIC}}\partial_t\boldsymbol E_\mathrm M\label{SICetaGovEqnLocalElectroneutrality}\\
  -\Delta \psi_\mathrm{d}^\mathrm{eq} &= -\left(\widehat E_\mathrm d^\circ+\frac{k_{\mathrm{B}} T}{e}\ln \left[\left(
  \frac{\rho_\mathrm{salt, mac}}{2\rho_+^o-\left(1+\frac{\rho_+^o}{\rho_p^o}\right)
{\rho_\mathrm{salt, mac}}}\right)\frac{ q_\mathrm{d}}{\rho_+^o}\frac{\rho_{i_\mathrm{d}}^o-\rho_{i_\mathrm{d}}}{\rho_{i_\mathrm{d}}}
\right]\right)\label{SICetaLocalElectroneutrality}.
\end{align}

For a regular quasistatic bulk limit the last term is higher order. At $t_{+,\mathrm
V}^{\mathrm{SIC}}=1$ this is only a formal bulk reduction; effective interfacial conditions must be
obtained by matching through the space-charge layers.

\clearpage

\begin{table}
  \caption{Governing equations and boundary conditions in the SIC model.}
  \label{tab:SICElectrolytetab}
  \setlength{\tabcolsep}{5pt}
  \scriptsize
  \begin{tabular}{@{}lll@{}}
    \toprule
    {Variable} & {Equation} &{Boundary condition}\\
    \midrule
    & Electrolyte & \\
    $\rho_{\mathrm{salt,mac}}$ & $\frac{\partial \rho_\mathrm{salt,mac}}{\partial t} = \nabla \cdot \left[ {D}_\mathrm{salt,mac}\nabla \rho_{\mathrm{salt,mac}} \right.$ & $\left[ {D}_\mathrm{salt,mac}\nabla \rho_{\mathrm{salt,mac}}\right.$\\
    & $\left.\qquad+\xi_\mathrm{SIC}{D}_\mathrm{salt,mac}\nabla \left(\frac{\rho_c}{e}\right) - \left(t_{+,\mathrm{V}}^\mathrm{SIC}-t_{p,\mathrm{V}}^\mathrm{SIC}\right) \frac{\boldsymbol{j}_{\rho_c}}{e}\right] $ &
    $\left.\qquad+\xi_\mathrm{SIC}{D}_\mathrm{salt,mac}\nabla \left(\frac{\rho_c}{e}\right) - \left(t_{+,\mathrm{V}}^\mathrm{SIC}-t_{p,\mathrm{V}}^\mathrm{SIC}\right)\frac{\boldsymbol{j}_{\rho_c}}{e}\right]
    \cdot \boldsymbol{n}_\mathrm{M}$\\
    & & $\qquad=-\frac{I_{ex,\mathrm{d}}}{e}\left[ \exp\left(-\frac{\alpha_\mathrm{d}e}{k_BT}\eta_\mathrm{d}\right)-\exp\left(\frac{\left(1-\alpha_\mathrm{d}\right)e}{k_BT}\eta_\mathrm{d}\right)\right]
    $ on $\Sigma_\mathrm{d}$\\
    $\rho_c$ &Equation~\eqref{rhoCSISIC} & {$\boldsymbol{j}_{\rho_c}\cdot \boldsymbol{n}_\mathrm{M} = -{I_{ex,\mathrm{d}}}\left[ \exp\left(-\frac{\alpha_\mathrm{d}e}{k_BT}\eta_\mathrm{d}\right)-\exp\left(\frac{\left(1-\alpha_\mathrm{d}\right)e}{k_BT}\eta_\mathrm{d}\right)\right]$} at $\Sigma_\mathrm
    d$ \\
    $\boldsymbol E_\mathrm M$ & Equation~\eqref{EMSISIC} & $\epsilon_\mathrm{M} \boldsymbol{E}_\mathrm{M}\cdot\boldsymbol{n}_\mathrm{M} + \epsilon_\mathrm{d} \boldsymbol{E}_\mathrm{d} \cdot \boldsymbol{n}_\mathrm{d} = 0$ at $\Sigma_\mathrm{d}$\\
      $\eta_\mathrm{d}$ & Equation~\eqref{eq:eleceqSISIC}& --\\
    & Electrode & \\
  $\rho_{i_\mathrm{d}}$ & Equation~\eqref{dimrhoI}& $\boldsymbol{J}_{i_\mathrm{d}} \cdot \boldsymbol{n}_\mathrm{d} = 0$ at $\Gamma_\mathrm{d}$
\\
& &$\boldsymbol{J}_{i_\mathrm{d}} \cdot \boldsymbol{n}_\mathrm{d} = \frac{I_{ex,\mathrm{d}}}{e}\left[ \exp\left(-\frac{\alpha_\mathrm{d}e}{k_BT}\eta_\mathrm{d}\right)-\exp\left(\frac{\left(1-\alpha_\mathrm{d}\right)e}{k_BT}\eta_\mathrm{d}\right)\right]$ at $\Sigma_\mathrm{d}$
\\
    $\rho_c$ & Equation~\eqref{rhoCElectrode}&
    $\boldsymbol{j}_{\rho_c} \cdot \boldsymbol{n}_\mathrm{L} = {-}{I_\mathrm{app}} $ at $\Gamma_\mathrm{L}$
    \\
    & & $\boldsymbol{j}_{\rho_c} \cdot \boldsymbol{n}_\mathrm{R} = \phantom{-}{I_\mathrm{app}} $ at $\Gamma_\mathrm{R}$
    \\
    $\boldsymbol{E}_\mathrm{d}$ & Equation~\eqref{Ed}&$\boldsymbol{E}_\mathrm{d}\cdot \boldsymbol{n}_\mathrm{d} = 0$ at $\Gamma_\mathrm{d}$\\
    \bottomrule
  \end{tabular}
\end{table}
\begin{table}
  \caption{{Formal bulk equations for the locally electroneutral SIC limiting case; interface
  entries require matching to space-charge layers when $t_{+,\mathrm V}^{\mathrm{SIC}}=1$, assuming
  $\sigma_\mathrm{SIC}/\sigma_\mathrm d\ll1$, negligible
  $\nabla\ln(\rho_{e_\mathrm{d}}/\rho_{e_\mathrm{d}}^{\mathrm{eq}})$, and
  $\epsilon_0\epsilon_\mathrm d/(\sigma_\mathrm d\tau)\ll1$.}}
  \label{tab:SICElectrolyteLocalElectroneutraltab}
  \begin{tabular}{@{}lll@{}}
    \toprule
    {Variable} & {Equation} &{Boundary condition}\\
    \midrule
    & Electrolyte & \\
    $\rho_{\mathrm{salt,mac}}$ & $\frac{\partial \rho_\mathrm{salt,mac}}{\partial t} = \nabla \cdot \left[ {D}_\mathrm{salt,mac}\nabla \rho_{\mathrm{salt,mac}} \right.$ & $\left[ {D}_\mathrm{salt,mac}\nabla \rho_{\mathrm{salt,mac}}- \left(t_{+,\mathrm{V}}^\mathrm{SIC}- t_{p,\mathrm{V}}^\mathrm{SIC}\right)\frac{\boldsymbol{j}_{\rho_c}}{e} \right]
    \cdot \boldsymbol{n}_\mathrm{M}$\\
    &- $\left.\left(t_{+,\mathrm{V}}^\mathrm{SIC}- t_{p,\mathrm{V}}^\mathrm{SIC}\right)\frac{\boldsymbol{j}_{\rho_c}}{e} \right]$&$=-\frac{I_{\mathrm{ex,d}}}{e}\left[ \exp\left(-\frac{\alpha_\mathrm{d}e}{k_BT}\eta_\mathrm{d}\right)-\exp\left(\frac{\left(1-\alpha_\mathrm{d}\right)e}{k_BT}\eta_\mathrm{d}\right)\right]
    $ at $\Sigma_\mathrm{d}$\\
    $\boldsymbol E_\mathrm M$ & Equation~\eqref{SICetaGovEqnLocalElectroneutrality}& $\epsilon_\mathrm{M} \boldsymbol{E}_\mathrm{M} \cdot \boldsymbol{n}_\mathrm{M}+\epsilon_\mathrm{d} \boldsymbol{E}_\mathrm{d} \cdot \boldsymbol{n}_\mathrm{d} = 0$ at $\Sigma_\mathrm{d}$\\
    $\eta_\mathrm{d}$ & Equation~\eqref{SICetaLocalElectroneutrality} & \\
    & Electrode & \\
  $\rho_{i_\mathrm{d}}$ & Equation~\eqref{dimrhoI}& $\boldsymbol{J}_{i_\mathrm{d}} \cdot \boldsymbol{n}_\mathrm{d} = 0$ at $\Gamma_\mathrm{d}$
\\
& &$\boldsymbol{J}_{i_\mathrm{d}} \cdot \boldsymbol{n}_\mathrm{d} = \frac{I_{ex,\mathrm{d}}}{e}\left[ \exp\left(-\frac{\alpha_\mathrm{d}e}{k_BT}\eta_\mathrm{d}\right)-\exp\left(\frac{\left(1-\alpha_\mathrm{d}\right)e}{k_BT}\eta_\mathrm{d}\right)\right]$ at $\Sigma_\mathrm{d}$
\\
    $\boldsymbol{E}_\mathrm{d}$ & Equation~\eqref{Ed}&$\boldsymbol{E}_\mathrm{d}\cdot \boldsymbol{n}_\mathrm{d} = 0$ at $\Gamma_\mathrm{d}$\\
    \toprule
  \end{tabular}
\end{table}

The overpotential $\eta_\mathrm d$ at each interface $\Sigma_\mathrm d$ is
part of the solution. It is determined by its definition in Equation~\eqref{eta},
the Butler--Volmer law~\eqref{reactionRateBV}, and the reaction-flux conditions.
Setting $\eta_\mathrm d=0$ would put the reaction at equilibrium and give zero
net reaction rate, $\nu_\mathrm d=0$. We therefore do not impose this condition
at an interface where charge transfer occurs.

The electroneutral SIC equations in Table~\ref{tab:SICElectrolyteLocalElectroneutraltab}
apply in the bulk electrolyte, away from the electrode interfaces. When
$t_{+,\mathrm V}^{\mathrm{SIC}}=1$, the polymer is immobile ($D_p=0$).
Cations can cross the interface, but the charged polymer cannot, so charge
separation must be allowed near the interface during charge transfer.
Poisson's equation allows the model to describe charge separation in this layer. To use an electroneutral
bulk model instead, one must derive boundary conditions that account for the
layer by matching the solutions inside and outside it. We have not derived
those conditions here. The interface conditions listed in the table apply when Poisson's equation
is solved alongside the species transport equations. They cannot simply be
carried over to a model that imposes local electroneutrality.

\clearpage
\makeatletter
\renewcommand{\thesubsubsection}{\thesubsection.\arabic{subsubsection}}
\renewcommand{\p@subsubsection}{}
\makeatother
\subsection{Conservation and numerical accuracy of the DIC and SIC simulations}
\label{sec:numericalChecksSI}
The checks below use the parameters and current protocols in
Section~\ref{sec:numerical-simulations}. They compare the 960-element base
mesh, a calculation with half the maximum time step on that mesh, and the
1920-element mesh used for the main-text figures. Conservation diagnostics
and refinement comparisons assess different aspects of numerical accuracy;
neither alone establishes full convergence of the interfacial layers.
Global electroneutrality is assessed by monitoring the absolute normalized
net cell charge at every accepted time step, using $Q^*$ for DIC and
$Q^*_{\rm SIC}$ for SIC as defined below. The maximum absolute residuals
are reported in Tables~\ref{tab:dicNumericalChecks} and~\ref{tab:sicNumericalChecks}.
These diagnostics check neutrality of the complete cell, while permitting
local space charge and nonzero net charge in individual materials.

Continuous linear finite elements are cosine-graded separately in each
material to resolve the interfacial layers. Both plotted solutions use
$320/1280/320$ elements in the left electrode, electrolyte, and right electrode,
respectively. The smallest element is $1.5060\times10^{-6}$ and the largest
is $1.2272\times10^{-3}$. Time integration uses adaptive BDF2 with initial
step $10^{-8}$ and maximum step $5\times10^{-5}$. Nonlinear absolute and
relative tolerances are $10^{-10}$ and $10^{-8}$. Automatic residual scaling
is recomputed during the calculation, and a reduced-space active-set Newton
method enforces the polymer lower bound. Fields are saved every
$6.25\times10^{-4}$, including the initial and final states.

Figures~\ref{fig:dicVerification} and~\ref{fig:sicVerification} compare
three discretizations of each model. The base calculation uses 960 elements
and maximum time step $5\times10^{-5}$. The temporal check keeps that mesh
and halves the maximum step to $2.5\times10^{-5}$; the spatial check doubles
the element count to 1920 while retaining the base maximum step. Because
time stepping is adaptive, these are step-size limits, not fixed steps.
All other parameters and the applied-current protocol are held fixed within
each three-run comparison.

The upper panels show mean right-electrode utilization, defined by
Equation~\eqref{eq:numericalUtilization}; the lower panels show the terminal
electrochemical potential difference from Equation~\eqref{valueECBCNonDim},
including the terminal electron-density contribution. Both are plotted
against nondimensional time. The vertical dotted line marks current
reversal at $\hat t=0.05$. The legend in the upper panel applies to both panels.
For each output, reported differences are maximum absolute differences
between the check and base histories over the saved times, rather than
errors relative to an exact solution. The utilization bounds quoted below
cover both electrodes, although only the right electrode is plotted.
Pointwise field comparisons interpolate the base solution onto the refined
coordinates separately within each material. Nearly coincident curves can
conceal small discrepancies at the plotted scale; the numerical differences
below quantify them.

\subsubsection{DIC conservation and resolution checks}\label{sec:dicNumericalChecksSI}
We monitor the nondimensional total charge
\begin{equation}
Q^*=\int_0^1(\phi_+-\phi_-)\,\mathrm d\hat x+
\sum_{\mathrm d\in\{\mathrm L,\mathrm R\}}
\int_{\hat\Omega_\mathrm d}(1.5-\phi_{e_\mathrm d})\,\mathrm d\hat x\label{QDIC}
\end{equation}
and the lithium inventory
$L^*=\int_0^1\phi_+\,\mathrm d\hat x+
\sum_\mathrm d\int_{\hat\Omega_\mathrm d}\phi_{i_\mathrm d}\,\mathrm d\hat x$.
Table~\ref{tab:dicNumericalChecks} compares the base calculation, its
half-time-step check, and the refined mesh used for the figures. All three
complete the cycle. Charge and lithium diagnostics cover all accepted
steps; polymer inventory is integrated over the saved finite-element fields.
The refined-mesh anion inventory error is at most $9.99\times10^{-15}$.
\begin{table*}
\caption{DIC conservation errors at current magnitude $1.6$.
$L^*=\int_0^1 \phi_+ \rm{d} \hat{x} + \sum_{\rm{d} \in \left\{\rm{L}, \rm {R}\right\}} \int_{\hat{\Omega}_{\rm{d}}} \phi_{i_{\rm{d}}} \rm{d} \hat{x}$, $P^*=\int_0^1\phi_p\,\mathrm d\hat x$, $\Delta P^*=P^*(\hat t)-P^*(0)$, and $Q^*$ is as defined in Equation~\eqref{QDIC}.
The base and half-step meshes have $160/640/160$ elements; the refined mesh
has $320/1280/320$.}
\label{tab:dicNumericalChecks}
\begin{ruledtabular}\begin{tabular}{lccccc}
Run & Elements & $\Delta\hat t_{\max}$ & $\max|Q^*|$ & $\max|L^*-L^*(0)|$ & $\max|\Delta P^*|/P^*(0)$\\
\midrule
Base & 960 & $5\times10^{-5}$ & $2.19\times10^{-14}$ & $2.00\times10^{-14}$ & $4.27\times10^{-3}$\\
Half step & 960 & $2.5\times10^{-5}$ & $1.89\times10^{-13}$ & $1.30\times10^{-13}$ & $4.27\times10^{-3}$\\
Refined mesh & 1920 & $5\times10^{-5}$ & $2.59\times10^{-14}$ & $2.00\times10^{-14}$ & $3.24\times10^{-5}$\\
\end{tabular}\end{ruledtabular}
\end{table*}

Halving the maximum time step on the base mesh changes either electrode's
mean utilization by at most $1.59\times10^{-5}$ at the saved times. The
maximum terminal electrochemical potential difference between those runs
is $0.5764$ at reversal and $0.04560$ when that exact sample is excluded.
Doubling the mesh changes mean utilization by at most $3.87\times10^{-5}$
and terminal electrochemical potential by $0.04457$.
Figure~\ref{fig:dicVerification} shows that the right-electrode utilization
histories nearly coincide, while the potential is more sensitive near the
abrupt current reversal. Excluding the sample exactly at reversal removes
only that sample, not a finite time interval around the switch; it does not
establish complete temporal resolution of the switching transient.
Pointwise differences between meshes reach $0.01202$, $0.01898$, and
$0.04382$ for the cation, anion, and polymer fractions, respectively.
Thus agreement of integrated utilization does not establish convergence
of the thin interfacial layers or the switching transient.
The polymer inventory drift decreases from $0.427\%$ on the base mesh to
$0.00324\%$ on the refined mesh. Because the polymer positivity constraint
is active, exact discrete polymer conservation is not claimed.
\begin{figure*}
\includegraphics[width=0.85\textwidth]{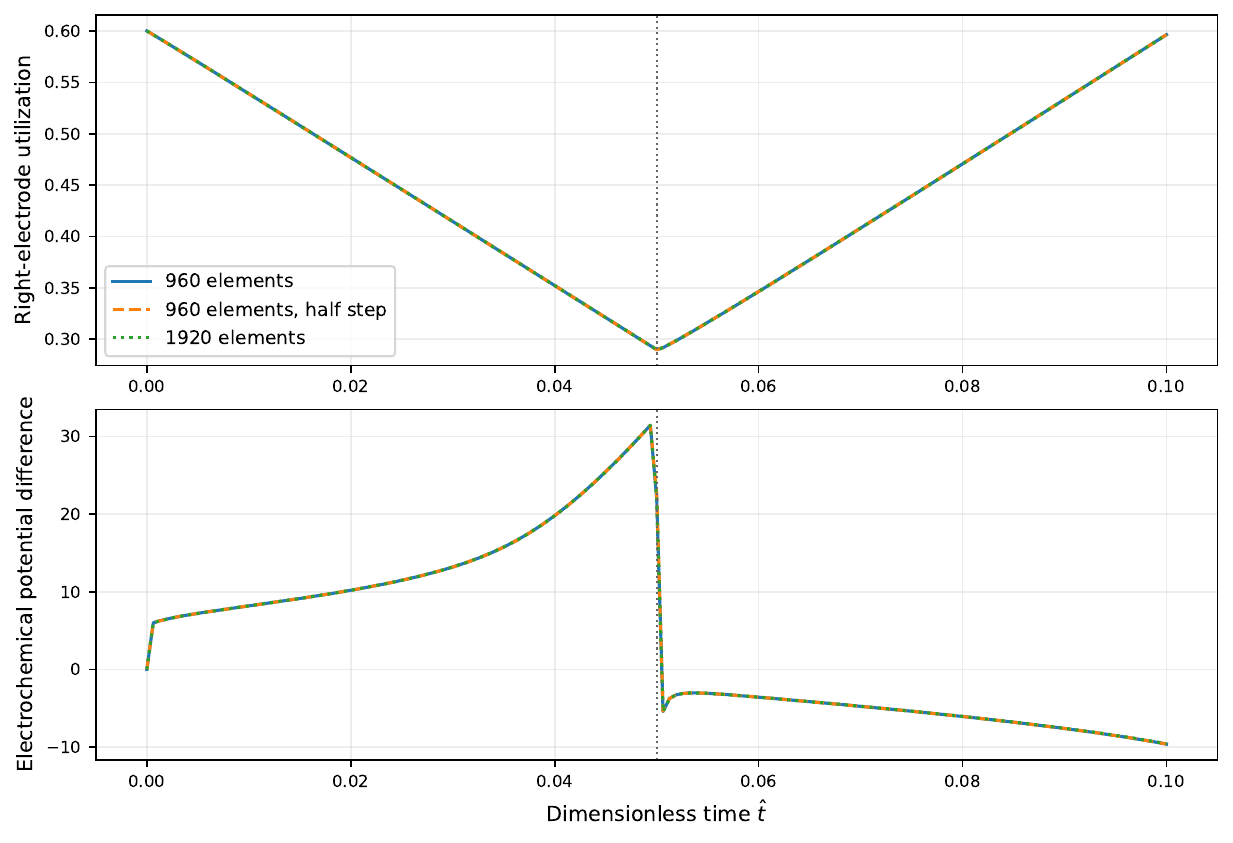}
\caption{DIC mesh and time-step checks at current magnitude $1.6$.
Upper panel: mean right-electrode utilization. Lower panel: terminal
electrochemical potential difference, including the electron-density
contribution. Solid curves use 960 elements with maximum time step
$5\times10^{-5}$; dashed curves use 960 elements with maximum step
$2.5\times10^{-5}$; dotted curves use 1920 elements with maximum step
$5\times10^{-5}$. The upper-panel legend applies to both panels.
The vertical gray dotted line marks current reversal at $\hat t=0.05$;
the cycle ends at $\hat t=0.10$. All other parameters are identical.
The histories nearly overlap at this scale; the text reports their maximum
absolute differences and the remaining sensitivity of the switching transient.}
\label{fig:dicVerification}
\end{figure*}

\clearpage
\subsubsection{SIC conservation and resolution checks}\label{sec:sicNumericalChecksSI}
For SIC, the normalized total charge is
\begin{equation}
Q^*_{\rm SIC}=\int_0^1\left(\phi_+-\frac{\rho_p^o}{\rho_+^o}\phi_p\right)\,\mathrm d\hat x+
\sum_{\mathrm d\in\{\mathrm L,\mathrm R\}}
\int_{\hat\Omega_\mathrm d}(1.5-\phi_{e_\mathrm d})\,\mathrm d\hat x. \label{QSIC}
\end{equation}
The lithium inventory is defined as in DIC, and the normalized polymer
inventory is $P^*=\int_0^1\phi_p\,\mathrm d\hat x$.
All three cycles completed without solver cutbacks. A separate zero-current
check through $\hat t=0.001$ preserved the uniform equilibrium fields,
with zero potential and maximum absolute charge $5.55\times10^{-17}$.
Table~\ref{tab:sicNumericalChecks} reports conservation over all accepted
steps. Unlike the DIC calculation, the polymer lower bound is not reached
in the saved SIC fields, and polymer inventory is conserved to solver accuracy.
\begin{table*}
\caption{SIC conservation errors at current magnitude $1.6$. $L^*=\int_0^1 \phi_+ \rm{d} \hat{x} + \sum_{\rm{d} \in \left\{\rm{L}, \rm {R}\right\}} \int_{\hat{\Omega}_{\rm{d}}} \phi_{i_{\rm{d}}} \rm{d} \hat{x}$, $P^*=\int_0^1\phi_p\,\mathrm d\hat x$, and $Q^*_{\rm {SIC}}$ is as defined in Equation~\eqref{QSIC}.
The base and half-step meshes have $160/640/160$ elements; the refined mesh
has $320/1280/320$.}
\label{tab:sicNumericalChecks}
\begin{ruledtabular}\begin{tabular}{lccccc}
Run & Elements & $\Delta \hat t_{\rm max}$ & $\max|Q^*_{\rm SIC}|$ & $\max|L^*-L^*(0)|$ & $\max|P^*-P^*(0)|$\\
\midrule
Base & 960 & $5 \times 10^{-5}$ & $2.76\times10^{-14}$ & $9.02\times10^{-15}$ & $2.00\times10^{-14}$\\
Half step & 960 & $2.5 \times 10^{-5}$& $1.92\times10^{-13}$ & $5.50\times10^{-14}$ & $5.00\times10^{-14}$\\
Refined mesh & 1920 & $5 \times 10^{-5}$ & $5.33\times10^{-14}$ & $9.02\times10^{-15}$ & $2.00\times10^{-14}$\\
\end{tabular}\end{ruledtabular}
\end{table*}
Halving the maximum time step changes mean utilization by at most
$1.72\times10^{-5}$ at the saved times and terminal electrochemical
potential by $2.1201$ ($0.5978$ excluding the exact reversal sample).
Doubling the mesh changes these quantities by at most
$7.53\times10^{-6}$ and $0.2296$, respectively.
Pointwise polymer-fraction differences reach $0.002801$ between meshes.
Figure~\ref{fig:sicVerification} shows nearly coincident right-electrode
utilization histories, including the delayed recovery after reversal.
The large potential peak persists in all three runs. Its small relative
variation can be difficult to distinguish on the full potential scale,
so the absolute differences above are needed to assess numerical sensitivity.
The comparison excluding reversal omits only the exact switching sample.
The strong polarization persists under these checks, but they do not
establish an asymptotic convergence rate or complete resolution of the
switching transient. Material calibration and assessment of the model
assumptions in the depleted regime are needed before quantitative physical
predictions are inferred.
\begin{figure*}
\includegraphics[width=0.85\textwidth]{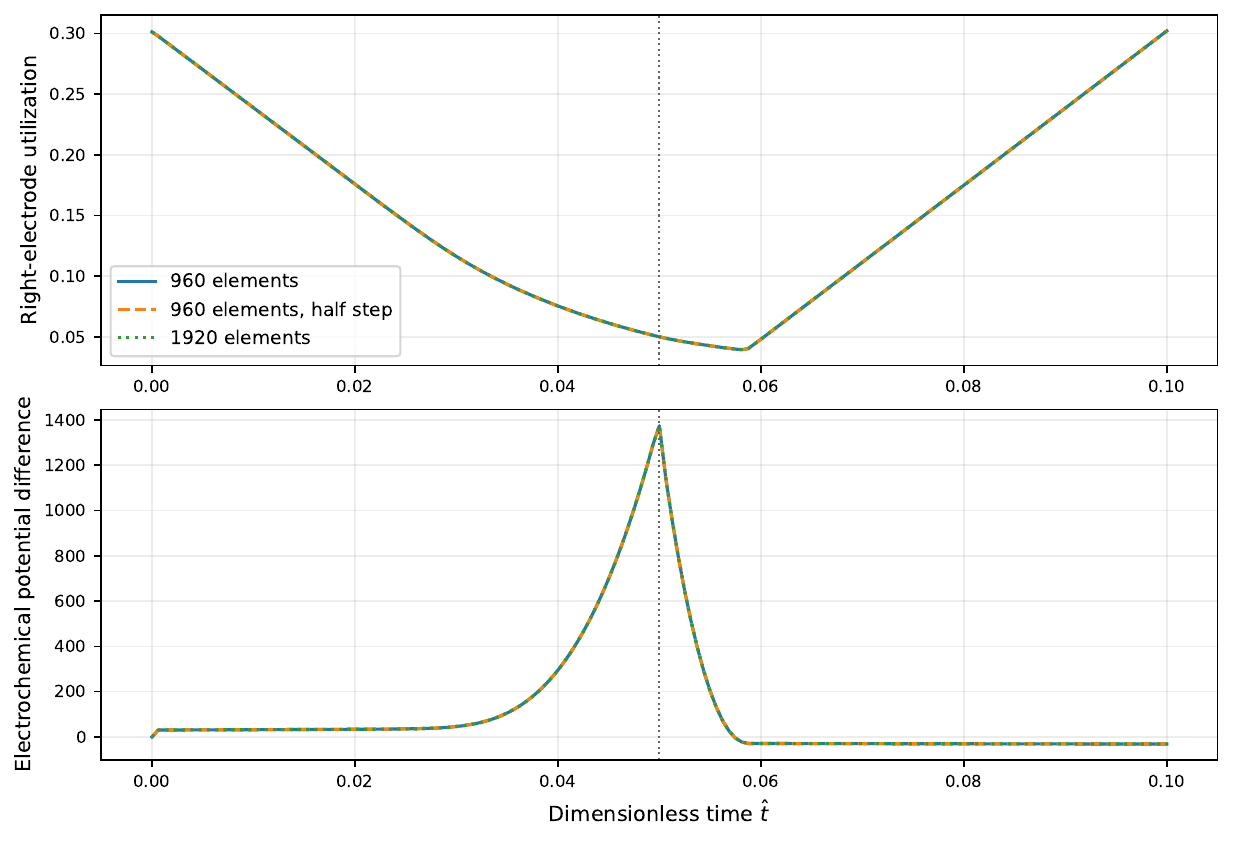}
\caption{SIC mesh and time-step checks at current magnitude $1.6$.
Upper panel: mean right-electrode utilization. Lower panel: terminal
electrochemical potential difference, including the electron-density
contribution. Solid curves use 960 elements with maximum time step
$5\times10^{-5}$; dashed curves use 960 elements with maximum step
$2.5\times10^{-5}$; dotted curves use 1920 elements with maximum step
$5\times10^{-5}$. The upper-panel legend applies to both panels.
The vertical gray dotted line marks current reversal at $\hat t=0.05$;
the cycle ends at $\hat t=0.10$. All other parameters are identical.
Delayed right-electrode recovery and the large potential peak persist
under both checks. Near-overlap does not establish full convergence;
absolute differences and limitations are discussed in the text.}
\label{fig:sicVerification}
\end{figure*}

\end{document}